\documentclass{jfm}

\usepackage{graphicx}
\usepackage{natbib}
\usepackage{hyperref}
\usepackage{epstopdf, epsfig,amsmath}
\usepackage[percent]{overpic}
\usepackage{subfigure}
\usepackage{array}
\usepackage{xcolor}
\usepackage{tikz}
\usepackage{caption}
\usepackage{color}

\usepackage{multirow}

\usepackage{lscape}

\hypersetup{
	colorlinks = true,
	urlcolor   = blue,
	citecolor  = blue,
}

\newcommand{\RomanNumeralCaps}[1]
\nolinenumbers

\definecolor{purple0}{rgb}{0.63,0.13,0.94}
\definecolor{c00}{rgb}{0,1,1}
\definecolor{m00}{rgb}{1,0,1}
\definecolor{g00}{rgb}{0,0.5,0}
\definecolor{brown0}{rgb}{0.8549,0.64706,0.12549}

\title{Reinforcement-learning-based control of turbulent channel flows at high Reynolds numbers}

\author{Zisong Zhou\aff{1},
  Mengqi Zhang\aff{2},
  \and Xiaojue Zhu\aff{1}
  \corresp{\email{zhux@mps.mpg.de}}}

\affiliation{
\aff{1}Max Planck Institute for Solar System Research, 37077 Göttingen, Germany
\aff{2}Department of Mechanical Engineering, National University of Singapore, 9 Engineering Drive 1, 117575, Republic of Singapore
}

\begin{document}

\maketitle

\begin{abstract}

Deep reinforcement learning (DRL) is employed to develop control strategies for drag reduction in direct numerical simulations (DNS) of turbulent channel flows at high Reynolds numbers. 
The DRL agent uses near-wall streamwise velocity fluctuations as input to modulate wall blowing and suction velocities. 
These DRL-based strategies achieve significant drag reduction, with maximum rates of $35.6\%$ at $Re_{\tau}\thickapprox180$, $30.4\%$ at $Re_{\tau}\thickapprox550$, and $27.7\%$ at $Re_{\tau}\thickapprox1000$, outperforming traditional opposition control methods.
Expanded range of wall actions further enhances drag reduction, although effectiveness decreases at higher Reynolds numbers.
The DRL models elevate the virtual wall through blowing and suction, aiding in drag reduction. 
However, at higher Reynolds numbers, the amplitude modulation of large-scale structures significantly increases the residual Reynolds stress on the virtual wall, diminishing the drag reduction. 
Analysis of budget equations provides a systematic understanding of the drag reduction dynamics behind. 
DRL models reduce skin friction by inhibiting the redistribution of wall-normal turbulent kinetic energy.
This further suppresses the wall-normal velocity fluctuations, reducing the production of Reynolds stress, thereby decreasing skin friction.
This study showcases the successful application of DRL in turbulence control at high Reynolds numbers and elucidates the nonlinear control mechanisms underlying the observed drag reduction.


\end{abstract}


\section{Introduction}\label{sec:intro}

Turbulent flows lead to significantly greater energy losses compared to laminar flows, presenting a major challenge in various engineering applications \citep{brunton2015closed}. 
For instance, wall friction contributes to approximately $50\%$ of total resistance in aircraft, up to $90\%$ in submarines, and nearly all resistance in pipeline flows \citep{gad1989selective}. 
These applications typically operate under high Reynolds number conditions, making the turbulent drag reduction at high Reynolds numbers both theoretically significant and practically valuable.

In wall-bounded turbulence, coherent structures are strongly associated with high skin friction \citep{kravchenko1993relation,choi1994active,xu2005transient}, leading to the concept that real-time control of these structures could effectively potentially suppress turbulence and reduce skin friction. 
At low Reynolds numbers, the near-wall region is dominated by velocity streaks and quasi-streamwise vortices.
These structures are cyclically generated through a self-sustaining process \citep{jimenez1991minimal,hamilton1995regeneration}, which can persist even without turbulence in the outer region \citep{jimenez1999autonomous}.
The work by \cite{choi1994active} pioneered an active control method targeting the streamwise vortices, known as the opposition control strategy. 
In this method, the wall-normal velocity fluctuations are monitored on a hypothetical detection plane in the near-wall region. 
Based on these detected signals, counteracting wall-normal blowing and suction velocities are applied at the wall to suppress the ejection and sweep events caused by streamwise vortices, thereby reducing Reynolds shear stress and achieving drag reduction. 
The effectiveness of opposition control was confirmed by DNS of turbulent channel flows, as demonstrated by \cite{choi1994active}, which showed a maximum drag reduction rate of approximately $25\%$ at a friction Reynolds number of $Re_{\tau} = 180$. 
Subsequent investigations by \cite{hammond1998observed} and \cite{chung2011effectiveness} further elucidated the mechanisms behind this drag reduction. 
They found that wall-normal blowing and suction significantly limited momentum transport toward the wall, effectively creating a `virtual wall' that hindered high-speed fluid motions towards the wall induced by streamwise vortices, thus reducing local high friction drag.
Building on the concept of opposition control, various other strategies have been developed. 
These include neural network-based control schemes \citep{lee1997application} and suboptimal control schemes \citep{lee1998suboptimal,fukagata2004suboptimal,hasegawa2011dissimilar}, which utilize measurable wall quantities to achieve drag reduction.

As Reynolds numbers increase, the efficacy of drag reduction schemes, such as opposition control, markedly declines.
For instance, in turbulent channel flows, the maximum drag reduction rate achieved by opposition control decreases from $25\%$ at $Re_{\tau} = 180$ to $18\%$ at $Re_{\tau} = 720$ \citep{iwamoto2002reynolds,chang2002viscous,pamies2007response}. 
As Reynolds numbers rise, large-scale structures and very-large-scale structures emerge in the logarithmic and outer regions \citep{jimctr98,kim1999very,del2003spectra,del2004spectra,guala2006large,balakumar2007large,hutchins2007evidence,monty2009comparison}. 
\cite{hwang2013near} suggested that these structures contribute to Reynolds shear stress, thereby diminishing drag reduction rates. 
Furthermore, \cite{mathis2009large} classified the influence of outer large-scale structures on near-wall turbulence into two effects: superposition and amplitude modulation.
The superposition effect, a linear process, represents the footprint of large-scale structures on near-wall turbulence \citep{hoyas06,hutchins2007large}. 
These large-scale structures extend deeply into the near-wall region, significantly contributing to turbulent kinetic energy \citep{hoyas06,mathis2009large,marusic2010high}. 
On the other hand, amplitude modulation is a nonlinear process that describes how small-scale turbulent fluctuations are intensified in large-scale high-speed regions and suppressed in low-speed regions.
\cite{deng2012influence} highlighted that the reduced drag reduction rate at high Reynolds numbers is primarily due to the decreased effectiveness of near-wall turbulence control, which is related to the amplitude modulation effect of large-scale structures.

In recent years, the extensive application of DRL has been highlighted in various domains such as video classification, voice recognition, and language processing.
In fluid mechanics, DRL has also been applied extensively to flow control problems \citep{gueniat2016statistical,rabault2019artificial,han2020active,paris2021robust,zeng2021symmetry,li2022reinforcement,varela2022deep,lee2023turbulence,guastoni2023deep,sonoda2023reinforcement,suarez2024flow}.
For instance, \cite{varela2022deep} demonstrated DRL's capability to extend control strategies across varying Reynolds numbers, adapting to different flow characteristics as the Reynolds number increases.
Additionally, \cite{suarez2024flow} leveraged multi-agent DRL to develop three-dimensional strategies as 3D instabilities emerged in the cylinder flow, achieving greater drag reduction than traditional methods.
These advancements, driven by artificial intelligence and data science, underscore DRL's robust capability to model complex interactions between inputs and outputs \citep{jordan2015machine}. 
Unlike traditional control methods that depend heavily on researchers' insights, neural network-based DRL can partially automate this process, constructing highly nonlinear models between input signals and output controls.
This makes DRL-based turbulence control particularly appealing, as it offers greater flexibility in selecting input signals, and potentially devises control strategies more attuned to the nonlinear mechanisms of turbulence, thereby enhancing drag reduction effects. 
The initial foray into using machine learning for drag reduction in channel flows can be traced back to \cite{lee1997application}, who employed a linear neural network with multiple neurons to predict wall-normal blowing and suction velocities based on spanwise wall shear stress, proposing a straightforward control scheme.
In more recent developments, \cite{han2020active} and \cite{lee2023turbulence} utilized reinforcement learning to predict wall-normal velocity fluctuations at the detection plane, effectively replicating opposition control based solely on wall measurements. 
Moreover, \cite{guastoni2023deep} and \cite{sonoda2023reinforcement} have achieved better control models and higher drag reduction rates with reinforcement learning compared to traditional opposition control methods.
Collectively, these studies demonstrate the significant potential of DRL in reducing drag in wall-bounded turbulence.
While DRL-based control strategies have shown great promise, their practical implementation still presents challenges.
Many current approaches rely on detailed flow-domain information, such as velocities at specific wall-normal locations, which may be difficult to measure in real-world settings.
This underscores the importance of developing strategies that can bridge the gap between numerical simulations and practical applications.

Higher Reynolds number studies also represent an essential step toward conditions more representative of real-world scenarios.
However, previous studies on DRL for turbulence control have been limited to low friction Reynolds numbers, with most $Re_{\tau}$ not exceeding $180$. 
Consequently, research on DRL-based control strategies at higher Reynolds numbers remains scarce. 
Additionally, there is a significant gap in understanding the drag reduction mechanisms underlying DRL models. 
This study aims to address these gaps by extending DRL-based control strategies to high Reynolds numbers. 
To the best of the authors' knowledge, this is the first study applying DRL control to turbulent channel flows with $Re_{\tau}$ larger than $500$. 
Our main purpose is to evaluate the effectiveness of DRL models in achieving drag reduction at high Reynolds numbers and to explore the underlying drag reduction mechanisms from both kinematic and dynamic perspectives.

The paper is organized as follows. 
The numerical methodologies, including DNS and DRL methods, are detailed in \S\ref{sec:method}.
\S\ref{sec:discussion} presents the DNS results and their subsequent discussions.
The performance of the DRL-based control strategy is evaluated in \S\ref{sec:performance}, while velocity statistics are elaborated upon in \S\ref{sec:basic}.
The analysis of the drag reduction mechanism is approached from both a kinematic perspective, based on virtual wall theory, and a dynamic perspective, using budget equations, in \S\ref{sec:virtual_wall} and \S\ref{sec:budget}, respectively.
Finally, the conclusions are summarized in \S\ref{sec:conc}.



\section{Numerical Methodology}\label{sec:method}

\subsection{DNS of the turbulent channel flows}\label{sec:dns}
We consider the turbulent channel flows established between two parallel plates separated by $2h$, driven by a pressure gradient.
The governing equations of the turbulent flow are the Navier-Stokes equations of an incompressible Newtonian fluid, written as:
\begin{equation}
	\frac{\partial u_{j}}{\partial x_{j}}=0,
	\label{eq:N-S-1}
\end{equation}
\begin{equation}
	\frac{\partial u_{i}}{\partial t}+u_{j}\frac{\partial u_{i}}{\partial x_{j}}=-\frac{1}{\rho}\frac{\partial p}{\partial x_{i}}+\nu\frac{\partial^{2}u_{i}}{\partial x_{j}\partial x_{j}}+f_{1}\delta_{i1},
	\label{eq:N-S-2}
\end{equation}
where $x_{i}(i=1,2,3)=(x,y,z)$ represents the coordinates in the streamwise, wall-normal, and spanwise directions, respectively, and $u_{i}(i=1,2,3)=(u,v,w)$ denotes the corresponding velocity components.
Here, $t$ is the time, $\rho$ is the density, $p$ is the pressure, and $\nu$ is the kinematic viscosity.
A body force $f_{1}$ is introduced to maintain constant momentum in the channel, ensuring the averaged bulk velocity $U_m$ in the channel flow.

The flow is assumed to be periodic in the streamwise and spanwise directions, with periods $L_x$ and $L_z$, respectively.
The upper wall imposes no-slip and no-penetration conditions, setting the velocities $u=v=w=0$. 
On the other hand, the lower wall adheres to the no-slip condition with $u=w=0$ and implements turbulent control through blowing and suctions.

The code AFiD \citep{verzicco1996finite,van2015pencil,zhu2018afid} was utilized to carried out the DNS of turbulent channel flows.
An energy-conserving second-order finite difference scheme is applied in the spatial discretization, with velocities on a staggered grid.
Time marching is performed using a third-order Runge–Kutta scheme, combined with a Crank-Nicholson scheme for the implicit terms.
The grids are uniformly distributed in both the streamwise and spanwise directions, with wall-normal grid refinement applied near the walls.

The computational parameters are listed in table \ref{tab:CP-cases} for the three Reynolds numbers $Re=U_{m}h/\nu$ considered in this study.
The friction velocity $u_{\tau}=\sqrt{\tau_{w}/\rho}$ and the friction Reynolds number $Re_{\tau}=h^{+}=u_{\tau}h/\nu$ define wall units in the following discussions, denoted by a `+' superscript, where $\tau_{w}$ is the skin friction.
Here, $y^{+}=y/\delta_{\nu}$, where $\delta_{\nu}=\nu/u_{\tau}=h/Re_{\tau}$ is the friction length.

\begin{table}
  \begin{center}
\def~{\hphantom{0}}
  \begin{tabular}{lccccccccc}
       Set of cases       & ${Re}$        & ${Re}_{\tau}$ & $L_{x}$  & $L_{z}$   & $\Delta_x^+$  & $\Delta_z^+$  & $(\Delta_y^+)_{min}$  & $(\Delta_y^+)_{max}$\\[3pt]
       C180               & $2800$        & $176.9$       & $2\pi h$ & $\pi h$   & $8.7$         & $4.3$         & $0.10$                & $4.3$               \\
       C550               & $10000$       & $544.3$       & $2\pi h$ & $\pi h$   & $13.4$        & $6.7$         & $0.19$                & $5.8$               \\
       C1000              & $20000$       & $983.4$       & $2\pi h$ & $\pi h$   & $12.1$        & $6.0$         & $0.19$                & $7.9$               \\
  \end{tabular}
\caption{Computational parameters. $\Delta_x$, $\Delta_y$ and $\Delta_z$ are the resolutions in the streamwise, wall-normal and spanwise directions, respectively.}
  \label{tab:CP-cases}
  \end{center}
\end{table}

\subsection{DRL Methodology}\label{sec:DRL_method}

\begin{figure}
	\centering
		\begin{overpic}
			[scale=0.3]{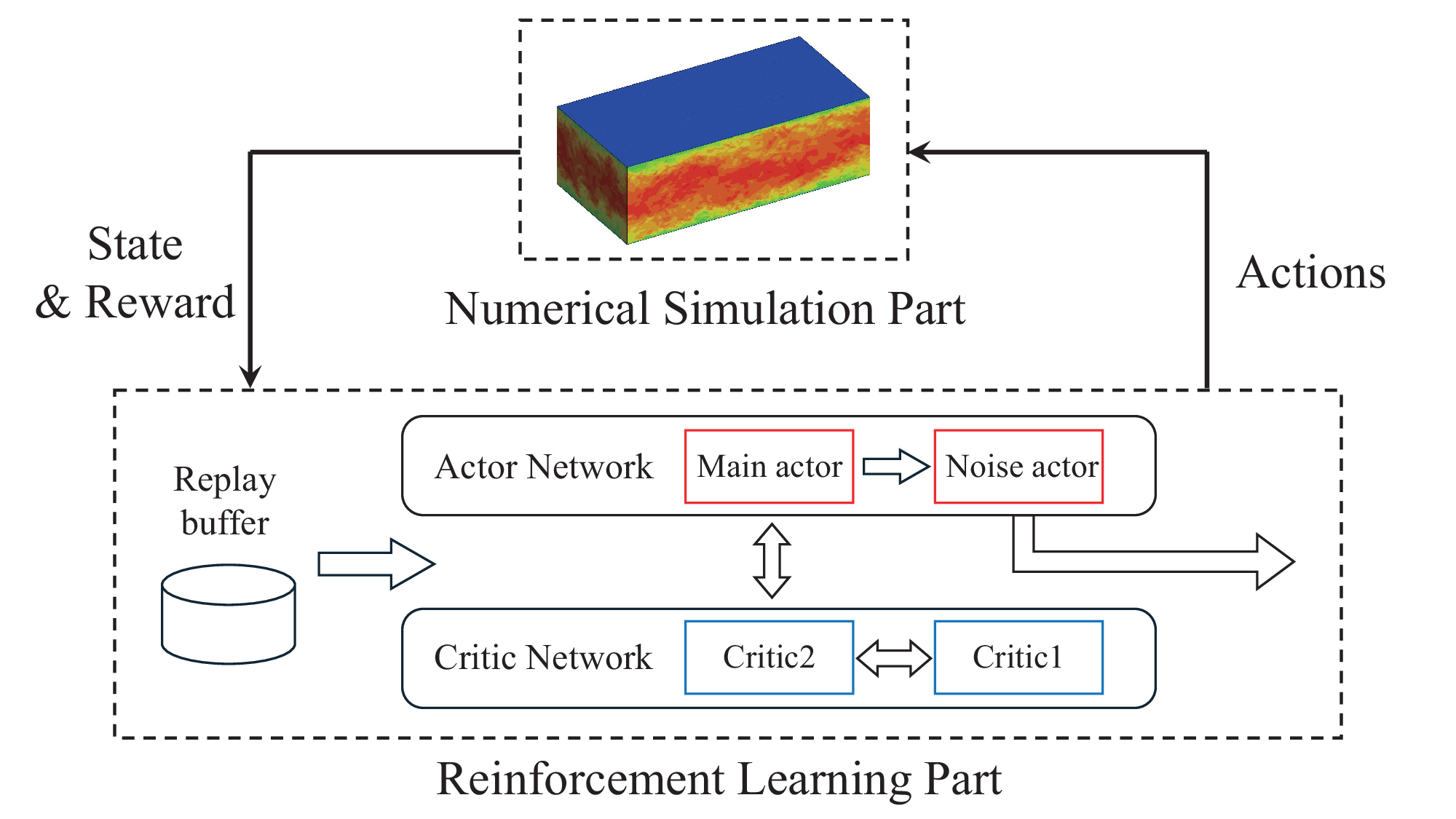}
		\end{overpic}

	\caption{
		The flow chart of reinforcement-learning-driven control in turbulent channel flows.
	}
	\label{fig:flow-chart}
	
\end{figure}

In order to control the turbulent channel flows, blowing and suctions based on the DRL predictions are applied to the lower wall. 
The flow chart of the control driven by reinforcement learning is shown in figure \ref{fig:flow-chart}.
Our current program mainly consists of two parts: the numerical simulation part and the reinforcement learning part. 
The numerical simulation part, as discussed in section \S\ref{sec:dns}, acts as the environment and outputs the state $s_{t}$ and reward $r_{t}$ obtained in the flow field. 
The reinforcement learning part, acting as the agent, receives these variables, optimizes the decision-making policy $\pi(s_{t})$ based on the reward, and outputs actions $a_{t}$ based on the state. 
The numerical simulation part then uses these actions to control the flow and advances the simulation in time. 
This creates a loop to achieve active control driven by reinforcement learning.
Here, we select the wall blowing and suction velocities $v_{w}^{\prime}$ as the actions, and we choose the streamwise velocity fluctuations $u^{\prime}(x,z)\mid_{y^{+}=15}$ in the near-wall region as the states, similar to those adopted by \cite{sonoda2023reinforcement}.
Velocity fluctuations are defined based on the mean velocity profile of each case, where $u^{\prime}=u(x,y,z)-U(y)$.
The mean wall blowing and suction velocity is set to zero.

The agent we adopted is based on the open-source code provided by \cite{lee2023turbulence}, which employs the twin-delayed deep deterministic policy gradient (TD3) model \citep{lillicrap2015continuous}, an actor-critic network structure.
The TD3 model offers improved stability and performance in learning by addressing overestimation bias, incorporating delayed updates, and implementing target smoothing.
It has been proven to be suitable for turbulence control optimization \citep{lee2023turbulence}.
In the TD3 model, the goal is to optimize the action value function $q_{\pi}(s_{t},a_{t})$ by satisfying the Bellman equation, where
\begin{equation}
	q_{\pi}(s_{t},a_{t})=\mathbb{E}[r_{t}^{d}+\gamma^{n}q_{\pi}(s_{t+n},\pi_{\phi}(s_{t+n})+\epsilon)].
	\label{eq:DRL-1}
\end{equation}
Here, $r_{t}^{d}=\sum_{j=1}^{n}\gamma^{j-1}r_{t+j}$ is the $n$-step reward, $\gamma$ is the discounted factor, $\pi_{\phi}(s_{t+n})$ is the delayed policy update, and $\epsilon$ is the clipped random noise.
We adopt $n=5$ and $\gamma=0.95$ in all cases, following \cite{lee2023turbulence}.
The expected cumulative reward is predicted by the critic networks, and the objective function for updating the parameters of the critic networks is given by
\begin{equation}
	J(\theta)=N^{-1}\sum\left[r_{t}^{d}+\gamma^{n}q_{\pi}(s_{t+n},\pi_{\phi}(s_{t+n})+\epsilon)-q(s_{t},a_{t})\right]^{2},
	\label{eq:DRL-2}
\end{equation}
where $N=64$ is the minibatch size, and $\theta$ is the weight parameter of the critic networks.
The actor network aims to find an optimal policy, guided by the policy objective function $J(\phi)$. 
Here $\phi$ represents the weight parameters of the actor (or critic) networks.
The objective function is updated by:
\begin{equation}
	\nabla_{\psi}J(\psi)=\mathbb{E}[\nabla_{a}q_{\pi}(s_{t},a_{t})\mid_{a=\pi(s_{t})}\nabla_{\psi}\pi_{\psi}(s_{t})].
	\label{eq:DRL-3}
\end{equation}
where $\psi$ is the weight parameter of the actor network.
The actor network includes $3$ convolutional layers, with the first two layers activated by the ReLU function.
The number of filter kernels for these layers is set to $64$, $32$, and $1$, respectively, with each filter kernel sized at $3\times3$.
In contrast, the critic network is structured with $6$ convolutional layers followed by $3$ fully connected layers, all activated by the ReLU function. 
Each convolutional layer contains $32$ filter kernels of size $3\times3$.
Additionally, an average pooling layer is applied after every two convolutional layers. 
The fully connected layers each consist of $32$ neurons, and the network ultimately outputs a $q$ value to evaluate the control policy.
Detailed hyperparameters can be referred to in \cite{lee2023turbulence}.

\begin{table}
	\begin{center}
		\def~{\hphantom{0}}
		\begin{tabular}{lccc}
			Set of cases       & $\Delta t^{+}$        & $\Delta TU_{m}/h$   & $N_{st}$     \\[3pt]
			C180               & $55.8$                & $100.0$         	 & $20$    		\\
			C550               & $59.3$       		   & $40.0$         	 & $20$     	\\
			C1000              & $48.4$      		   & $20.0$         	 & $20$     	\\
		\end{tabular}
		\caption{
			Parameters of state steps and episodes. 
			$\Delta t$ and $\Delta T$ are the time lengths of each state step and episode, respectively, while $N_{st}$ is the number of state steps in one episode.
		}
		\label{tab:time_length}
	\end{center}
\end{table}

The choice of reward $r_{t}$ is crucial for the effectiveness of the training outcomes.
Inspired by the optimal control \citep{bewley2001dns}, we define the reward $r=1-e/e_{0}$ as the reduction of integrated turbulent kinetic energy (TKE) in the lower half channel at the end of each state step.
Here, $e$ is the integrated turbulent kinetic energy with control, defined as
\begin{equation}
	e=\frac{1}{L_{x}L_{z}hU_{m}^{2}}\int_{0}^{L_{z}}\int_{0}^{h}\int_{0}^{L_{x}}\left(\frac{1}{2}u_{i}^{\prime}u_{i}^{\prime}\right)\mathrm{d}x\mathrm{d}y\mathrm{d}z,
	\label{eq:DRL-reward}
\end{equation}	
and $e_0$ is the integrated turbulent kinetic energy without control.
It is important to note that the reward is only used during the model training process and is no longer needed once the model has converged.

Furthermore, the time lengths of each state step and episode for all cases are shown in table \ref{tab:time_length}.
First, $\Delta t$ needs to be sufficiently long to allow the changes in the control strategy to fully develop.
Therefore, $\Delta t^{+}\approx50$ is selected, which also meets the requirement for the prediction horizon $\Delta t^{+}>25$ in optimal control \citep{bewley2001dns}.
Additionally, given that the maximum ${Re}_{\tau}$ in our cases reaches $1000$, it is crucial to train a control strategy that remains effective under large-scale structure evolution. 
Consequently, we ensure that $\Delta T$ is at least $20h/U_{m}$, which exceeds the time required for large-scale structures to advect streamwise across the entire channel, approximately $L_x/U_m\thickapprox6h/U_{m}$.

In summary, a comparison of our computational method with previous DRL-based studies on turbulent channel control is shown in table \ref{tab:CP_method}.
Our work uses the TD3 algorithm, similar to that of \cite{lee2023turbulence}, with streamwise velocity fluctuations $u^{\prime}$ as input states and turbulent kinetic energy (TKE) reduction rate as the reward.
We have extended DRL-based wall blowing and suction control to higher Reynolds numbers, reaching ${Re}_{\tau}=1000$.

\begin{table}
	\begin{center}
		\def~{\hphantom{0}}
		\begin{tabular}{lllll}
			Study                           & DRL algorithm     & Input states                                                                 & Rewards                   &  ${Re}_{\tau}$      \\[4pt]
			\cite{guastoni2023deep}         & DDPG              & $(u^{\prime},v^{\prime})$                                                    & Drag reduction (DR) rate  &  $180$              \\
			\cite{sonoda2023reinforcement}  & DDPG              & $(u^{\prime},v^{\prime})$                                                    & DR rate \& control cost   &  $150$              \\
			\cite{lee2023turbulence}        & TD3               & $(\frac{\partial u}{\partial y},\frac{\partial w}{\partial y})\mid_{wall}$   & DR rate   		           &  $180,360$          \\
			Our work                        & TD3               & $u^{\prime}$                                                                 & TKE reduction rate   	   &  $180,550,1000$     \\
		\end{tabular}
		\caption{Comparison of computational details in DRL-based turbulent channel control studies. 
			DDPG denotes the deep deterministic policy gradient algorithm.
			In all the studies above, the output actions selected are the wall blowing and suction velocities, $v_{w}^{\prime}$.}
		\label{tab:CP_method}
	\end{center}
\end{table}

\section{DNS results and discussions}\label{sec:discussion}

\subsection{Performance of the DRL models}\label{sec:performance}

In this study, we focus on the DRL-optimized control models under different blowing and suction intensities and their impact on the flow mechanism. 
The DNS cases we utilized are detailed in table \ref{tab:cases_DR}. 
Among these, cases with the suffix ‘-0’ and ‘-opp’ did not use DRL models.
The former denotes cases with no blowing and suctions, whereas the latter represents cases with opposition control as suggested by \cite{choi1994active}. 
Cases with the suffixes ‘-1’, ‘-2’, and ‘-3’ are based on DRL-optimized control models, where the magnitude of wall blowing and suction $v_{w}^{\prime}$ is limited to $\left[-u_{\tau}^{0},u_{\tau}^{0}\right]$, $\left[-2u_{\tau}^{0},2u_{\tau}^{0}\right]$, and $\left[-3u_{\tau}^{0},3u_{\tau}^{0}\right]$, respectively.
Here, the superscript ‘0’ denotes variables before the application of turbulence control.

\begin{table}
	\begin{center}
		\def~{\hphantom{0}}
		\begin{tabular}{llcccccc}
			Cases     & Range of $v_{w}^{\prime}$ & ${Re}_{\tau}$ & $DR(\%)$ & $P_{S}/P_{I}$ & $\Delta U_{s}^{+}$   & $y_{vw}^{+}$  & $-\left\langle u^{\prime}v^{\prime}\right\rangle _{vw}^{+}$      \\[4pt]
			C180-0    & $v_{w}^{\prime}=0$                                            & $176.9$       & $0$    	    & --    	   & $0$               & $0$           & $0$             \\
			C180-opp  & $v_{w}^{\prime}=-v^{\prime}\mid_{y^{+}=15}$                   & $154.2$       & $24.0$      & $27.3$       & $2.92$   		   & $7.1$         & $0.029$         \\
			C180-1    & $v_{w}^{\prime}\in\left[-u_{\tau}^{0},u_{\tau}^{0}\right]$    & $153.9$       & $24.9$      & $12.4$ 	   & $3.30$   		   & $9.5$         & $0.144$         \\
			C180-2    & $v_{w}^{\prime}\in\left[-2u_{\tau}^{0},2u_{\tau}^{0}\right]$  & $143.1$       & $34.6$      & $22.6$ 	   & $5.17$   		   & $12.8$        & $0.025$         \\
			C180-3    & $v_{w}^{\prime}\in\left[-3u_{\tau}^{0},3u_{\tau}^{0}\right]$  & $142.0$       & $35.6$      & $26.3$ 	   & $5.18$   		   & $12.7$        & $0.023$         \\[4pt]
			C550-0    & $v_{w}^{\prime}=0$                                            & $544.3$       & $0$    	    & --    	   & $0$               & $0$           & $0$             \\
			C550-opp  & $v_{w}^{\prime}=-v^{\prime}\mid_{y^{+}=15}$                   & $480.7$       & $22.0$      & $11.4$ 	   & $3.00$   		   & $7.3$         & $0.054$         \\
			C550-1    & $v_{w}^{\prime}\in\left[-u_{\tau}^{0},u_{\tau}^{0}\right]$    & $484.3$       & $20.8$      & $12.1$ 	   & $3.21$   		   & $7.4$         & $0.125$         \\
			C550-2    & $v_{w}^{\prime}\in\left[-2u_{\tau}^{0},2u_{\tau}^{0}\right]$  & $460.0$       & $28.6$      & $13.2$ 	   & $4.23$   		   & $8.9$         & $0.069$         \\
			C550-3    & $v_{w}^{\prime}\in\left[-3u_{\tau}^{0},3u_{\tau}^{0}\right]$  & $454.0$       & $30.4$      & $13.5$ 	   & $4.28$   		   & $8.7$         & $0.074$         \\[4pt]
			C1000-0   & $v_{w}^{\prime}=0$                                            & $983.4$       & $0$    	    & --    	   & $0$               & $0$           & $0$             \\
			C1000-opp & $v_{w}^{\prime}=-v^{\prime}\mid_{y^{+}=15}$                   & $887.8$       & $18.5$      & $9.2$   	   & $2.64$   		   & $6.7$         & $0.044$         \\
			C1000-1   & $v_{w}^{\prime}\in\left[-u_{\tau}^{0},u_{\tau}^{0}\right]$    & $870.7$       & $21.6$      & $5.0$   	   & $2.71$   		   & $7.5$         & $0.306$         \\
			C1000-2   & $v_{w}^{\prime}\in\left[-2u_{\tau}^{0},2u_{\tau}^{0}\right]$  & $862.9$       & $23.0$      & $5.1$   	   & $3.14$   		   & $8.4$         & $0.299$         \\
			C1000-3   & $v_{w}^{\prime}\in\left[-3u_{\tau}^{0},3u_{\tau}^{0}\right]$  & $836.2$       & $27.7$      & $4.7$   	   & $3.27$   		   & $10.2$        & $0.242$         \\[4pt]
		\end{tabular}
		\caption{DNS cases and drag reduction results. 
			$DR$ represents the drag reduction rate. 
			$P_{S}/P_{I}$ denotes the power saving ratio, where $P_{S}$ is the power saving, and $P_{I}$ is the power input.
			$\Delta U_{s}^{+}$ denotes the shift of the mean velocity profile in the logarithmic region. 
			$y_{vw}$ indicates the height of the virtual wall, and $-\left\langle u^{\prime}v^{\prime}\right\rangle _{vw}$ is the averaged residual Reynolds stress on the virtual wall.}
		\label{tab:cases_DR}
	\end{center}
\end{table}

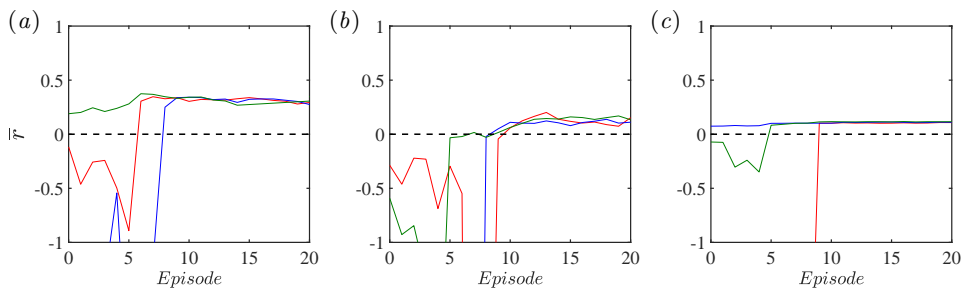
\begin{figure}
	\centering
	
	\subfigure{
		\begin{overpic}
			[scale=0.23]{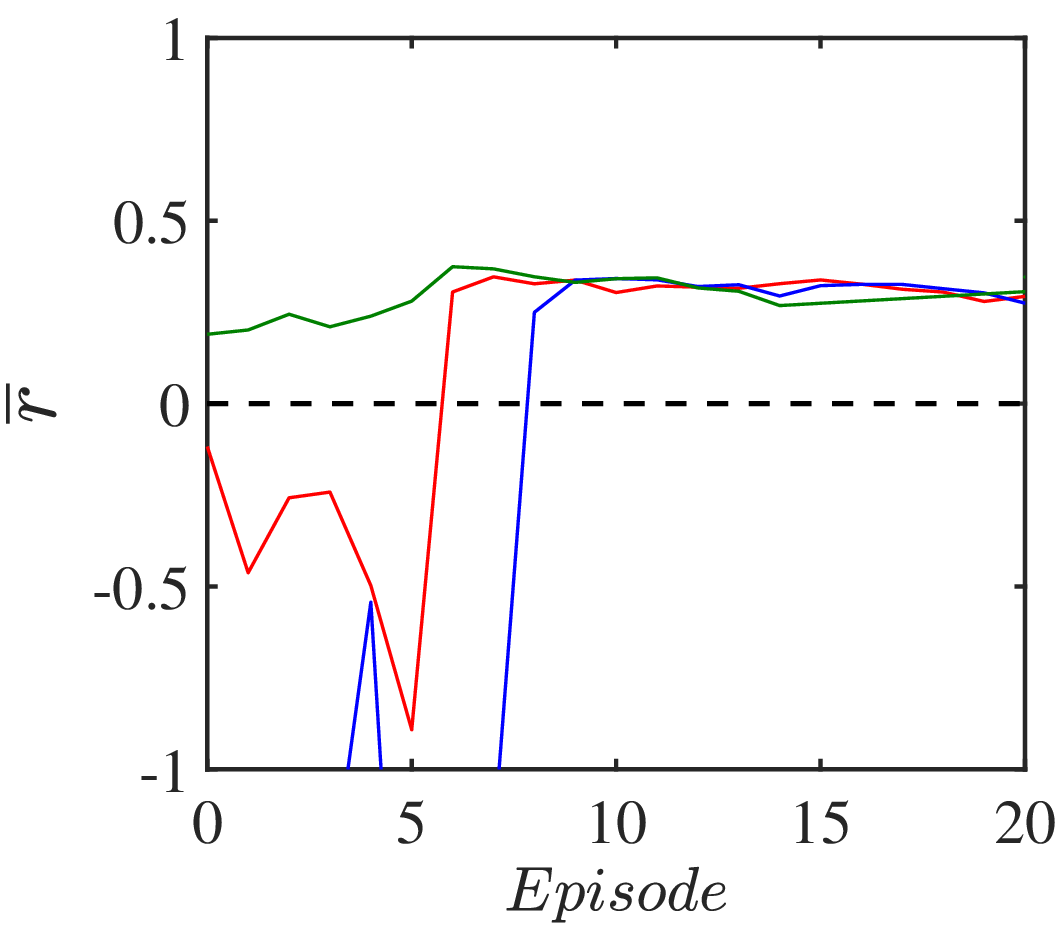}
			\put(0,84){(\textit{a})}
		\end{overpic}
		\begin{overpic}
			[scale=0.23]{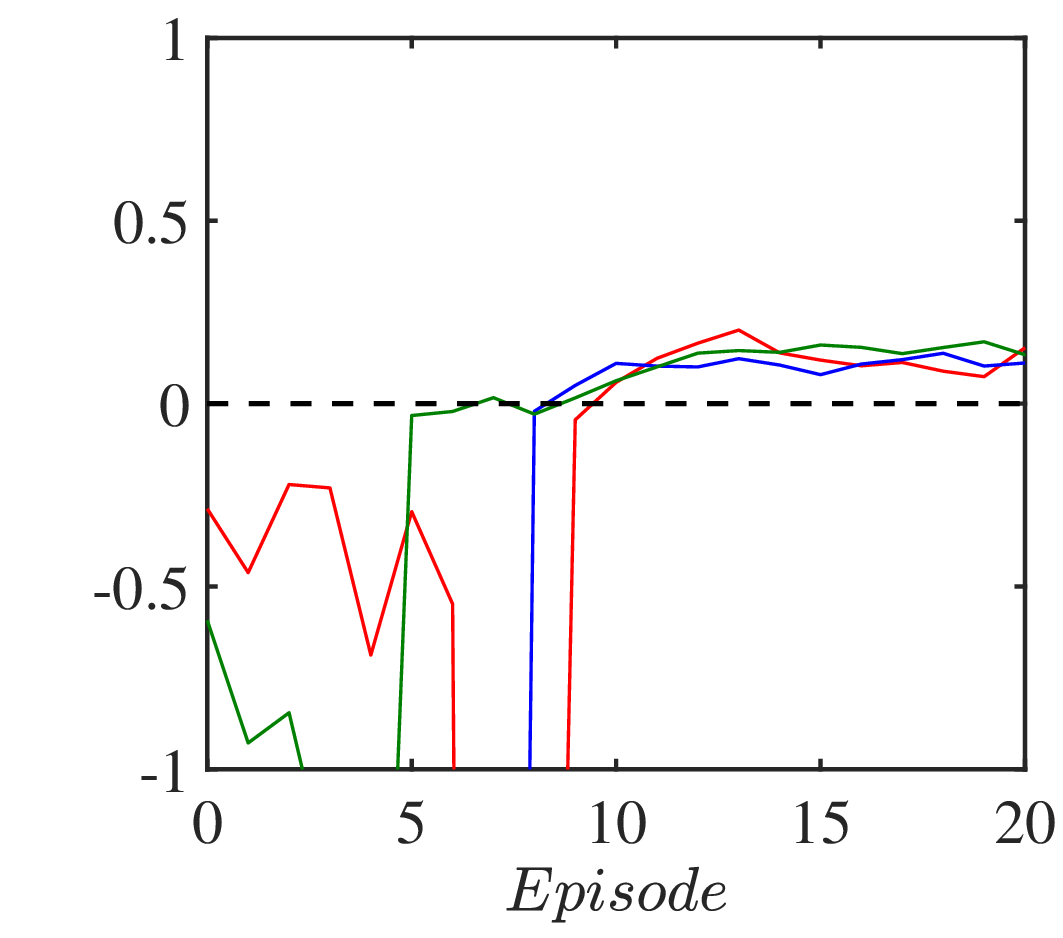}
			\put(0,84){(\textit{b})}
		\end{overpic}
		\begin{overpic}
			[scale=0.23]{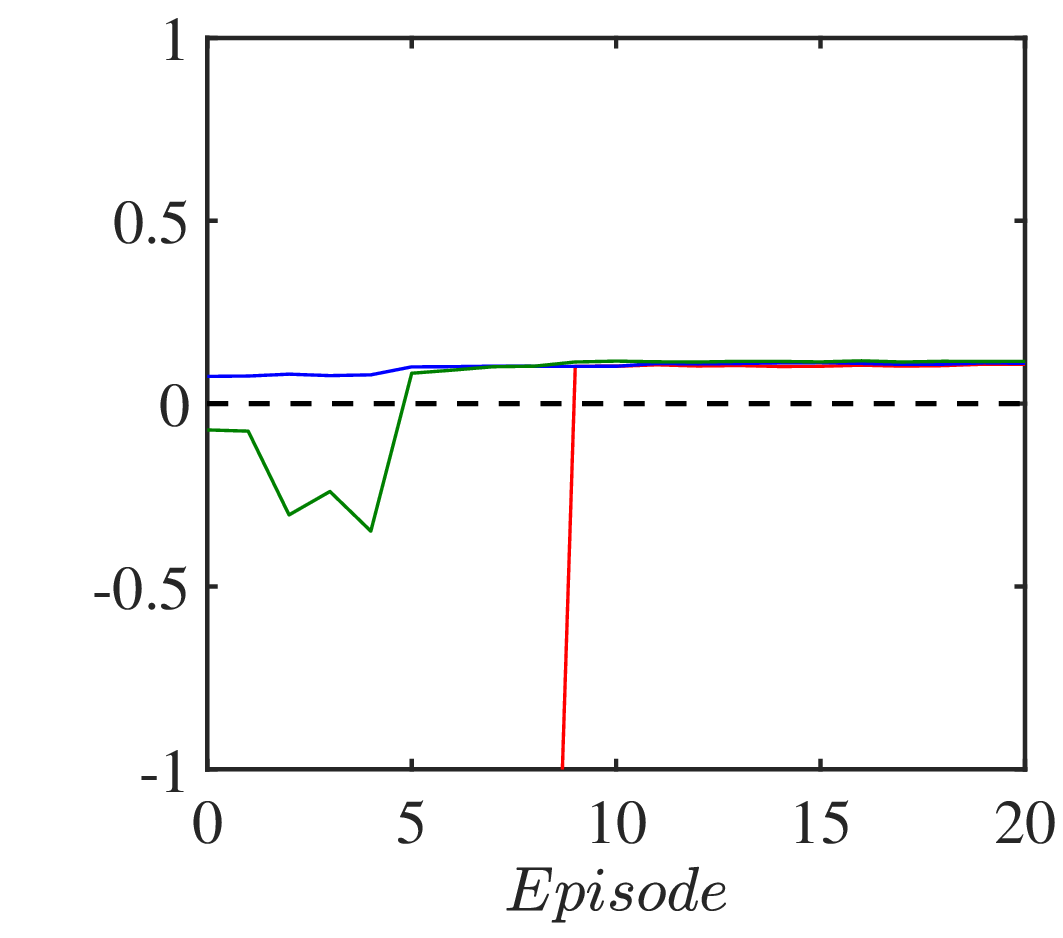}
			\put(0,84){(\textit{c})}
		\end{overpic}
	}
	\DeclareRobustCommand\mylabela{\tikz[baseline]{\draw[solid, red, thick] (0,0.5ex) -- (0.8,0.5ex);}}
	\DeclareRobustCommand\mylabelb{\tikz[baseline]{\draw[solid, blue, thick] (0,0.5ex) -- (0.8,0.5ex);}}
	\DeclareRobustCommand\mylabelc{\tikz[baseline]{\draw[solid, g00, thick] (0,0.5ex) -- (0.8,0.5ex);}}

	\caption{
		The evolution of the normalized reward over episodes during the training process.
		(\textit{a}) C180, (\textit{b}) C550, (\textit{c}) C1000.
		\mylabela: Cases with suffix '-1'; \mylabelb: with suffix '-2'; \mylabelc: with suffix '-3'.		
	}
	\label{fig:reward}
	
\end{figure}

Before further analysis, it is essential to confirm the training status of the current DRL models.
The normalized reward $\overline{r}$, defined as $\overline{r}=\sum_{j=1}^{n}\gamma^{j-1}r_{t+j}/\sum_{j=1}^{n}\gamma^{j-1}$, serves as an indicator of learning performance during the training of the control strategy.
The evolution of $\overline{r}$ over episodes for different cases is illustrated in figure \ref{fig:reward}.
Significant oscillations are primarily observed within the first $10$ episodes, while the rewards for all analyzed cases gradually converge after $10$ episodes, indicating stabilization of the DRL models. 
Consequently, the model at $20$ episodes will be uniformly utilized as the control strategy for subsequent analysis.
Additionally, we observed that further training beyond $20$ episodes does not significantly enhance learning performance, although not shown in figure \ref{fig:reward}.
As suggested by \cite{lee2023turbulence}, prolonged training can lead to issues such as catastrophic forgetting or overfitting, potentially causing the training process to fail.
Therefore, the models at $20$ episodes are deemed appropriate for our purposes.

The drag reduction effects under different blowing and suction intensities are shown in table \ref{tab:cases_DR}.
As the range of $v_{w}^{\prime}$ is expanded, the drag reduction rates ($DR$) achieved using DRL models continuously improve. 
When $v_{w}^{\prime}\in\left[-3u_{\tau},3u_{\tau}\right]$, the drag reduction effect significantly surpasses that of the traditional opposition control method, including high-Reynolds-number cases.
Specifically, the drag reduction rate is $35.6\%$ in case C180-3, $30.4\%$ in case C550-3, and $27.7\%$ in case C1000-3.
On the other hand, it is noted that the maximum drag reduction rate achieved using the DRL model decreases as the Reynolds number increases.
This trend is similar to that observed with the opposition control method \citep{iwamoto2002reynolds,chang2002viscous,pamies2007response,touber2012near,hwang2013near}.
We also tested several cases with modified input states or rewards, along with the performance of trained models under different resolutions and $Re_{\tau}$; see Appendix \ref{sec:supplementary}.

Furthermore, as proposed by \cite{bewley2001dns}, the energy efficiency of the active control policy can be quantified by the ratio of power saving to power input, $P_{S}/P_{I}$, as shown in Table \ref{tab:cases_DR}. 
The power input $P_{I}$ is calculated as
\begin{equation}
	P_{I}=\left\langle \left|p_{w}^{\prime}v_{w}\right|\right\rangle +\rho\left\langle \left|v_{w}^{3}\right|\right\rangle /2,
	\label{eq:P_I}
\end{equation}	
where $p_{w}^{\prime}$ represents pressure fluctuations on the wall, and $\rho$ denotes the fluid density. 
The power saving $P_{S}$, in turn, is given by $P_{S} = (\tau_{w}^{0} - \tau_{w}) U_{m}$, where $\tau_{w}^{0}$ is the wall shear stress before control.
For cases with ${Re}_{\tau} \leq 550$, the power saving ratio $P_{S}/P_{I}$ shows an increasing trend as the range of $v_{w}^{\prime}$ expands. 
Specifically, at ${Re}_{\tau} = 180$, the power saving ratio achieved by DRL-based control is lower than that achieved by opposition control. 
On the other hand, at ${Re}_{\tau} = 550$, DRL-based control demonstrates more effective energy savings compared to opposition control.
As the Reynolds number increases, the power efficiency ratio $P_{S}/P_{I}$ gradually declines, eventually dropping around $5$ at ${Re}_{\tau} = 1000$. 
Notably, at ${Re}_{\tau} = 1000$, adjustments in the range of $v_{w}^{\prime}$ have only a minor impact on the energy savings achieved by DRL-based control.

\subsection{Velocity statistics}\label{sec:basic}

Further investigation is needed to understand the impact of DRL-based control on flow field statistics and the underlying mechanisms affecting the flow.
Therefore, this subsection will compare the velocity statistics across different cases.

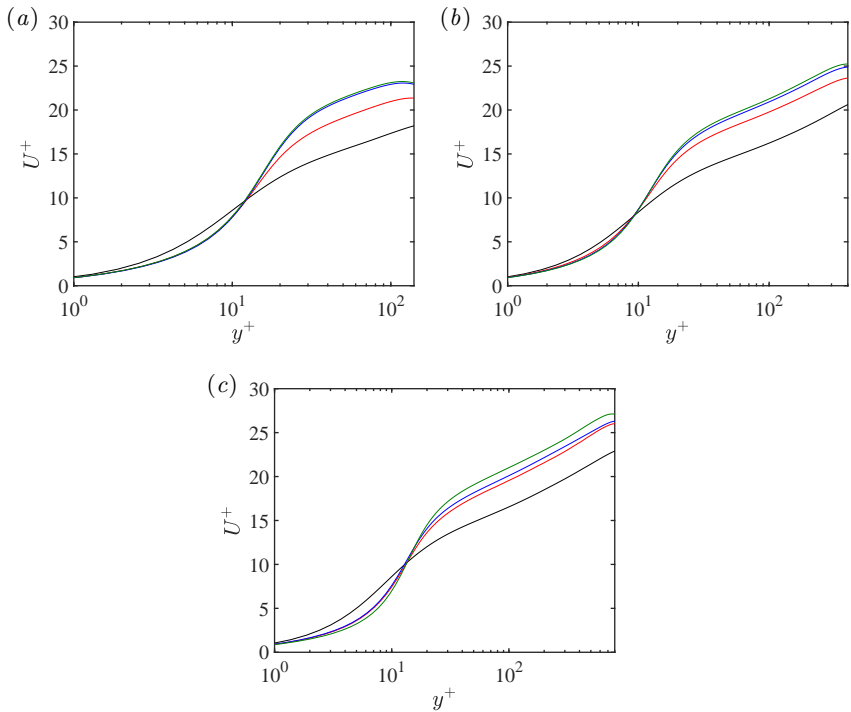
\begin{figure}
	\centering
	
	\subfigure{
		\begin{overpic}
			[scale=0.25]{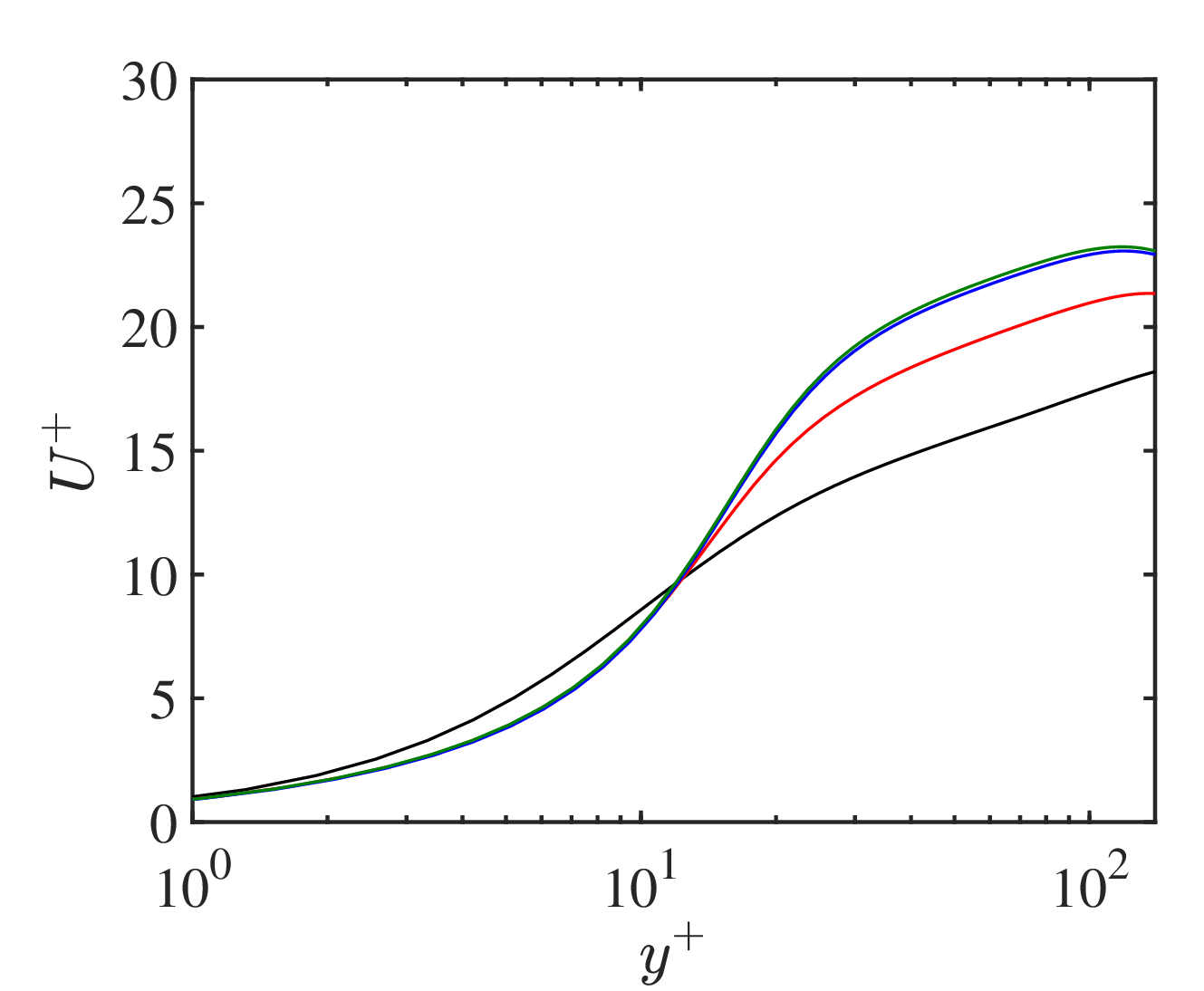}
			\put(0,75){(\textit{a})}
		\end{overpic}
		\begin{overpic}
			[scale=0.25]{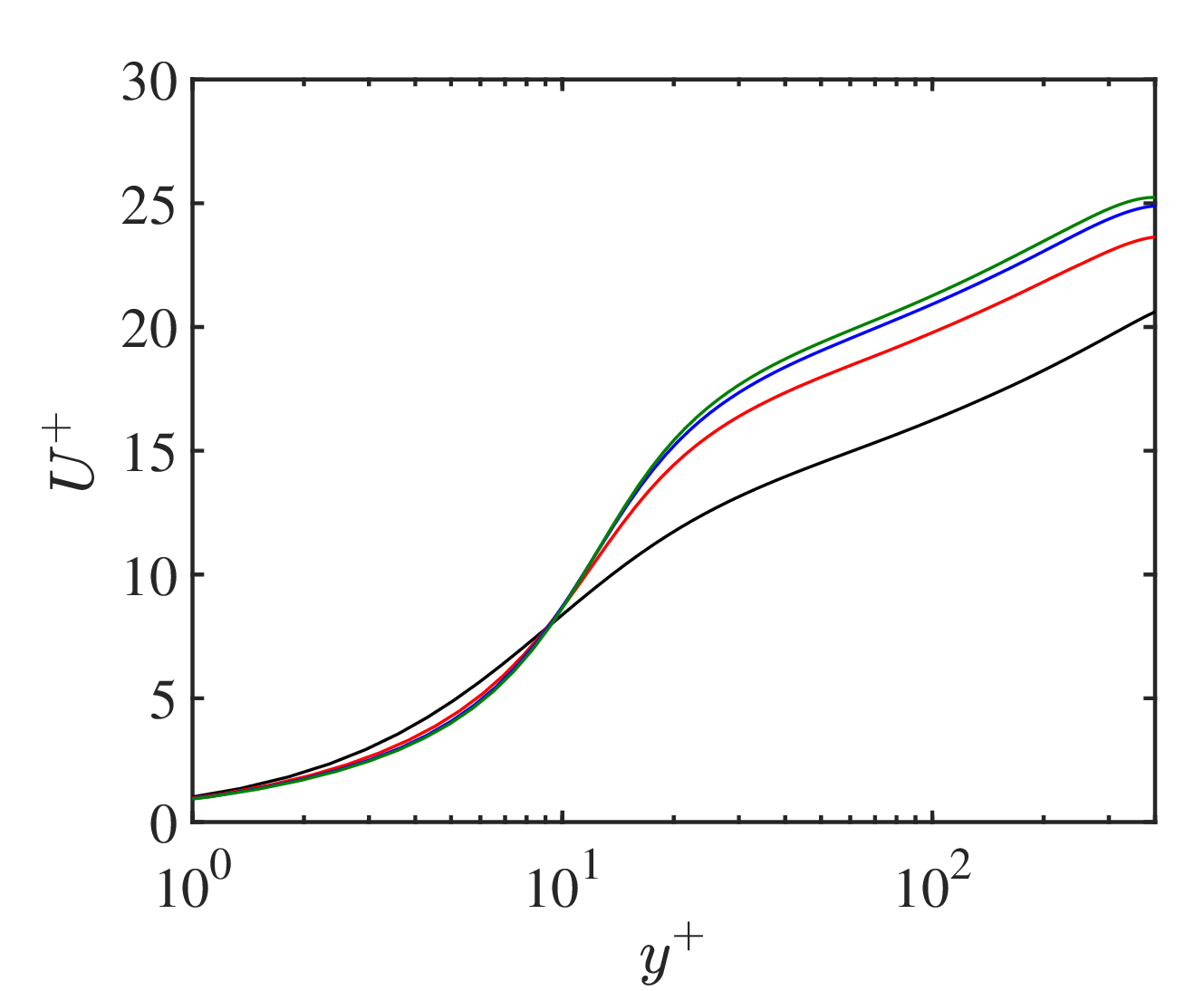}
			\put(0,75){(\textit{b})}
		\end{overpic}
	}
		\begin{overpic}
			[scale=0.25]{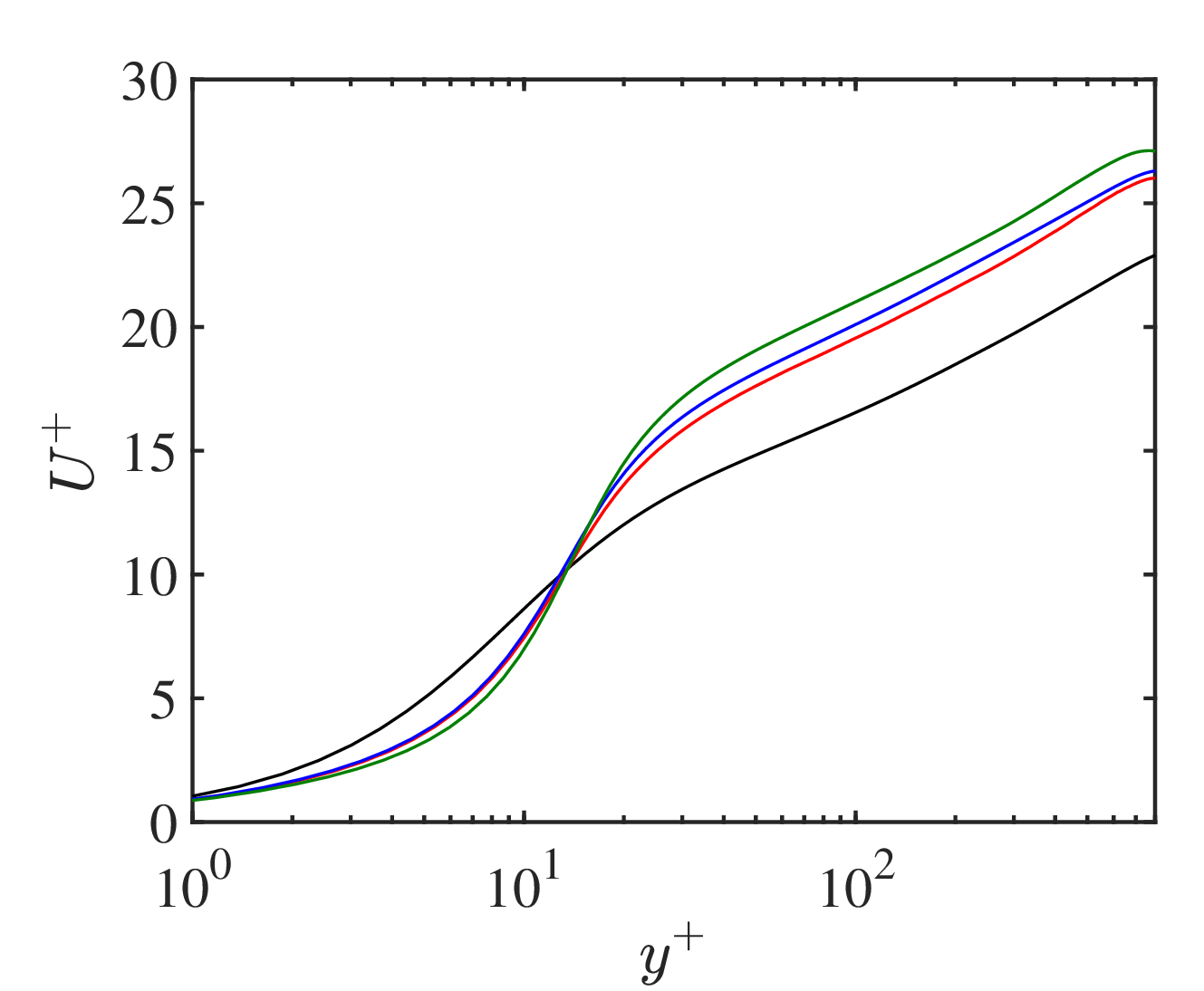}
			\put(0,75){(\textit{c})}
		\end{overpic}
	\DeclareRobustCommand\mylabeld{\tikz[baseline]{\draw[solid, black, thick] (0,0.5ex) -- (0.8,0.5ex);}}
	\DeclareRobustCommand\mylabela{\tikz[baseline]{\draw[solid, red, thick] (0,0.5ex) -- (0.8,0.5ex);}}
	\DeclareRobustCommand\mylabelb{\tikz[baseline]{\draw[solid, blue, thick] (0,0.5ex) -- (0.8,0.5ex);}}
	\DeclareRobustCommand\mylabelc{\tikz[baseline]{\draw[solid, g00, thick] (0,0.5ex) -- (0.8,0.5ex);}}

	\caption{
		Mean velocity profile under different control strategies.
		(\textit{a}) C180, (\textit{b}) C550, (\textit{c}) C1000.
		\mylabeld, Cases with suffix '-0'; \mylabela: with suffix '-1'; \mylabelb: with suffix '-2'; \mylabelc: with suffix '-3'.
	}
	\label{fig:U-avg}
	
\end{figure}

Figure \ref{fig:U-avg} shows the wall-normal distributions of the mean velocity profile, $U^{+}$.
In the viscous sublayer below $y^+=5$, the wall blowing and suctions based on the DRL model result in a decrease in the mean velocity compared to the uncontrolled case.
In the logarithmic region, the velocity profiles continue to follow the logarithmic law even in the presence of control, but with an upward shift relative to the uncontrolled case.
This behavior is analogous to what is observed with opposition control \citep{choi1994active}.
The profile shift $\Delta U_{s}$ is detailed in table \ref{tab:cases_DR}, with $\triangle U_{s}$ calculated as the averaged vertical shift between $y^+ = 50$ and $y/h = 0.5$.
As the range of $v_{w}^{\prime}$ is progressively extended, the drag reduction rate consistently increases, resulting in a corresponding rise in the mean velocity profile in the logarithmic region and a gradual increase in $\triangle U_{s}$. 
Furthermore, $\triangle U_{s}$ at higher Reynolds numbers gradually decreases, corresponding to a decline in the drag reduction rate.

\begin{figure}
	\centering
	
	\subfigure{
		\begin{overpic}
			[scale=0.25]{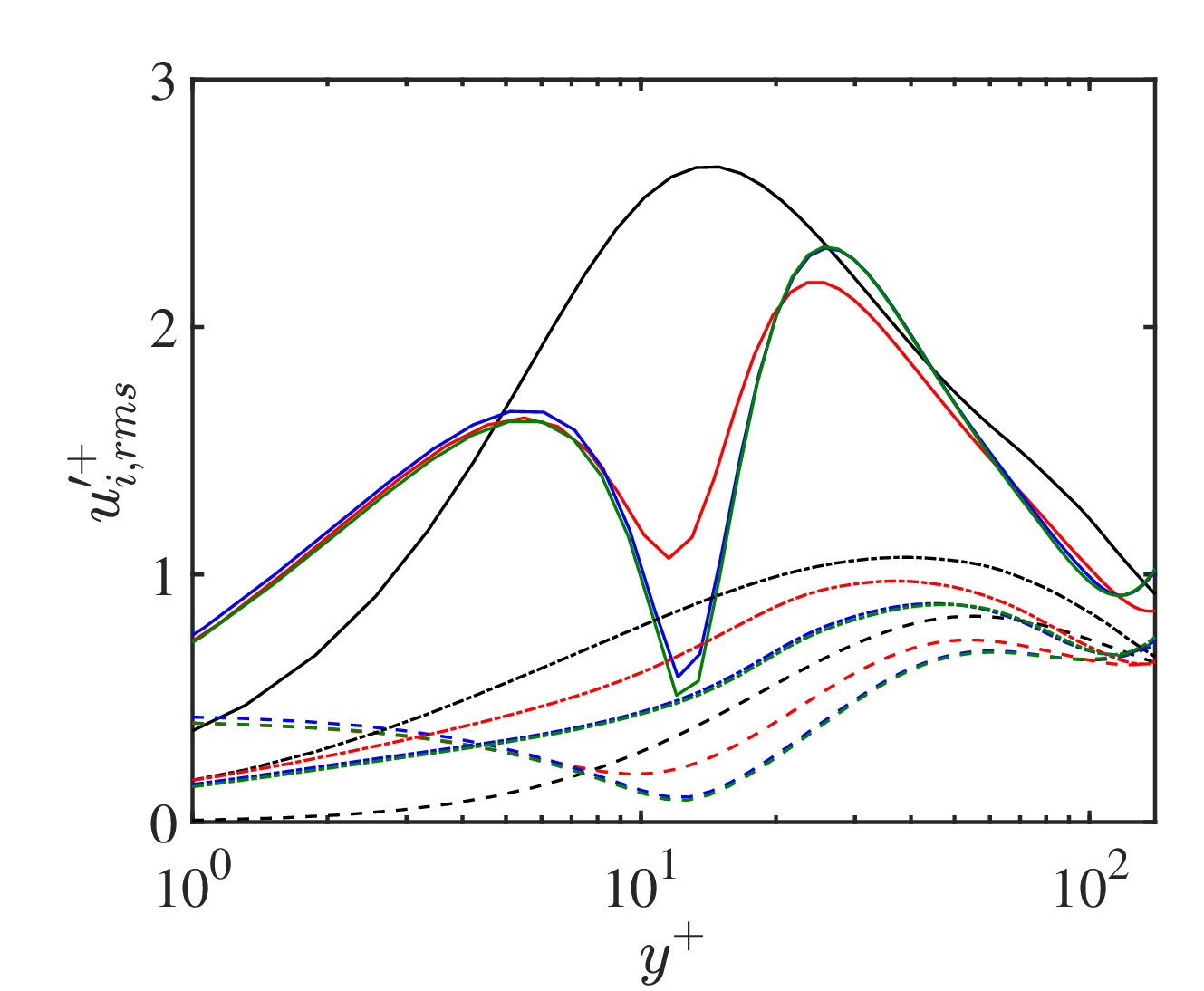}
			\put(2,74){(\textit{a})}
		\end{overpic}
		\begin{overpic}
			[scale=0.25]{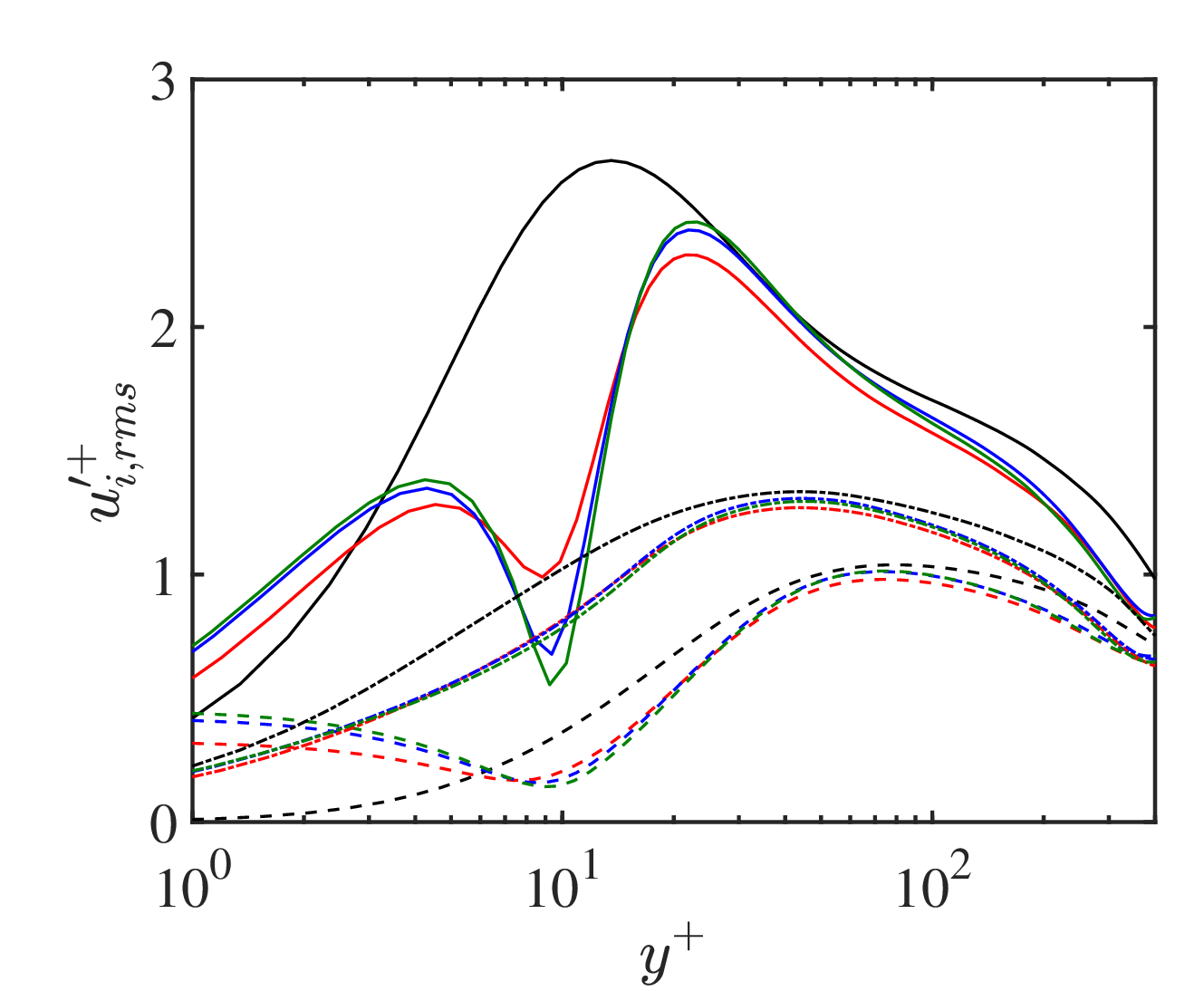}
			\put(2,74){(\textit{b})}
		\end{overpic}
	}
	\begin{overpic}
		[scale=0.25]{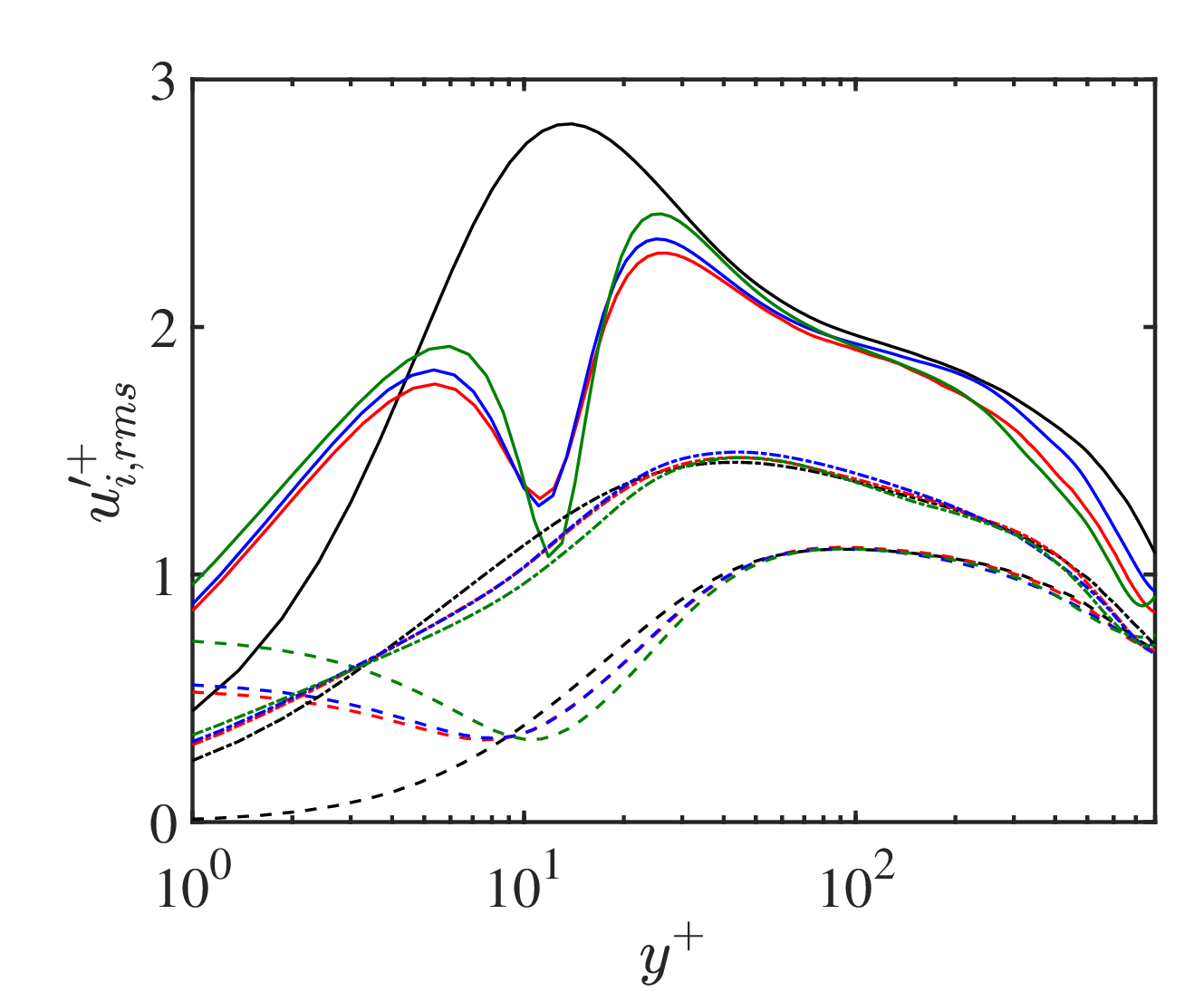}
		\put(2,74){(\textit{c})}
	\end{overpic}
	\DeclareRobustCommand\mylabeld{\tikz[baseline]{\draw[solid, black, thick] (0,0.5ex) -- (0.8,0.5ex);}}
	\DeclareRobustCommand\mylabela{\tikz[baseline]{\draw[solid, red, thick] (0,0.5ex) -- (0.8,0.5ex);}}
	\DeclareRobustCommand\mylabelb{\tikz[baseline]{\draw[solid, blue, thick] (0,0.5ex) -- (0.8,0.5ex);}}
	\DeclareRobustCommand\mylabelc{\tikz[baseline]{\draw[solid, g00, thick] (0,0.5ex) -- (0.8,0.5ex);}}

	\caption{
		Wall-normal distributions of the velocity fluctuations under different control strategies.
		Solid, dashed, and dash-dotted lines represent $u_{rms}^{\prime}$, $v_{rms}^{\prime}$, and $w_{rms}^{\prime}$, respectively.
		(\textit{a}) C180, (\textit{b}) C550, (\textit{c}) C1000.
		\mylabeld, Cases with suffix '-0'; \mylabela: with suffix '-1'; \mylabelb: with suffix '-2'; \mylabelc: with suffix '-3'.
	}
	\label{fig:u-rms}
	
\end{figure}

Wall-normal distributions of velocity fluctuations under different control strategies are illustrated in figure \ref{fig:u-rms}.
After applying control, the streamwise velocity fluctuations $u_{rms}^{\prime}$, indicated by solid lines, show a significant increase in the viscous sublayer below $y^+=5$.
In contrast, at higher positions, particularly around $y^+=10$ in the near-wall region, the peak of $u_{rms}^{\prime}$ vanishes.
The reduction in streamwise velocity fluctuations becomes more pronounced as the range of blowing and suction velocities is further expanded.
This trend is consistently observed across different Reynolds numbers. 
However, it is noteworthy that, for $Re_{\tau}^{0}\thickapprox550$ and $1000$, the impact of wall blowing and suctions is trivial in the outer region, as depicted in figures \ref{fig:u-rms}(\textit{b})(\textit{c}). 
The current DRL-based control strategy hardly affects the outer region, which is dominated by large-scale structures.
On the other hand, the application of control significantly increases the wall-normal velocity fluctuations $v_{rms}^{\prime}$ within the viscous sublayer, due to the direct impact of blowing and suction.
Conversely, the spanwise velocity fluctuations $w_{rms}^{\prime}$ below $y^+=5$ exhibit minimal changes.
As the range of blowing and suction velocities is further extended, both $v_{rms}^{\prime}$ and $w_{rms}^{\prime}$ within $10<y^{+}<30$ gradually decrease. 
This trend is particularly evident at the low Reynolds number $Re_{\tau}^{0}\thickapprox180$, as shown in figure \ref{fig:u-rms}(a), but becomes less pronounced at higher Reynolds numbers.
Additionally, in the controlled cases, $v_{rms}^{\prime}$ in the near-wall region initially decreases and then increases with height.
The point of minimum $v_{rms}^{\prime}$ in the near-wall region can be defined as the position of the virtual wall $y_{vw}$ \citep{hammond1998observed}. 
The virtual wall and the residual fluctuations on it will be further discussed in the following \S\ref{sec:virtual_wall}.

\begin{figure}
	\centering
	
	\subfigure{
		\begin{overpic}
			[scale=0.25]{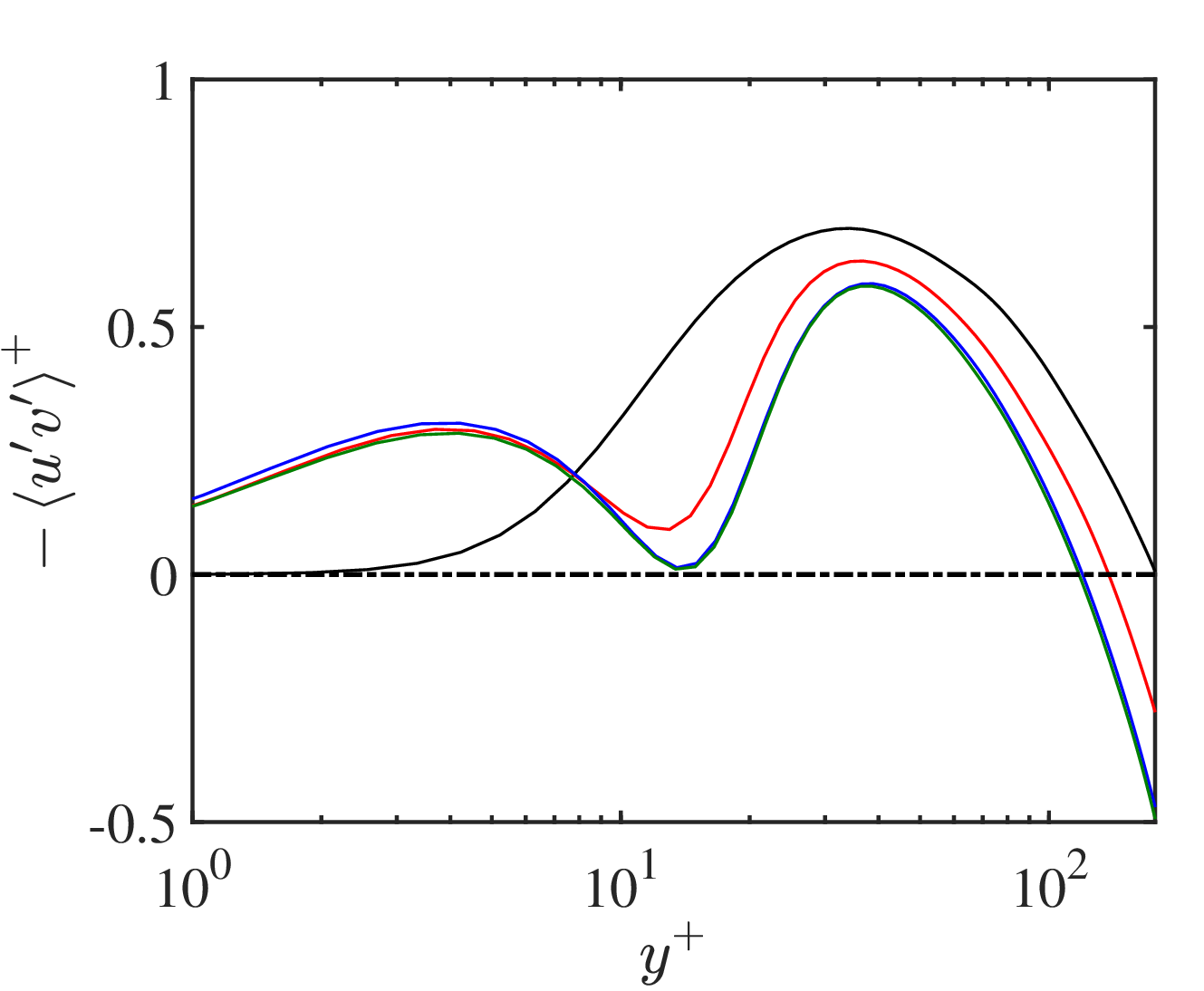}
			\put(0,75){(\textit{a})}
		\end{overpic}
		\begin{overpic}
			[scale=0.25]{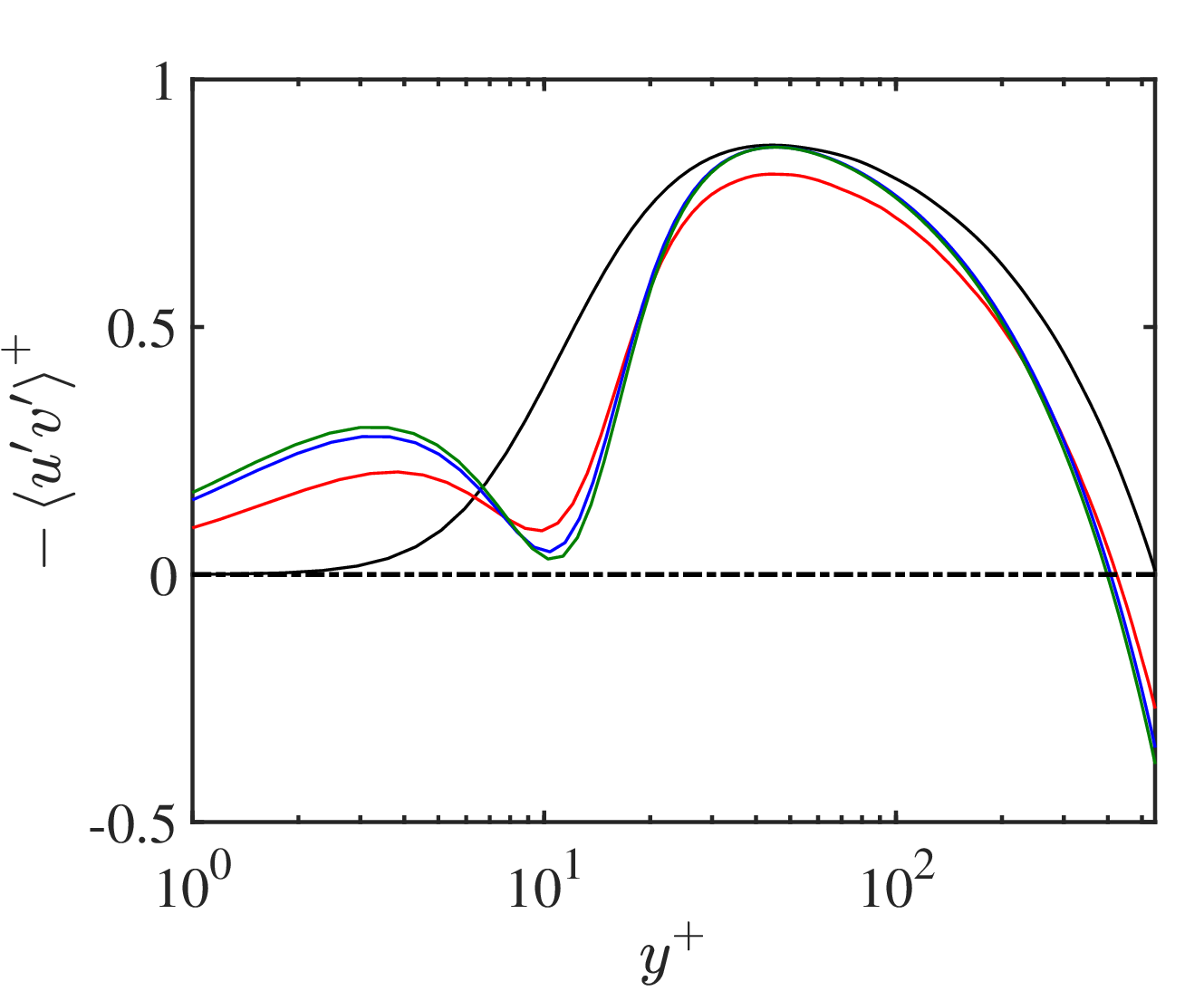}
			\put(0,75){(\textit{b})}
		\end{overpic}
	}
	\begin{overpic}
		[scale=0.25]{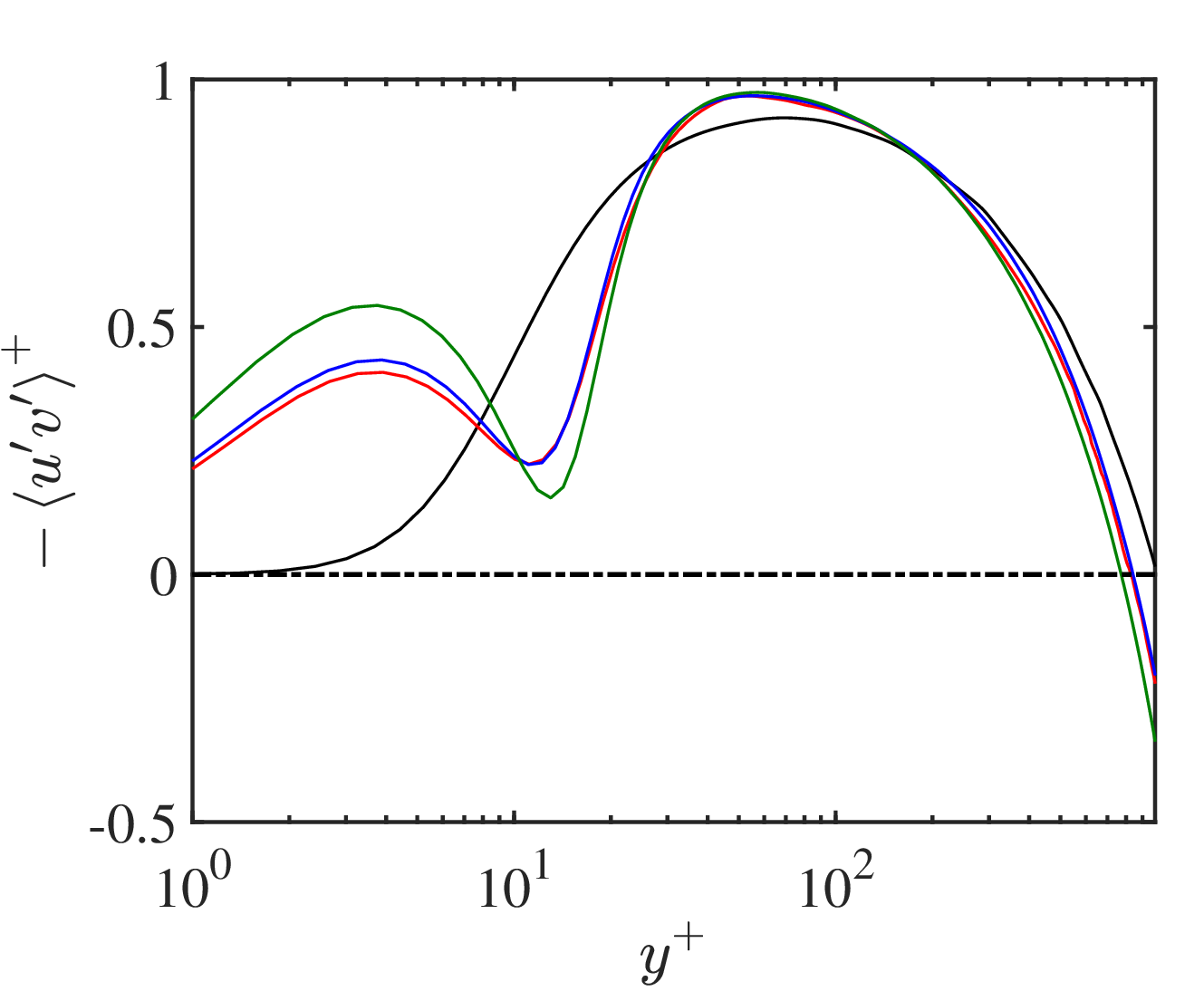}
		\put(0,75){(\textit{c})}
	\end{overpic}
	\DeclareRobustCommand\mylabeld{\tikz[baseline]{\draw[solid, black, thick] (0,0.5ex) -- (0.8,0.5ex);}}
	\DeclareRobustCommand\mylabela{\tikz[baseline]{\draw[solid, red, thick] (0,0.5ex) -- (0.8,0.5ex);}}
	\DeclareRobustCommand\mylabelb{\tikz[baseline]{\draw[solid, blue, thick] (0,0.5ex) -- (0.8,0.5ex);}}
	\DeclareRobustCommand\mylabelc{\tikz[baseline]{\draw[solid, g00, thick] (0,0.5ex) -- (0.8,0.5ex);}}

	\caption{
		Wall-normal distributions of the averaged Reynolds shear stress under different control strategies.
		(\textit{a}) C180, (\textit{b}) C550, (\textit{c}) C1000.
		\mylabeld, Cases with suffix '-0'; \mylabela: with suffix '-1'; \mylabelb: with suffix '-2'; \mylabelc: with suffix '-3'.
	}
	\label{fig:uv-avg}
	
\end{figure}

Figure \ref{fig:uv-avg} presents the wall-normal distributions of the averaged Reynolds shear stress, where $\left\langle \varphi\right\rangle$ denotes the variable $\varphi(x,y,z,t)$ averaged over the streamwise, spanwise, and temporal directions.
Although not shown in the figure, the Reynolds stress at the wall is always zero, confined by the boundary conditions.
The Reynolds stress in the viscous sublayer is higher in the controlled case compared to the uncontrolled case, due to the application of blowing and suction at the wall.
In the controlled case, the Reynolds stress slightly increases with height, reaching a peak near $y^+=5$, before rapidly decreasing and forming a trough around $y^{+}=10\sim12$.
Compared to the uncontrolled case, the Reynolds stress with control significantly decreases in the region of $10<y^{+}<20$.
This decreasing trend becomes more pronounced as the range of $v_{w}^{\prime}$ is further extended.
According to the FIK identity proposed by \cite{fukagata2002contribution}, this reduction in Reynolds stress also directly leads to a decrease in the skin friction.
At a low Reynolds number, $Re_{\tau}^{0}\thickapprox180$, the Reynolds stress in the logarithmic region is lower with control, as shown in figure \ref{fig:uv-avg}(\textit{a}). 
However, this trend gradually disappears at higher Reynolds numbers, corresponding to a decrease in drag reduction rate.

\begin{figure}
	\centering
	\subfigure{%
		\begin{overpic}
			[scale=0.25]{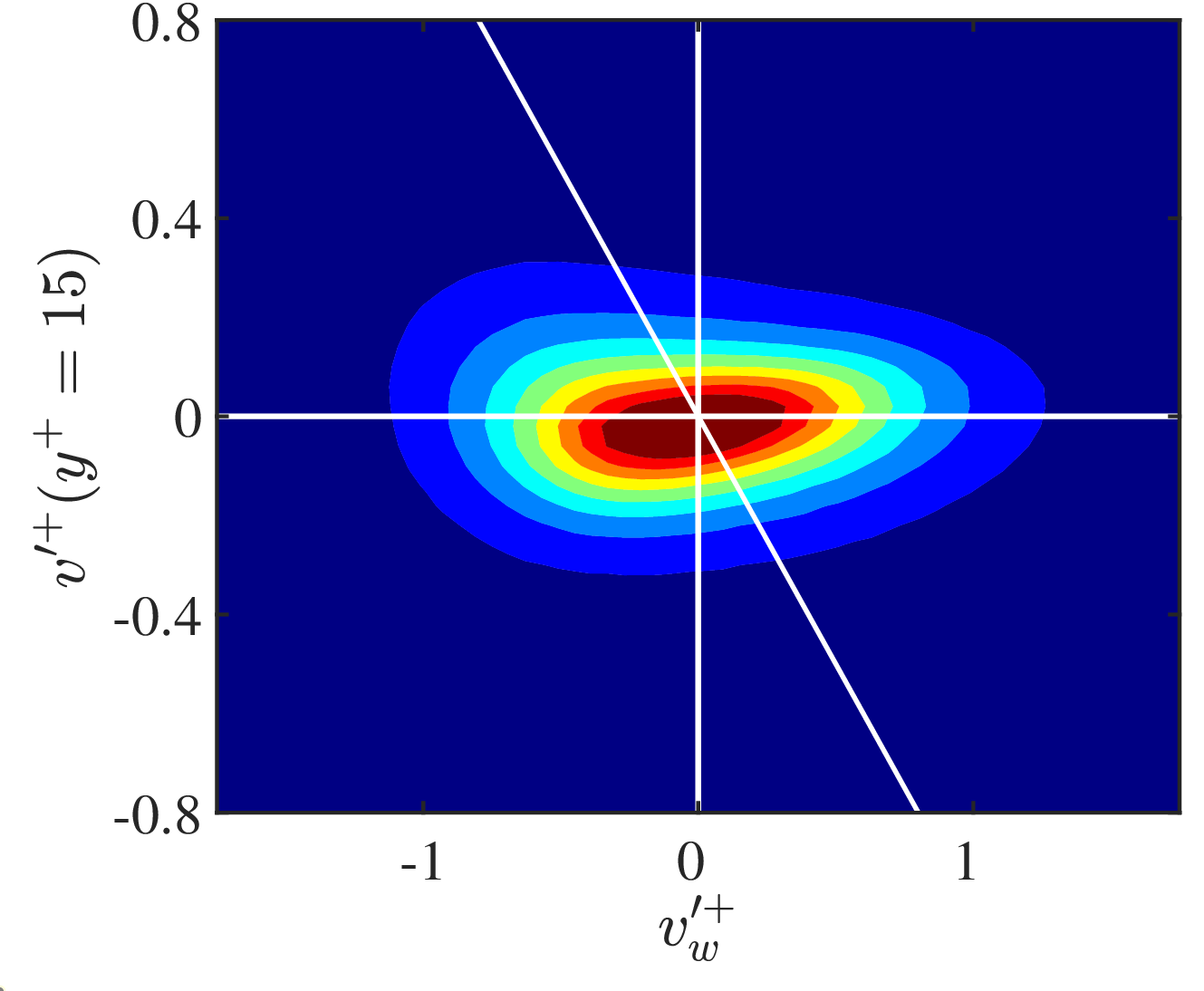}
			\put(0,80){(\textit{a})}
		\end{overpic}
		\begin{overpic}
			[scale=0.25]{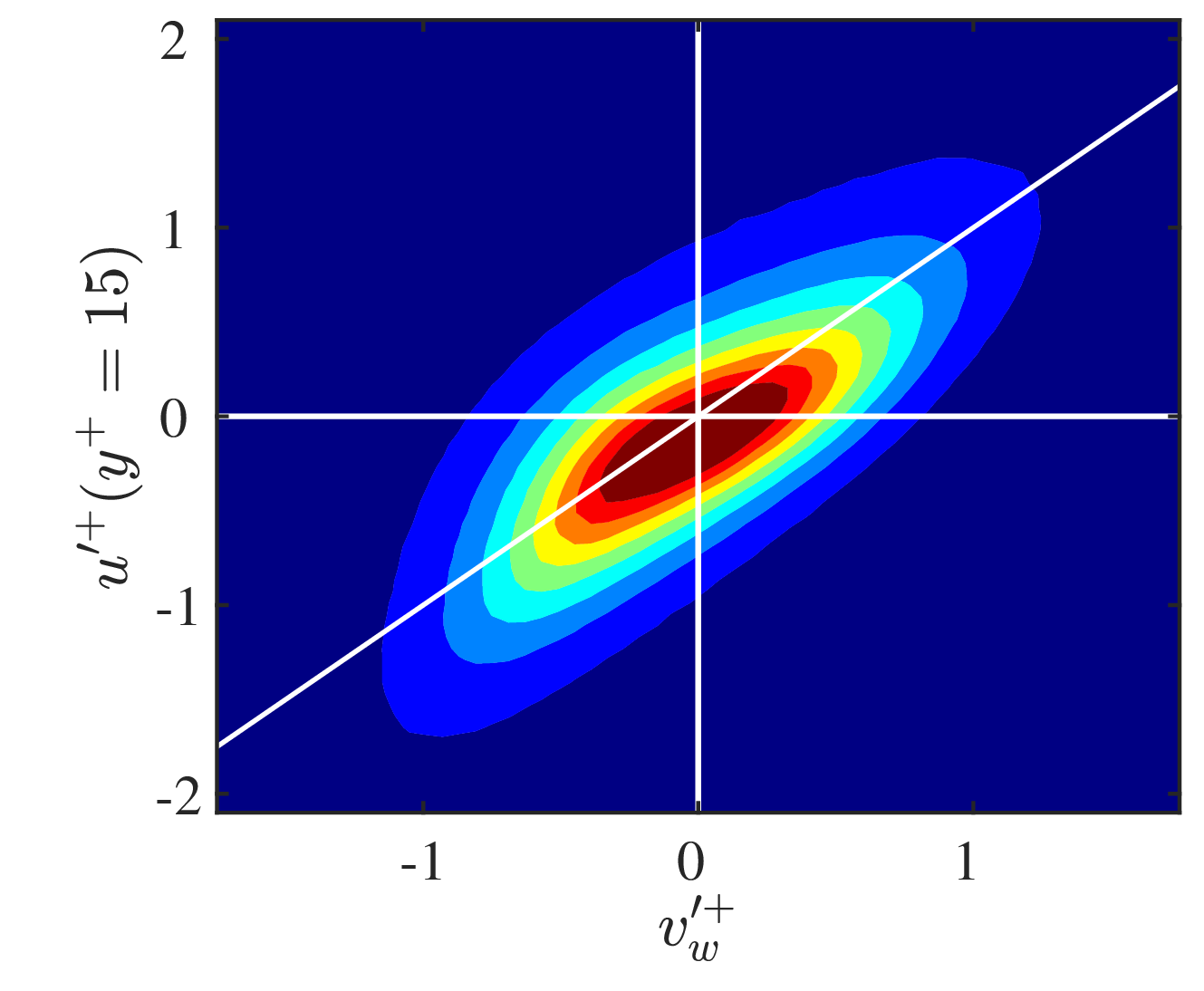}
			\put(2,80){(\textit{b})}
		\end{overpic}
	}
	
	\caption{%
		Joint probability density function of the wall blowing and suctions $v_{w}^{\prime}$ with (\textit{a}) $v^{\prime}$ and (\textit{b}) $u^{\prime}$ at $y^+=15$ in case C1000-3. 
		The white diagonals denote (\textit{a}) $v^{\prime}=-v_{w}^{\prime}$ and (\textit{b}) $u^{\prime}=v_{w}^{\prime}$, respectively.
		Contour levels are $0.1 (0.1) 0.8$ of the maximum probability density.
	}
	\label{fig:pdf_vw_15}
\end{figure}

The relationship between wall blowing and suction velocity and the velocity fluctuations at the detection plane is a crucial aspect of flow control.
Traditional opposition control employs blowing and suction with equal magnitudes but opposite directions to the wall-normal velocity fluctuations at the detection plane.
Consequently, the correlation $R$ between the blowing and suction velocity $v_{w}^{\prime}$ and $v^{\prime}$ at $y^+=15$, defined as
\begin{equation}
	R(v_{w}^{\prime},v^{\prime})=\frac{\left\langle v_{w}^{\prime}v^{\prime}\right\rangle }{\sqrt{\left\langle v_{w}^{\prime}v_{w}^{\prime}\right\rangle \left\langle v^{\prime}v^{\prime}\right\rangle }},
	\label{eq:R00}
\end{equation}
would be strictly $-1$.
Unlike opposition control, the current DRL-based control strategy is based on the streamwise velocity fluctuations $u^{\prime}$ at the detection plane $y^+=15$ as the input state.
The joint probability density function (p.d.f.) of wall blowing and suctions $v_{w}^{\prime}$ with velocity fluctuations at the near-wall detection plane is shown in figure \ref{fig:pdf_vw_15}, using the results from case C1000-3 as an example.
Here, $v_{w}^{\prime}$ has a relatively weak correlation with $v^{\prime}$ at the detection plane $y^+=15$, as illustrated in figure \ref{fig:pdf_vw_15}(\textit{a}), where $R(v_{w}^{\prime},v^{\prime})=-0.10$.
Conversely, the joint p.d.f. between $v_{w}^{\prime}$ and $u^{\prime}$ at $y^+=15$ in figure \ref{fig:pdf_vw_15}(\textit{b}) is predominantly aligned with the first and third quadrants.
This alignment indicates that wall blowing tends to occur beneath high-speed regions near the wall, while suction is more likely beneath low-speed regions.
Furthermore, the correlation $R(v_{w}^{\prime},u^{\prime})=0.71$ is positive.
This behavior closely resembles the mechanism identified by \cite{lee2023turbulence}, where their DRL models, using wall streamwise shear stress as input states, exhibited a similar trend of applying blowing beneath high-speed regions and suction beneath low-speed streaks.
They also observed a strong correlation between wall actuation and streamwise wall shear stress, similar to figure \ref{fig:pdf_vw_15}(\textit{b}), suggesting that these DRL models effectively reduce drag through direct control of sweep and ejection events.
Despite not being shown in the figure, this pattern consistently appears across all cases in our current work, implying a stronger connection between the DRL-based control strategies and the streamwise velocity fluctuations as the input state.


\subsection{Kinematic analysis of drag reduction based on virtual wall theory}\label{sec:virtual_wall}

The DRL-based control strategy could lead to larger drag reduction compared to the traditional opposition control, however, the underlying mechanism requires further investigation.
This subsection utilizes the virtual wall theory proposed by \cite{hammond1998observed} to analyze the drag reduction mechanism from a kinematic perspective.

According to the virtual wall theory by \cite{hammond1998observed}, wall blowing and suction create a virtual wall between the actual wall and the detection plane. 
This virtual wall hinders streamwise vortices from bringing high-speed fluid to the wall, which would otherwise create local high friction zones, thereby resulting in drag reduction.
The drag reduction effect is mainly influenced by two factors: the height of the virtual wall $y_{vw}$ and the magnitude of the residual Reynolds stress on the virtual wall $-\left\langle u^{\prime}v^{\prime}\right\rangle_{vw}$.
Specifically, the higher the virtual wall, the better the drag reduction effect.
And the lower the residual Reynolds stress, the stronger the virtual wall's ability to impede wall-normal momentum transport, resulting in better drag reduction.

The height of the virtual wall and the residual Reynolds stress under different control strategies are detailed in table \ref{tab:cases_DR}.
At the low Reynolds number $Re_{\tau}^{0}\thickapprox180$, as the range of blowing and suction velocities is further expanded, the height of the virtual wall gradually increases, and the residual Reynolds stress on the virtual wall gradually decreases.
Both these changes correspond to an improvement in drag reduction effect.
The values of $y_{vw}$ and $-\left\langle u^{\prime}v^{\prime}\right\rangle _{vw}$ for C180-2 and C180-3 are similar, resulting in comparable drag reduction rates for both cases.
Compared to the traditional opposition control method, the DRL-based control strategy in C180-3 does not show a significant reduction in residual stress on the virtual wall. 
However, its primary benefit is the ability to further elevate the virtual wall.
As the Reynolds number increases, the height of the virtual wall under the DRL-based control strategy is significantly lower for $Re_{\tau}^{0}\thickapprox550$ and $1000$ compared to the results for $Re_{\tau}^{0}\thickapprox180$. 
Additionally, the residual stress $-\left\langle u^{\prime}v^{\prime}\right\rangle _{vw}$ on the virtual wall rapidly increases with the rising Reynolds number. 
In case C550-3, $-\left\langle u^{\prime}v^{\prime}\right\rangle _{vw}$ is approximately three times that of case C180-3, and in case C1000-3, $-\left\langle u^{\prime}v^{\prime}\right\rangle _{vw}$ exceeds that of case C180-3 by more than ten times. 
These two factors together lead to a decrease in the drag reduction efficiency of the DRL-based control strategy.
Moreover, at higher Reynolds numbers, the control strategy optimized through DRL is less effective at impeding wall-normal momentum transport compared to traditional opposition control, as evidenced by the contrast of residual Reynolds stress. 
The main advantage of DRL optimization lies in its ability to effectively plan the wall blowing and suction in an expanded range, thereby elevating the virtual wall to a higher position and achieving better drag reduction efficiency.


\begin{figure}
	\centering
	
	\subfigure{
		\begin{overpic}
			[scale=0.18]{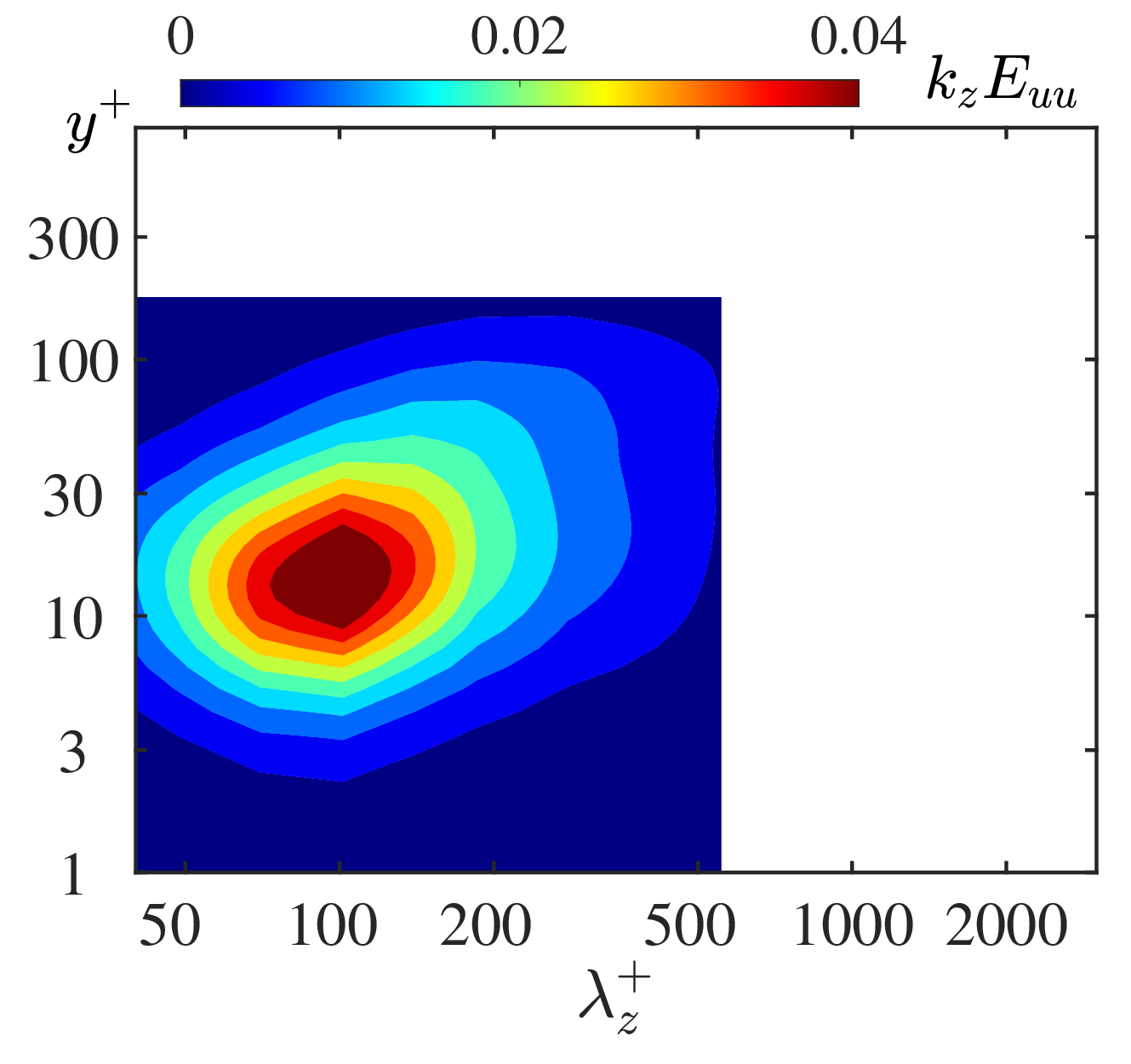}
			\put(-3,89){(\textit{a})}
		\end{overpic}
		\begin{overpic}
			[scale=0.18]{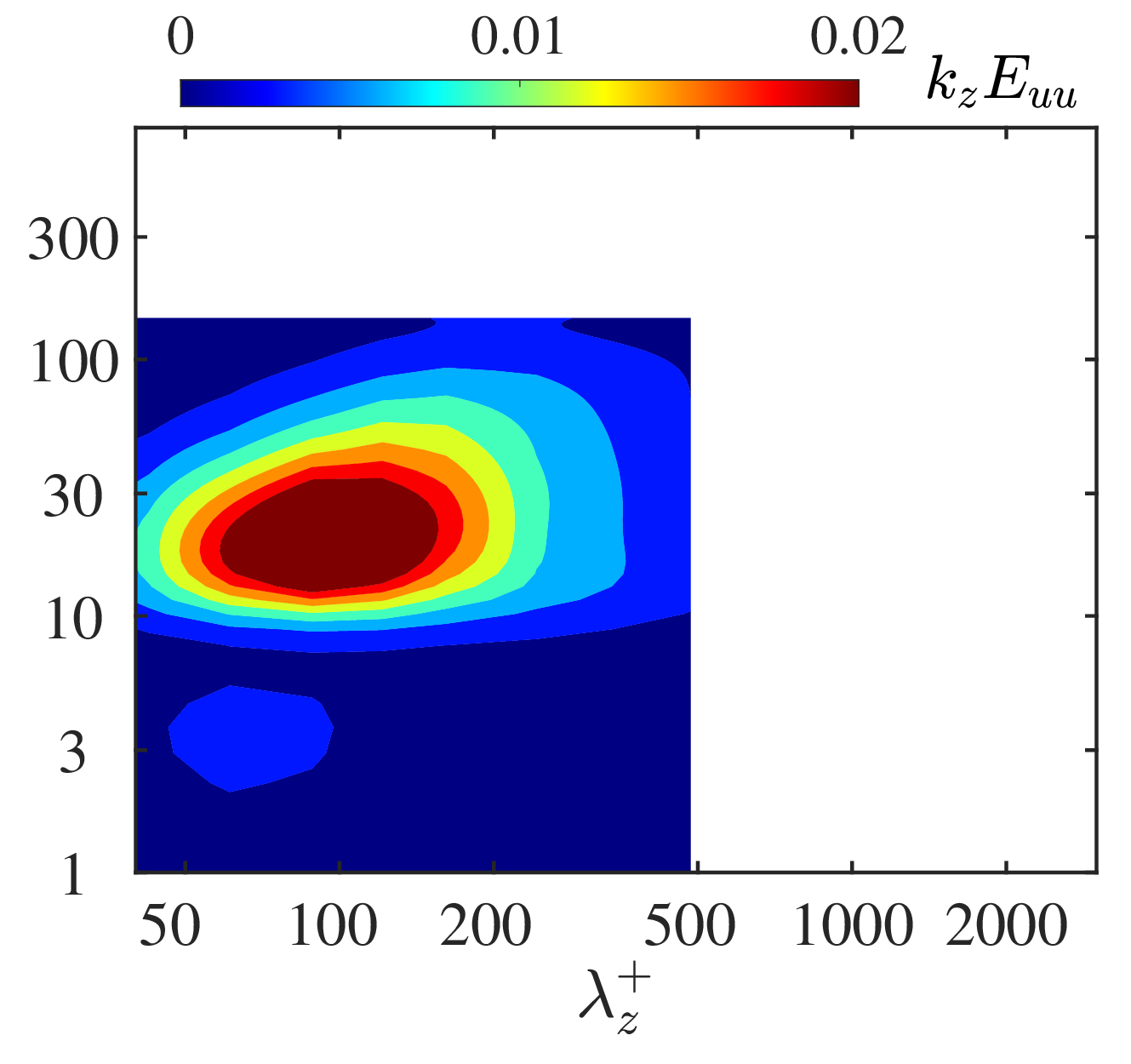}
			\put(0,89){(\textit{b})}
		\end{overpic}
		\begin{overpic}
			[scale=0.18]{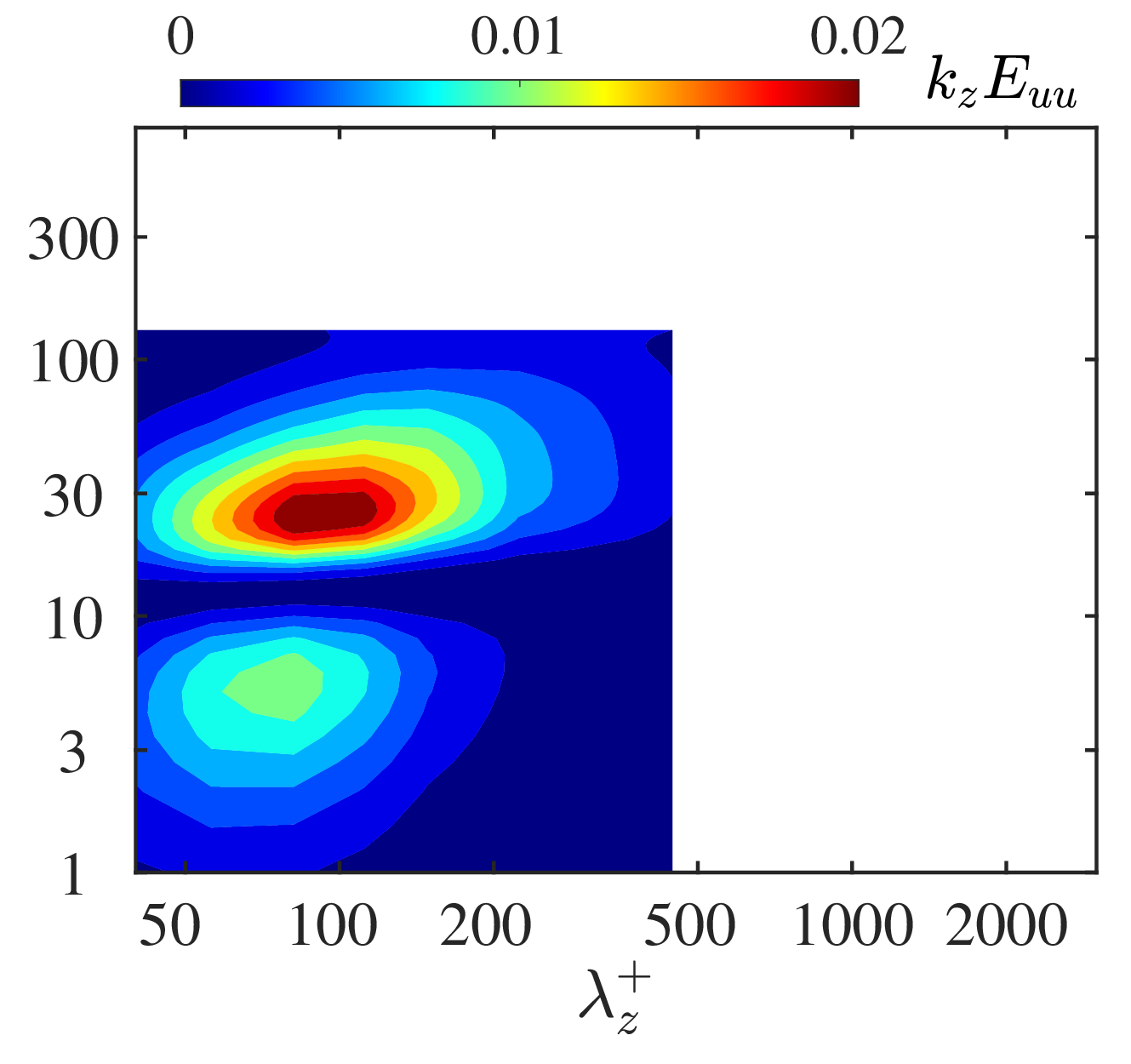}
			\put(0,89){(\textit{c})}
		\end{overpic}
	}
	\subfigure{
		\begin{overpic}
			[scale=0.18]{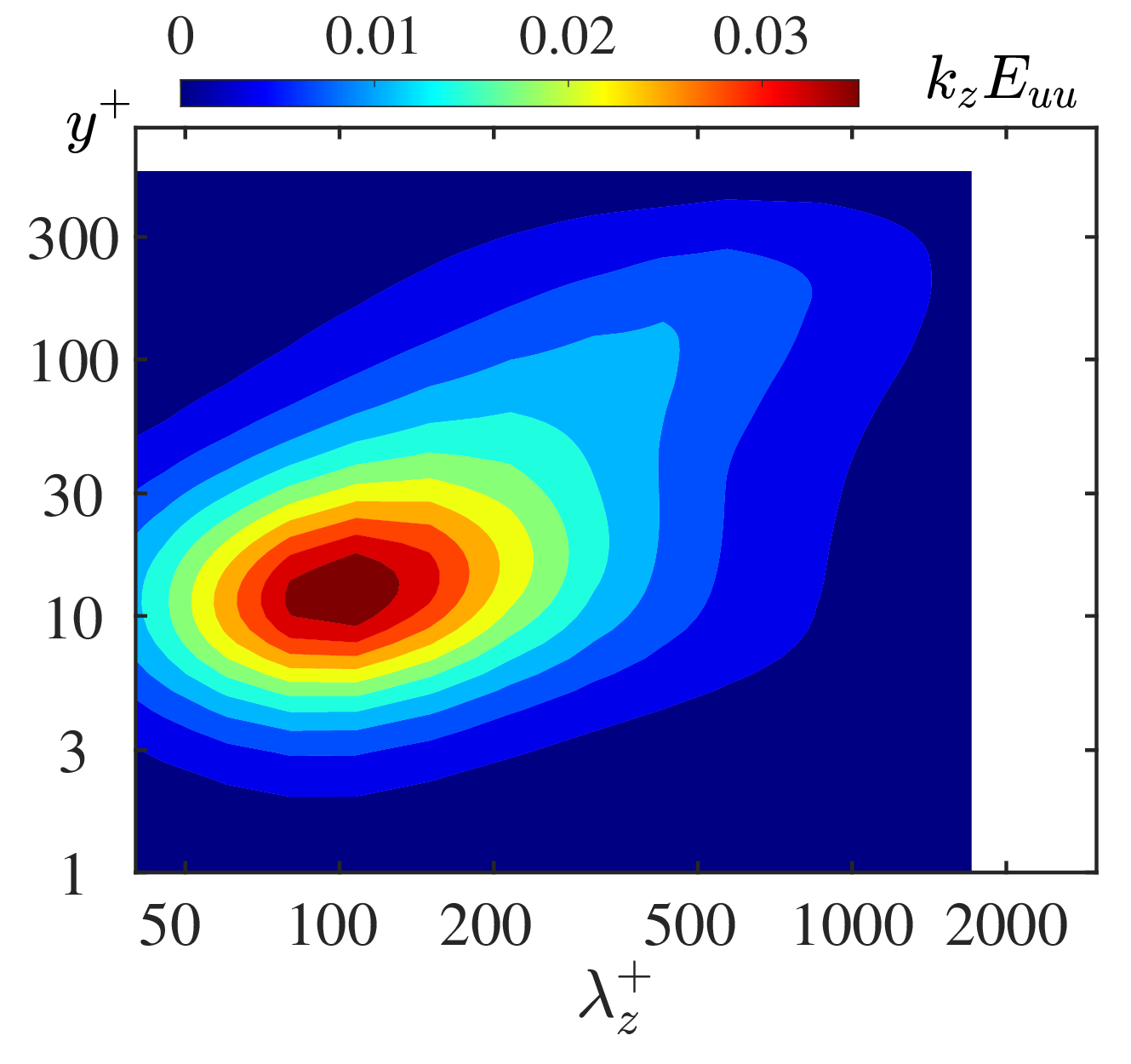}
			\put(-3,89){(\textit{d})}
		\end{overpic}
		\begin{overpic}
			[scale=0.18]{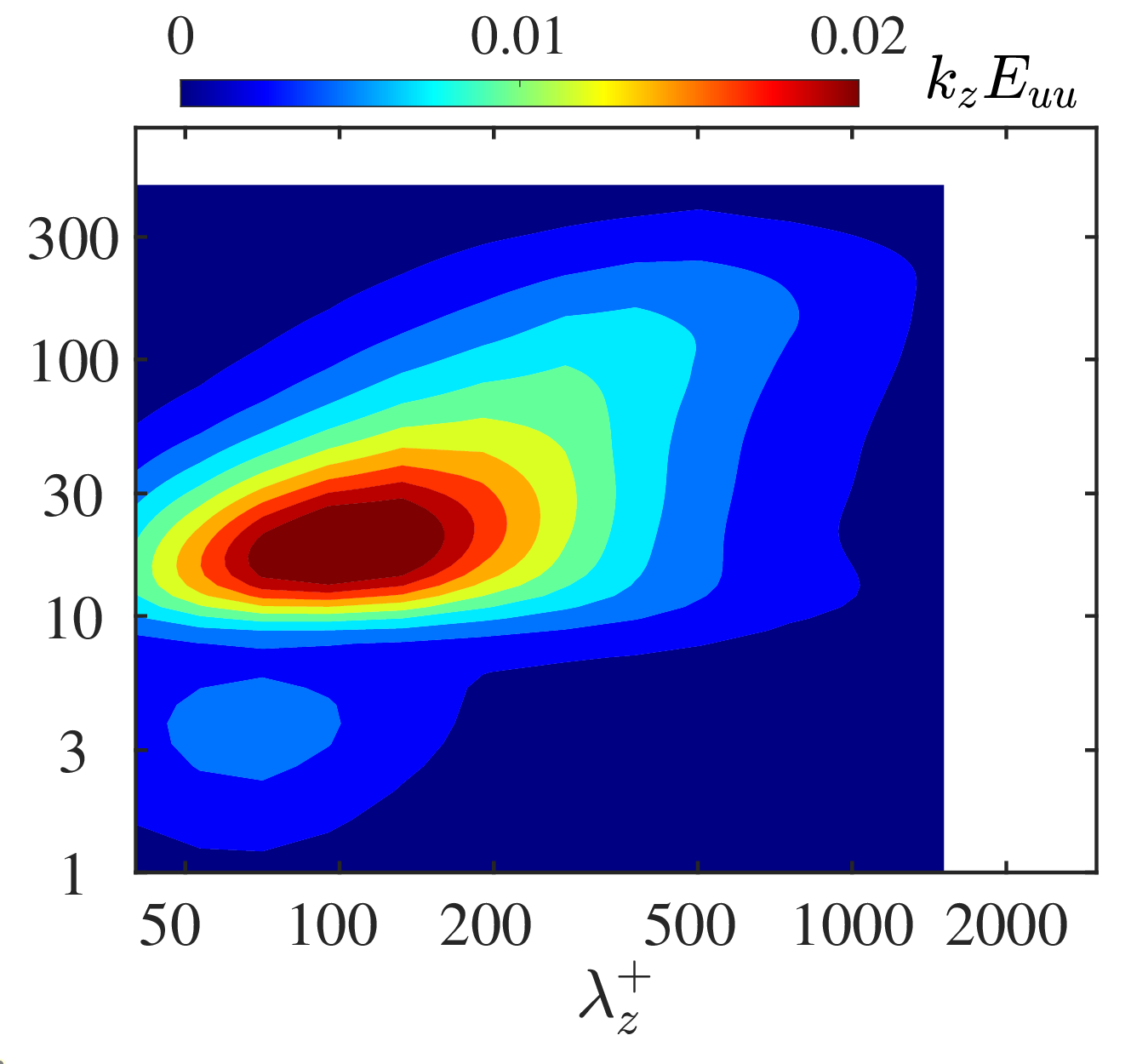}
			\put(0,89){(\textit{e})}
		\end{overpic}
		\begin{overpic}
			[scale=0.18]{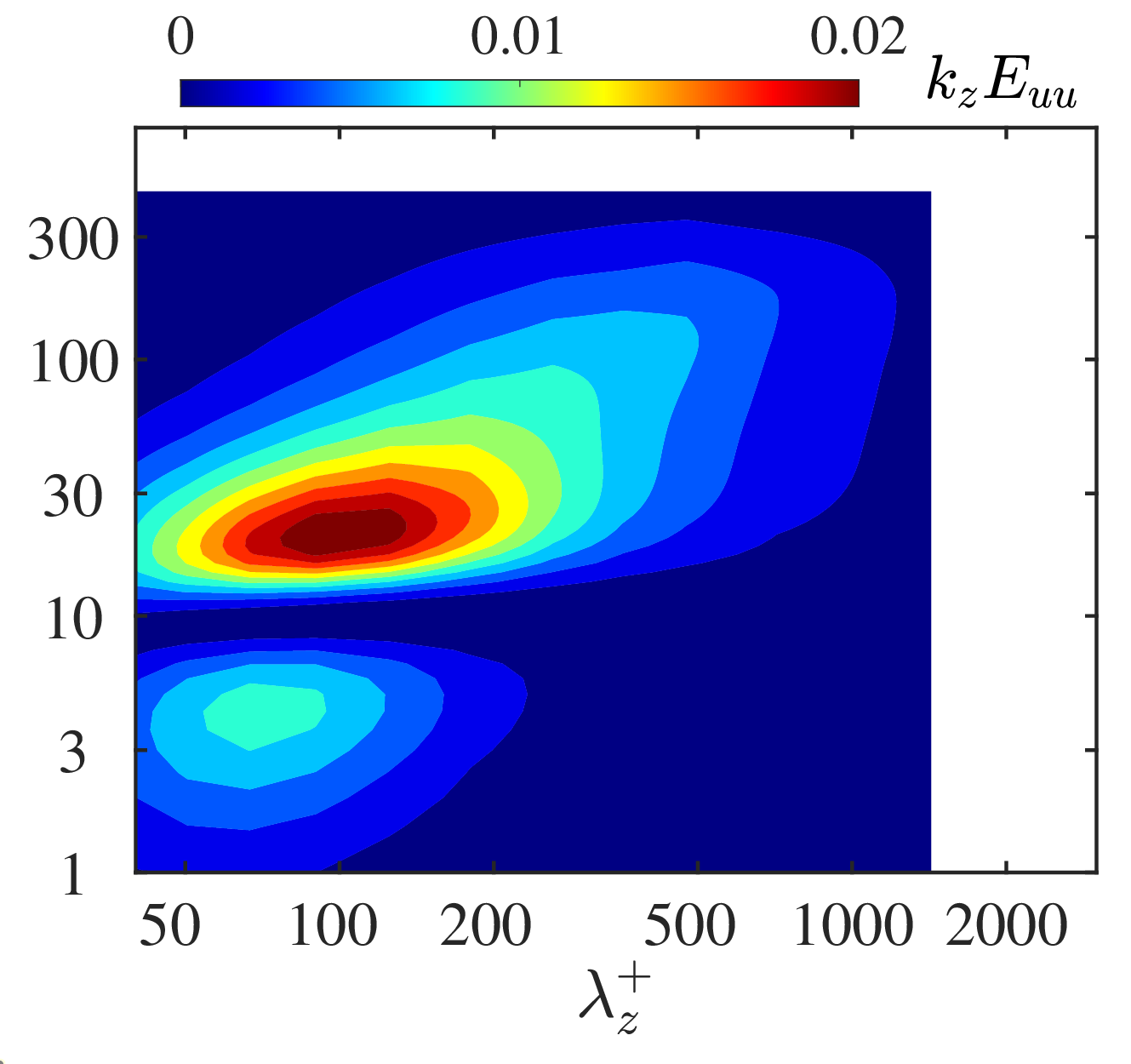}
			\put(0,89){(\textit{f})}
		\end{overpic}
	}
	\subfigure{
		\begin{overpic}
			[scale=0.18]{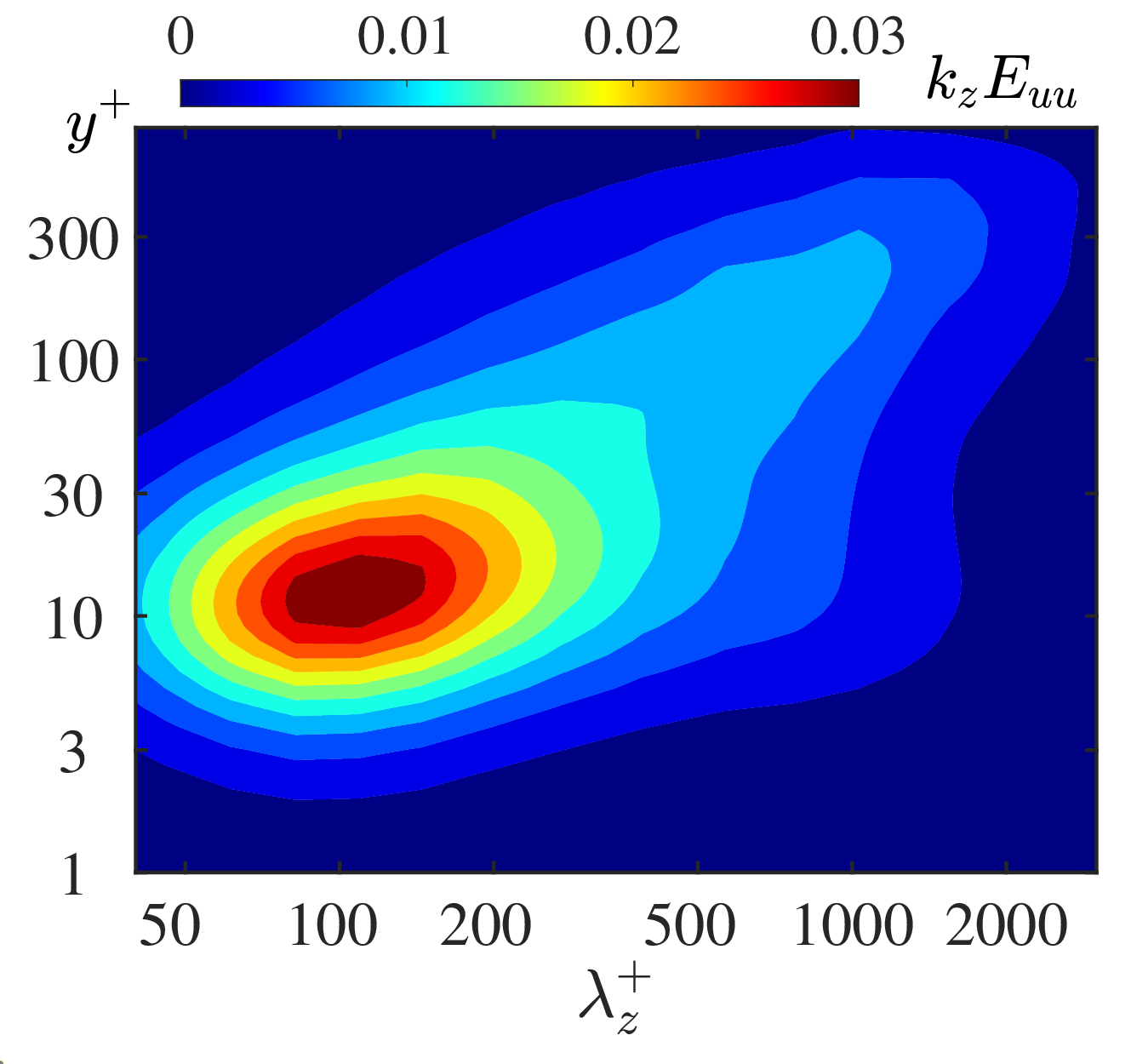}
			\put(-3,89){(\textit{g})}
		\end{overpic}
		\begin{overpic}
			[scale=0.18]{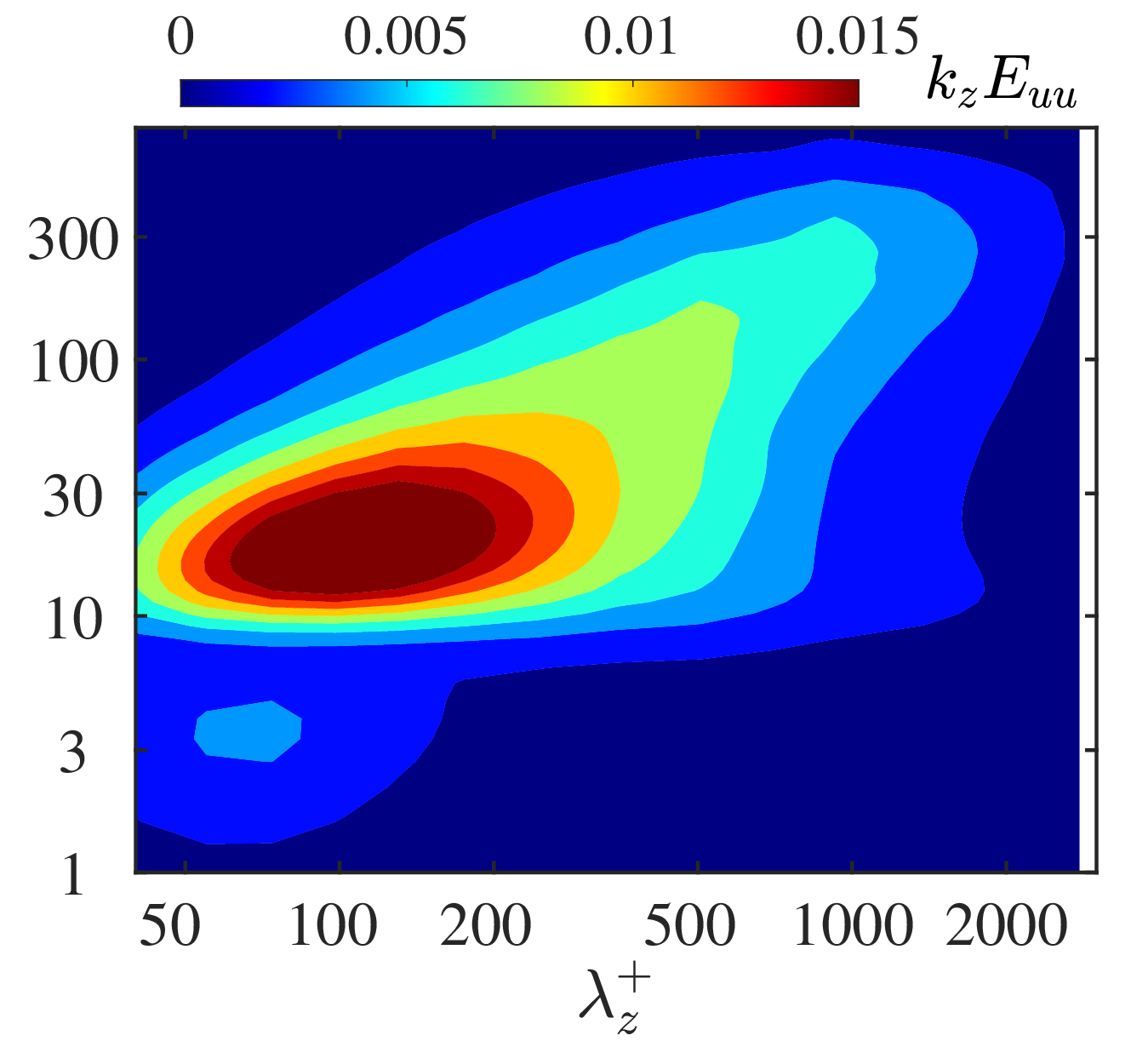}
			\put(0,89){(\textit{h})}
		\end{overpic}
		\begin{overpic}
			[scale=0.18]{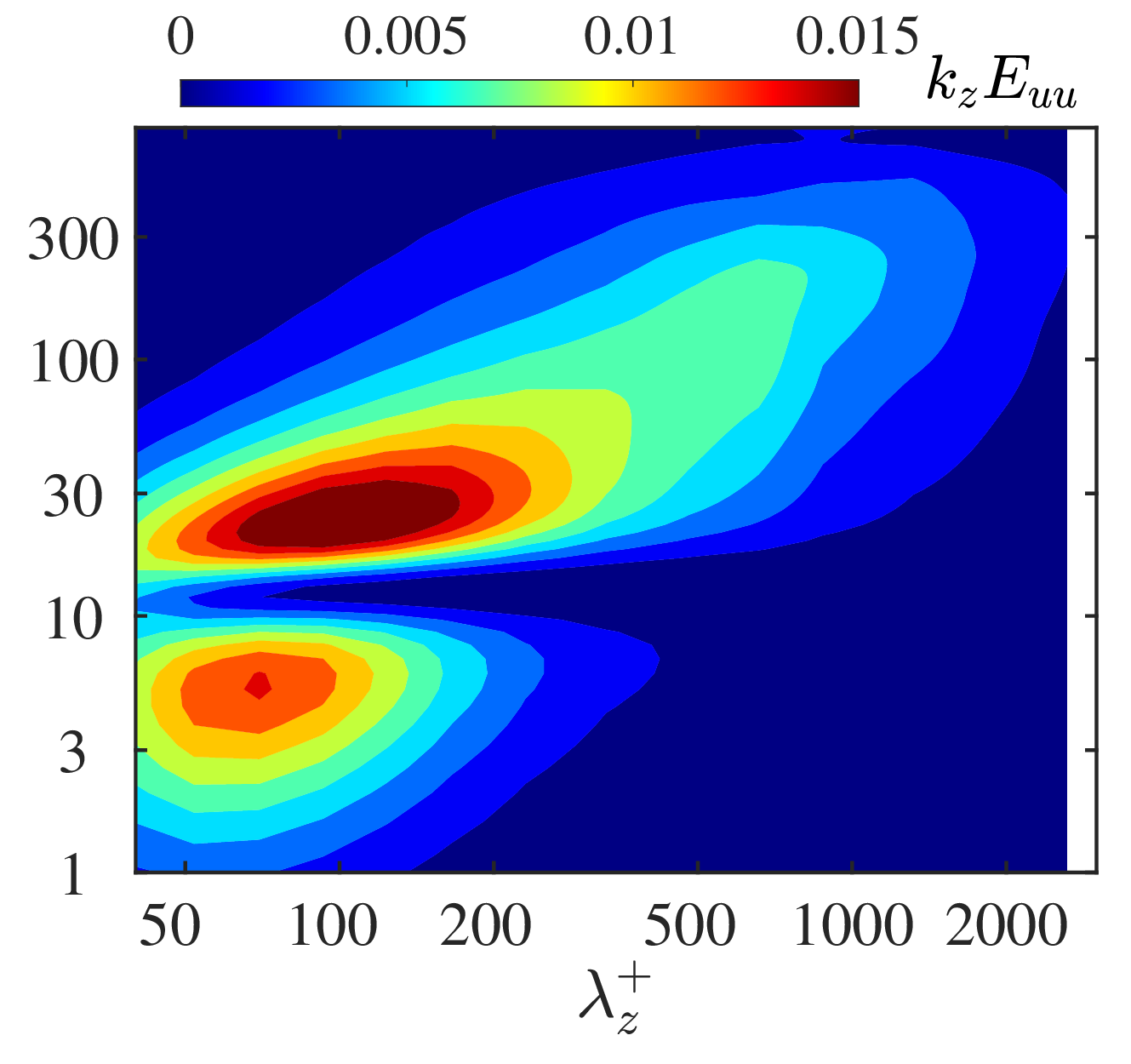}
			\put(0,89){(\textit{i})}
		\end{overpic}
	}

	\caption{
		Premultiplied spanwise energy spectra $k_{z}E_{uu}$ of streamwise velocity fluctuations $u^{\prime}$ under different control strategies.
		For $Re_{\tau}^{0}\thickapprox180$,  (\textit{a}) C180-0,  (\textit{b}) C180-opp,  (\textit{c}) C180-3.
		For $Re_{\tau}^{0}\thickapprox550$,  (\textit{d}) C550-0,  (\textit{e}) C550-opp,  (\textit{f}) C550-3.
		For $Re_{\tau}^{0}\thickapprox1000$, (\textit{g}) C1000-0, (\textit{h}) C1000-opp, (\textit{i}) C1000-3.
	}
	\label{fig:spectra}
	
\end{figure}


To further quantify the impact of control strategies on the scales of the structures at different heights, especially the flow structures near the virtual wall, figure \ref{fig:spectra} presents the premultiplied energy spectra $k_{z}E_{uu}$ of $u^{\prime}$.
Here, $k_z$ is the spanwise wavenumber, and $\lambda_z=2\pi/k_z$ is the corresponding wavelength.
In the near-wall region, the flow is dominated by streaks with a spanwise scale of $\lambda_{z}^{+}\approx100$ and a wall-normal height concentrated around $y^{+}=15$, as depicted in figures \ref{fig:spectra}(\textit{a})(\textit{d})(\textit{g}).
After applying wall blowing and suction control, the peak velocity fluctuations in the near-wall region shift to a higher position, and a second spectral peak emerges in the viscous sublayer. 
These two peaks are separated by the virtual wall, as also suggested by \cite{hammond1998observed}.
Notably, in cases utilizing the DRL-based control strategy (C180-3, C550-3, and C1000-3), the virtual wall is significantly higher than in those using traditional opposition control, corroborating the conclusions drawn from table \ref{tab:cases_DR}.
The peak of velocity fluctuations corresponding to the near-wall streaks in the buffer layer also rises to a higher position.
Furthermore, the intensity of velocity fluctuations in the viscous sublayer significantly increases after applying the DRL-based control strategy.

On the other hand, the characteristic scales of flow structures remain largely unchanged under different control strategies.
The spanwise sizes of the near-wall streaks are consistently around $\lambda_{z}^{+}\approx100$, while the peak of velocity fluctuations in the viscous sublayer due to wall blowing and suction stays within $\lambda_{z}^{+}=60\sim80$, slightly smaller than the spanwise sizes of the streaks.
At high Reynolds numbers, wall blowing and suction have a trivial effect on the spanwise sizes of the outer large-scale structures, which remain around $\lambda_{z}\thickapprox O(h)$.
In the cases without control, the footprint of outer large-scale structures penetrates deeply into the near-wall region, as shown by the near-wall large-scale components in figures \ref{fig:spectra}(\textit{d})(\textit{g}).
This phenomenon, known as the superposition effect \citep{hoyas06,hutchins2007large,mathis2009large,marusic2010high}, is noteworthy.
After applying wall control, however, the footprint of outer large-scale structures cannot penetrate the virtual wall to reach the viscous sublayer or contribute to the residual velocity fluctuations on the virtual wall.
This is particularly evident in cases C550-3 and C1000-3 using DRL models, as shown in figures \ref{fig:spectra}(\textit{f})(\textit{i}).
Thus, the superposition effect does not directly cause the increasing residual Reynolds stress on the virtual wall at rising Reynolds numbers.
Its impact on the decreasing drag reduction rate of the DRL models at high Reynolds numbers is also trivial.

\begin{figure}
	\centering
	
	\subfigure{
		\begin{overpic}
			[scale=0.21]{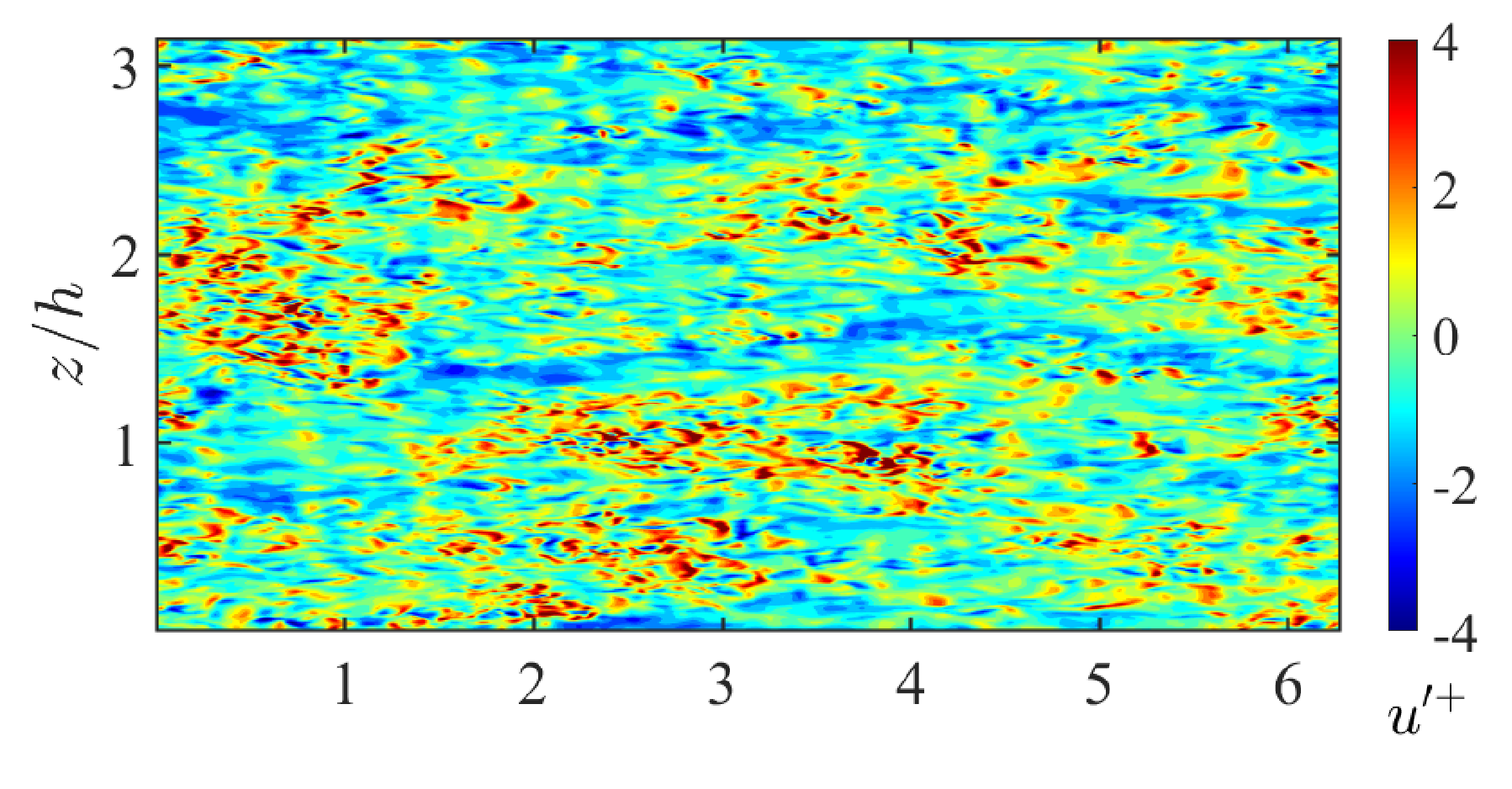}
			\put(1,52){(\textit{a})}
		\end{overpic}
		\begin{overpic}
			[scale=0.21]{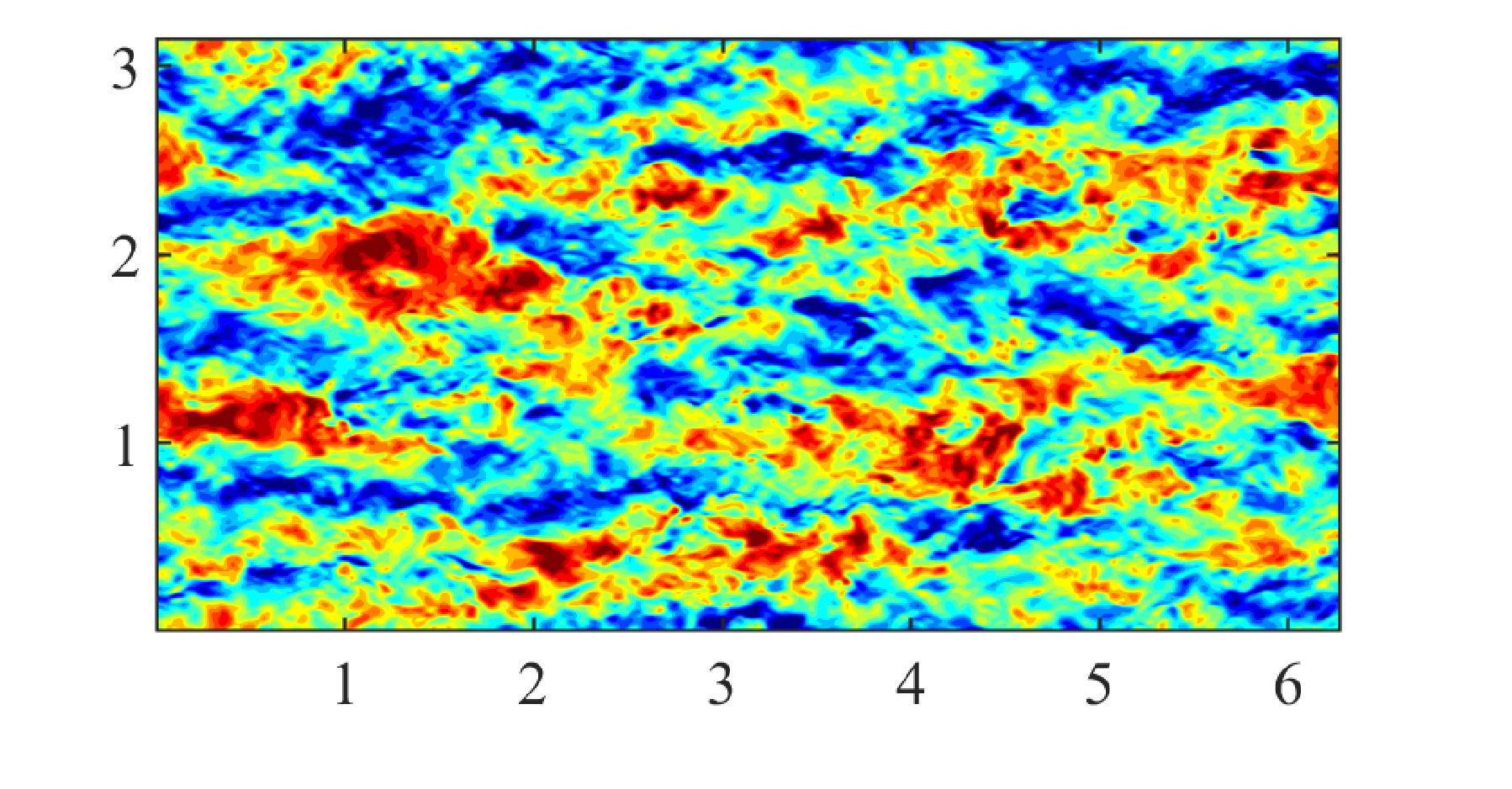}
			\put(1,52){(\textit{b})}
		\end{overpic}
	}
	
	\subfigure{
		\begin{overpic}
			[scale=0.21]{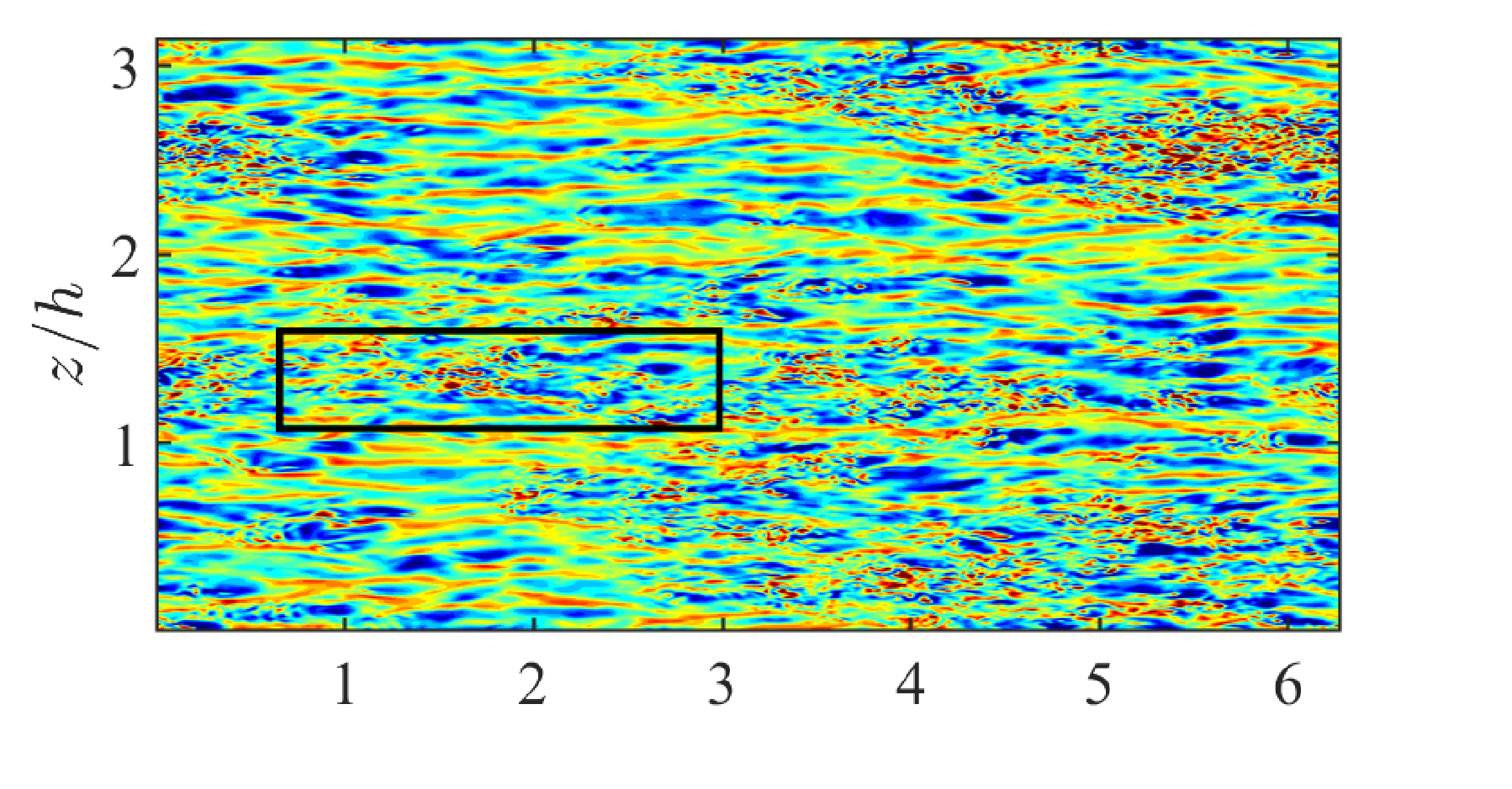}
			\put(1,52){(\textit{c})}
		\end{overpic}
		\begin{overpic}
			[scale=0.21]{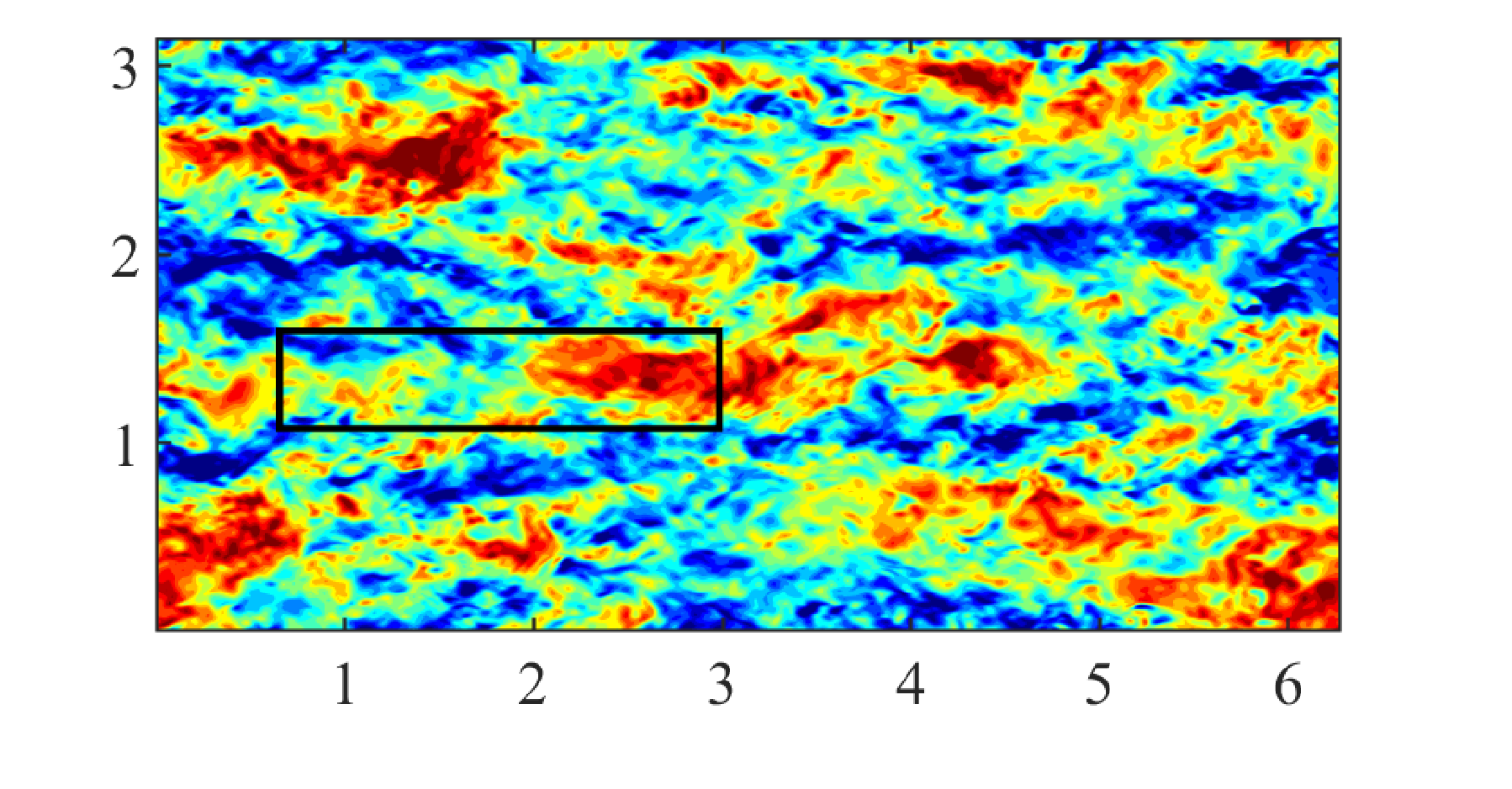}
			\put(1,52){(\textit{d})}
		\end{overpic}
	}

	\subfigure{
		\begin{overpic}
			[scale=0.21]{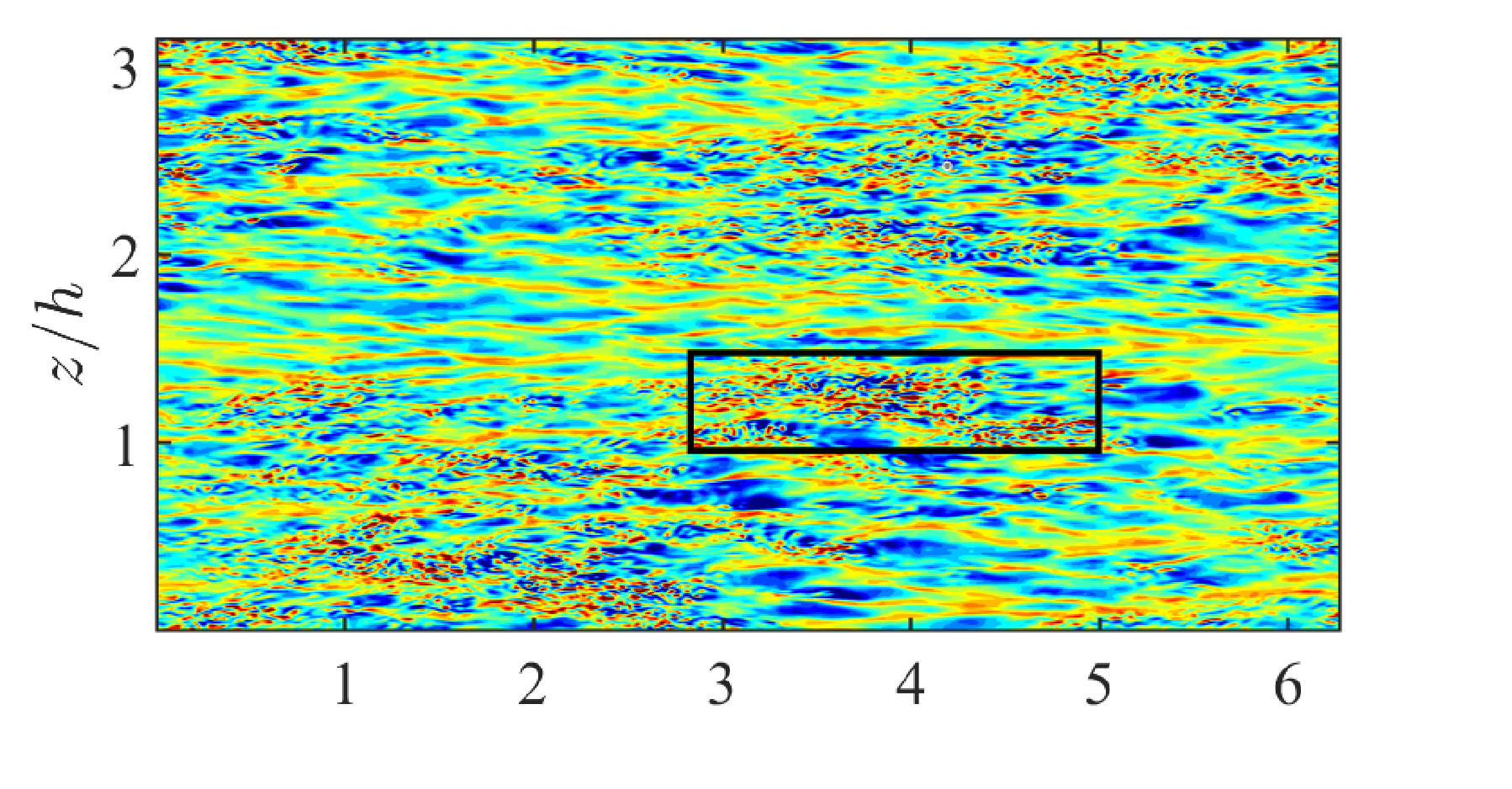}
			\put(1,52){(\textit{e})}
		\end{overpic}
		\begin{overpic}
			[scale=0.21]{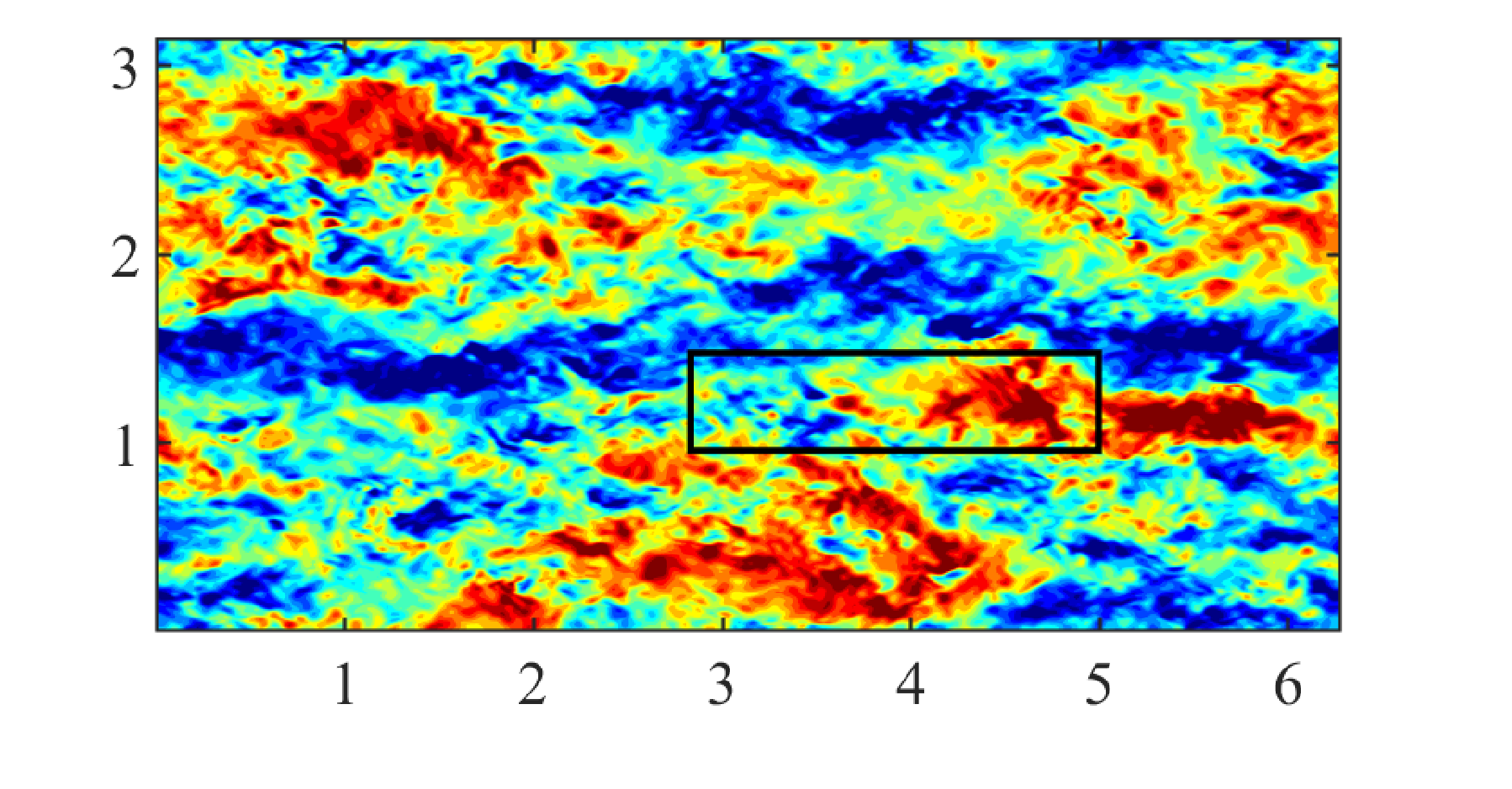}
			\put(1,52){(\textit{f})}
		\end{overpic}
	}

	\subfigure{
		\begin{overpic}
			[scale=0.21]{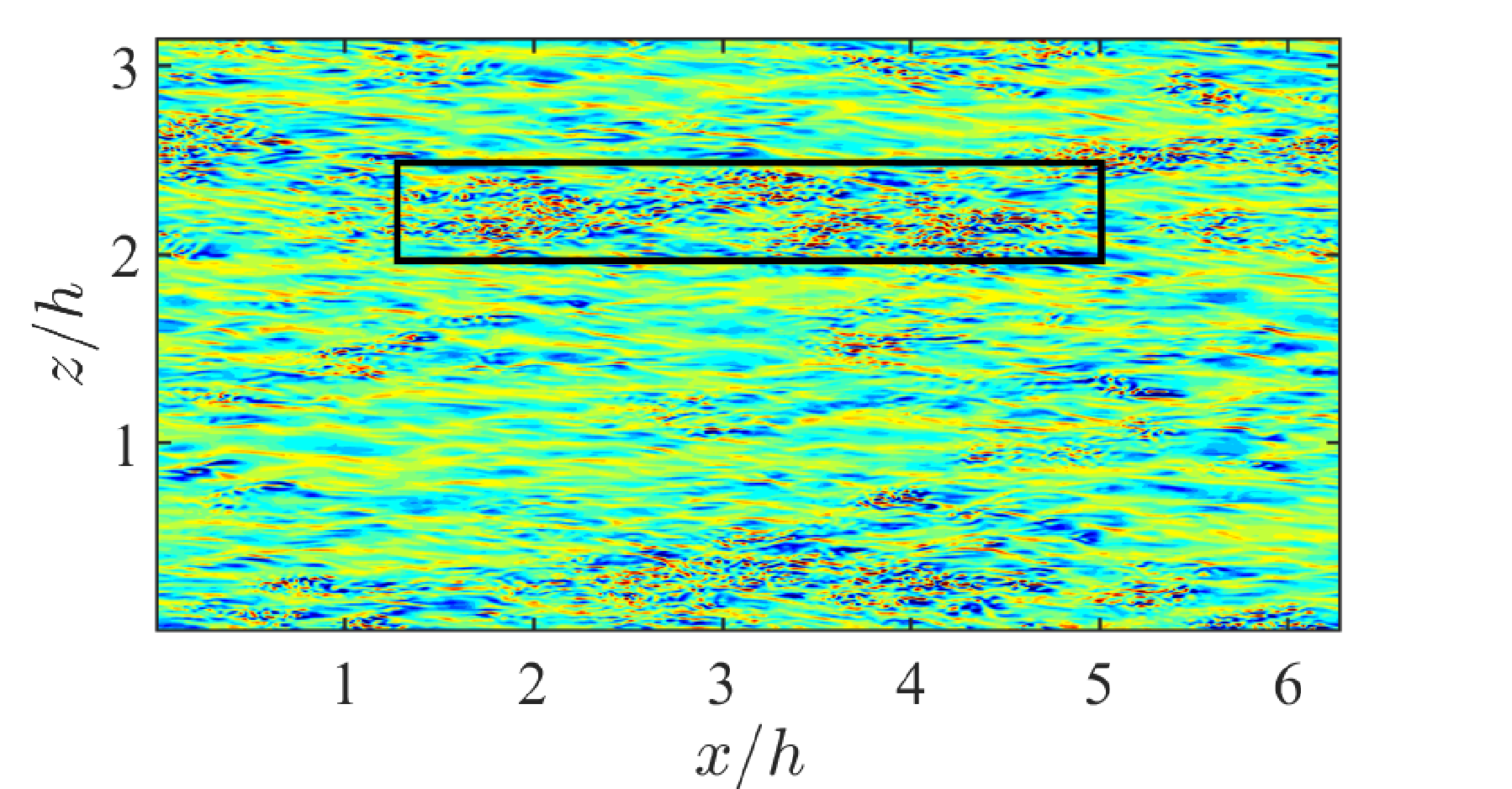}
			\put(1,52){(\textit{g})}
		\end{overpic}
		\begin{overpic}
			[scale=0.21]{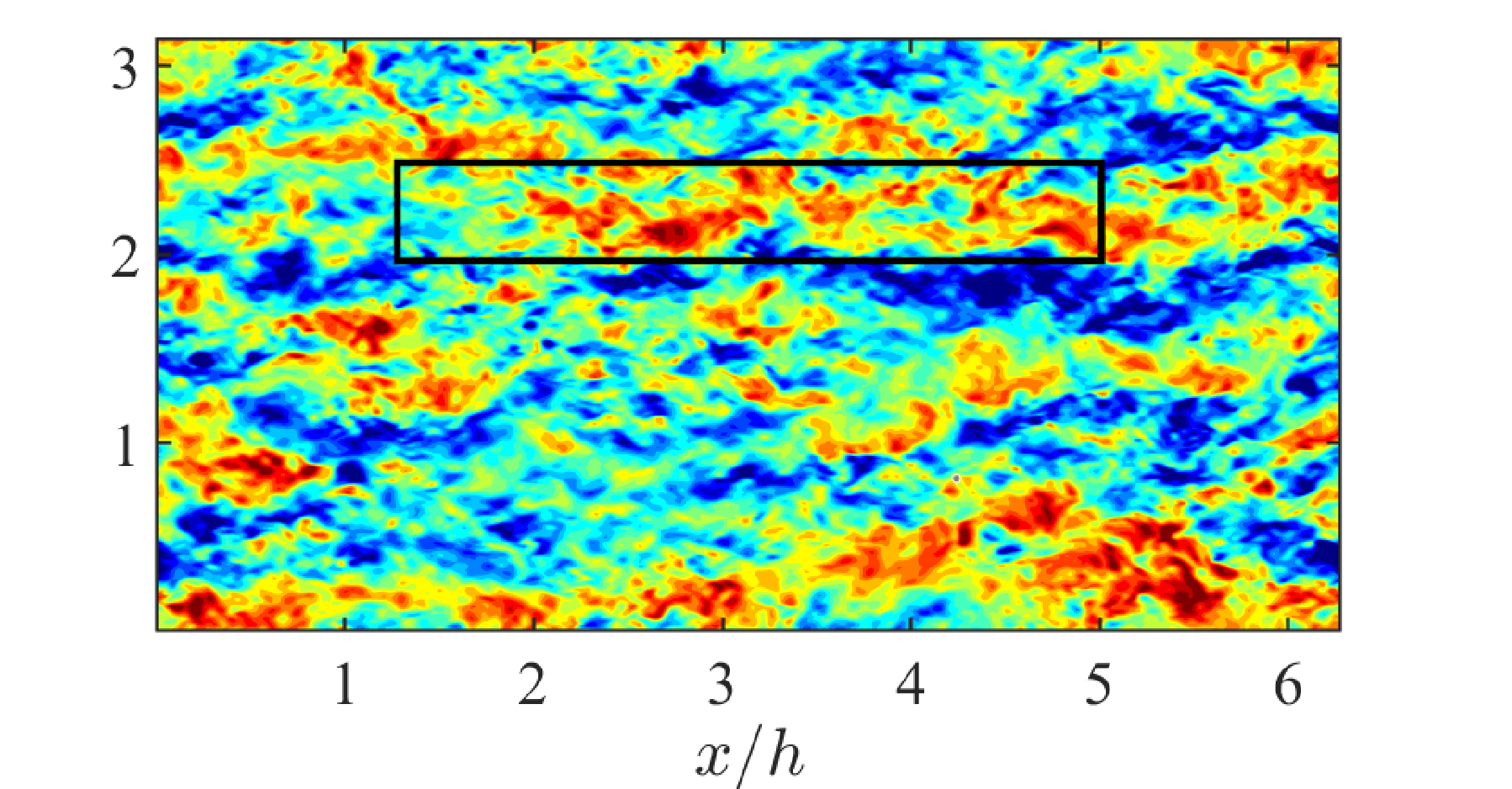}
			\put(1,52){(\textit{h})}
		\end{overpic}
	}

	\caption{
		Instantaneous distributions of $u^{\prime}$ on $(x,z)$ plane at (\textit{a})(\textit{c})(\textit{e})(\textit{g}) $y^{+}=y_{vw}^{+}$ and (\textit{b})(\textit{d})(\textit{f})(\textit{h}) $y^{+}=150$.
		(\textit{a})(\textit{b}) case C1000-opp, (\textit{c})(\textit{d}) C1000-1, (\textit{e})(\textit{f}) C1000-2, (\textit{g})(\textit{h}) C1000-3.
		The black rectangles represent some sample areas on the virtual wall where velocity fluctuations are stronger.
	}
	\label{fig:u-xz-1000}
\end{figure}

To identify the source of $-\left\langle u^{\prime}v^{\prime}\right\rangle _{vw}$ at high Reynolds numbers, figure \ref{fig:u-xz-1000} illustrates the distributions of streamwise velocity fluctuations at $y^{+}=y_{vw}^{+}$ and $y^{+}=150$.
The DRL-based control strategy reveals strong fluctuations on the virtual wall, characterized by clustered small-scale fluctuations concentrated in specific areas.
Although these fluctuations are mitigated when the range of blowing and suction velocities is expanded, they remain stronger than those observed after opposition control.
It shall be noted that these fluctuations are much smaller in size compared to the outer large-scale structures. 
Therefore, they are unlikely to be induced by the linear superposition effect, but are more plausibly related to the nonlinear amplitude modulation mechanism of the large-scale structures.
As indicated by the black rectangles in figure \ref{fig:u-xz-1000}, regions of strong fluctuations on the virtual wall often share similar spanwise locations with the outer large-scale high-speed regions, further supporting this point.
In the streamwise direction, areas with clustered fluctuations are frequently situated upstream of the large-scale high-speed regions. 
This phenomenon can be attributed to the inclination angle of the large-scale coherent structures, as suggested by the near-wall fluctuation predictive models proposed by \cite{marusic2010predictive} and \cite{mathis2011predictive}.

Further statistical evidence is required to support the relationship between the amplitude modulation of outer large-scale structures and the residual Reynolds stress at the virtual wall.
The streamwise velocity fluctuations, $u_{O}^{\prime}$, at the center of the logarithmic region $y_{O}^{+}\approx3.9\sqrt{Re_{\tau}}$, can be utilized to characterize outer large-scale structures \citep{mathis2009large,mathis2011predictive}. 
A positive $u_{O}^{\prime}$ indicates a large-scale high-speed region, while a negative $u_{O}^{\prime}$ denotes a low-speed region.
On the other hand, the residual fluctuations at the virtual wall exhibit a clustered distribution, as illustrated in figure \ref{fig:u-xz-1000}. 
In areas with strong fluctuations, the streamwise and spanwise scales of the fluctuations are smaller, and the spatial alternation between positive and negative values is more pronounced. 
Considering the impact of spatial alternation, we select the envelope of the Reynolds stress at the virtual wall, denoted as $\left|\mathcal{H}\left(\left\langle u^{\prime}v^{\prime}\right\rangle _{vw}\right)\right|$, to measure the strength of the residual stress fluctuations, where $\mathcal{H}$ represents the operator of the two-dimensional Hilbert transform.
Additionally, the inclination angle $\theta_{L}$ of the large-scale structures should be considered. 
The outer large-scale structures affecting the near-wall region are located downstream of this region.
Hence, we will primarily examine the relationship between the virtual wall fluctuations and $u^{\prime}_{O}\left(\Delta x_{m}\right)$ at a downstream displacement $\Delta x_{m}$.
Here, $\Delta x_{m}=(y_{O}-y_{vw})/\tan(\theta_{L})$, and $\theta_{L}=11^{\circ}\sim15^{\circ}$ according to \cite{mathis2011predictive}.
We select $\theta_{L}=13^{\circ}$ for the subsequent discussions, noting that the results are robust within the range of $\theta_{L}=11^{\circ}\sim15^{\circ}$.

\begin{figure}
	\centering
	\subfigure{
		\begin{overpic}
			[scale=0.26]{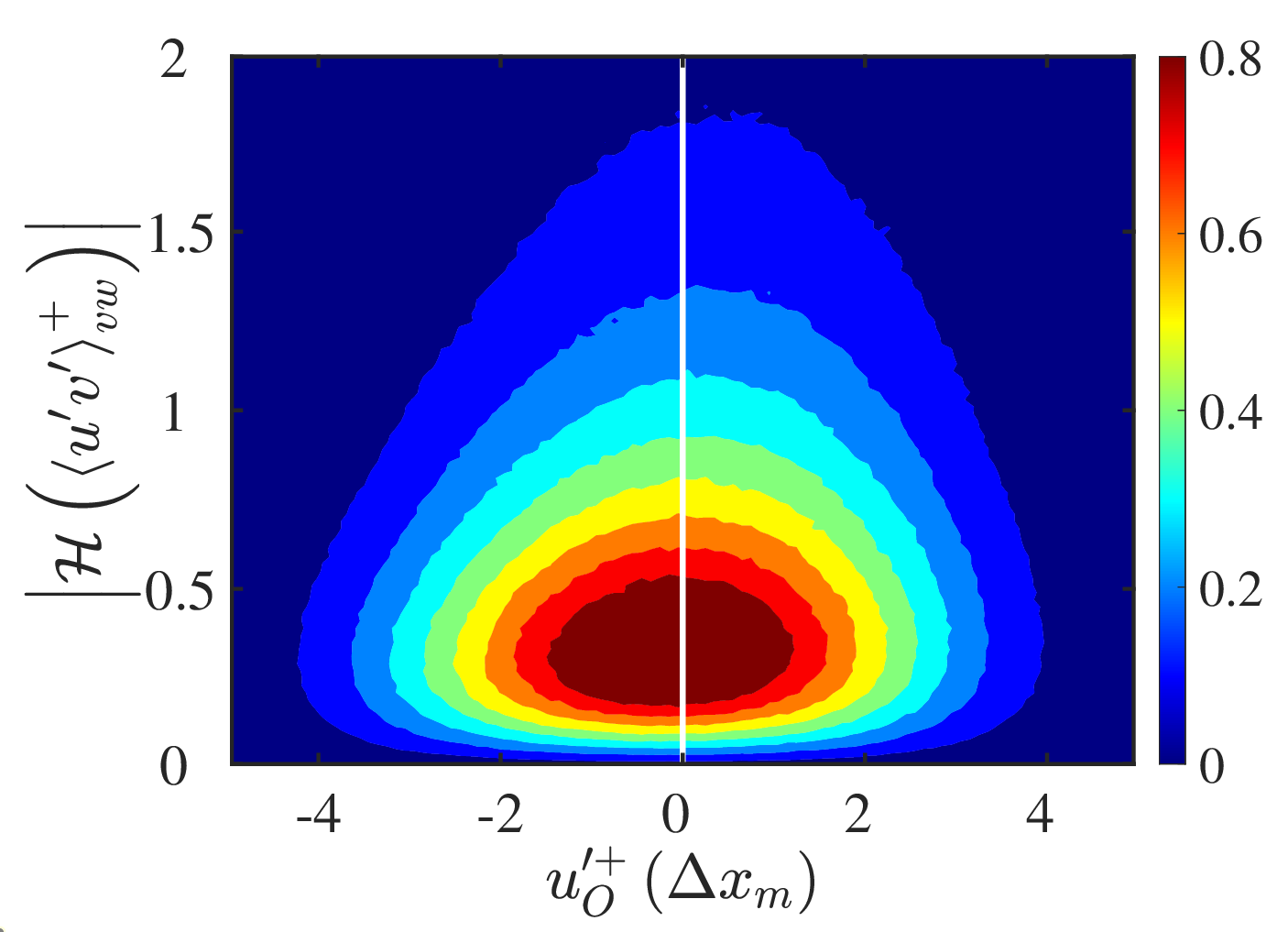}
			\put(2,72){(\textit{a})}
		\end{overpic}
		\begin{overpic}
			[scale=0.26]{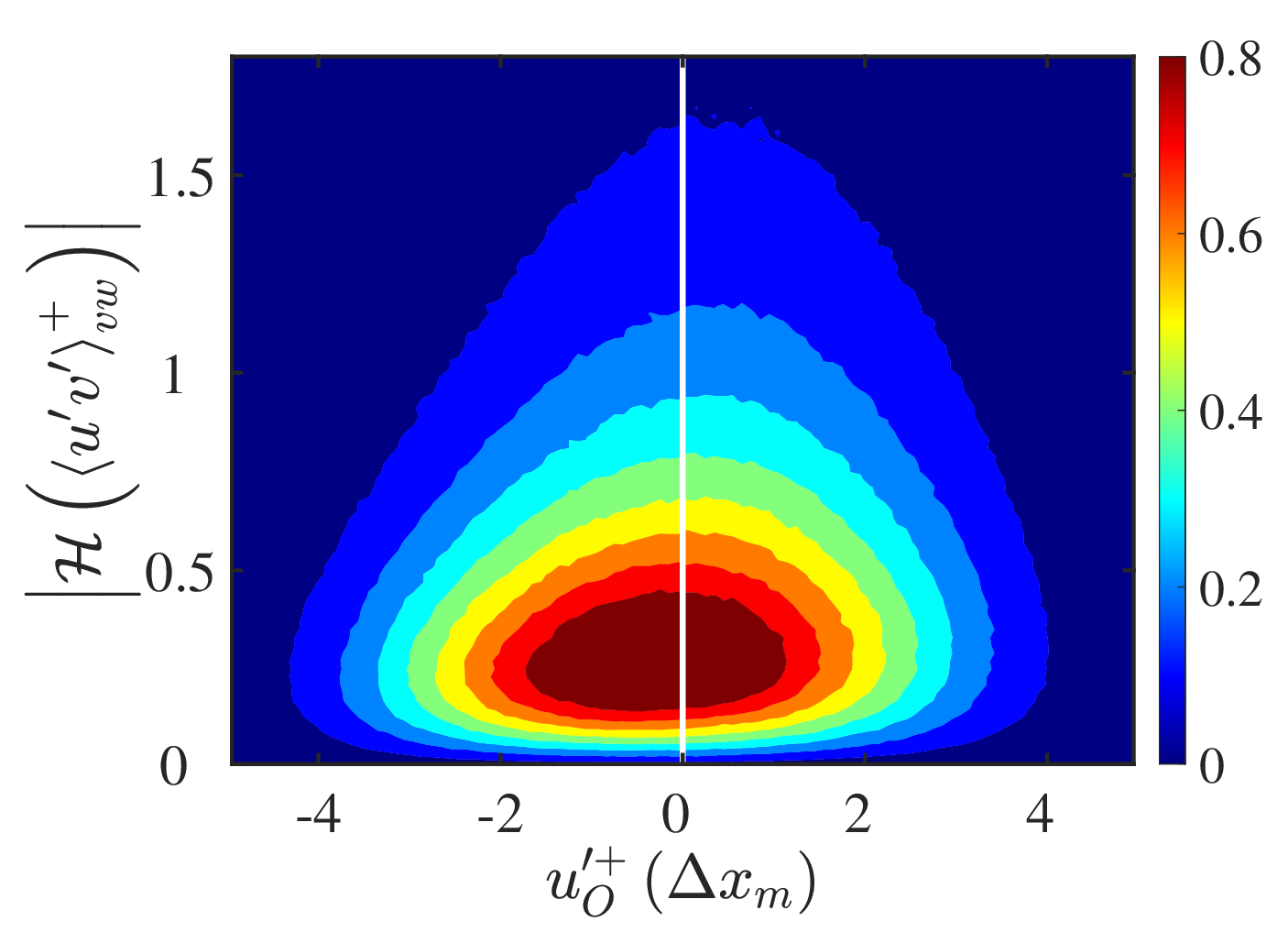}
			\put(3,72){(\textit{b})}
		\end{overpic}
	}
	\begin{overpic}
		[scale=0.26]{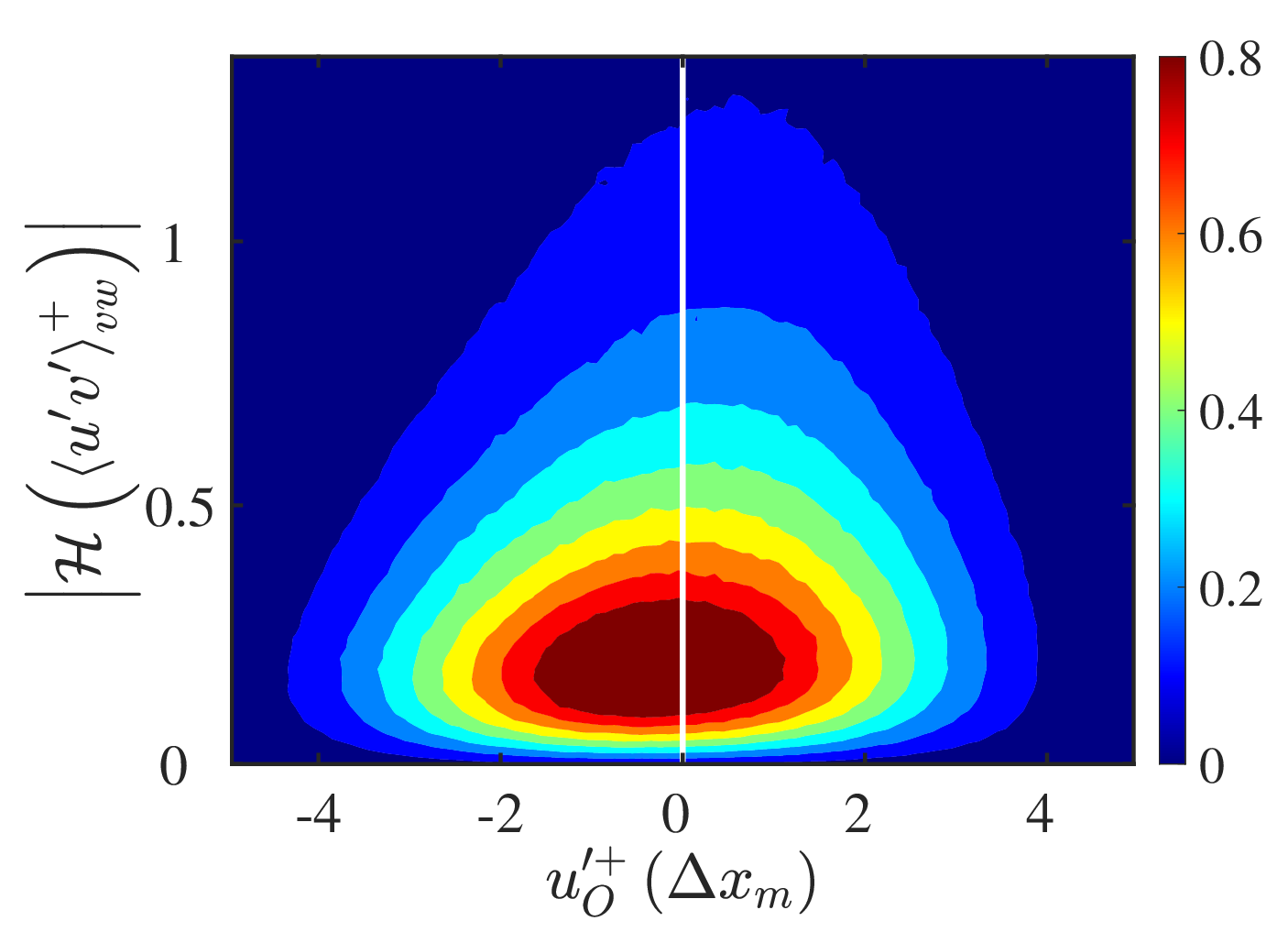}
		\put(4,69){(\textit{c})}
	\end{overpic}

	\caption{%
		Joint probability density function of the streamwise velocity fluctuations $u^{\prime}_{O}\left(\Delta x_{m}\right)$ at the center of the logarithmic region and the envelope of Reynolds stress $\left|\mathcal{H}\left(\left\langle u^{\prime}v^{\prime}\right\rangle _{vw}\right)\right|$ at the virtual wall.
		(\textit{a}) case C1000-1, (\textit{b}) C1000-2, (\textit{c}) C1000-3.
		Contour levels are $0.1 (0.1) 0.8$ of the maximum probability density.
	}
	\label{fig:pdf_LSM}
\end{figure}

The joint p.d.f. between the outer $u^{\prime}_{O}\left(\Delta x_{m}\right)$ and the envelope of Reynolds stress at the virtual wall is depicted in figure \ref{fig:pdf_LSM}.
At the position where $\left|\mathcal{H}\left(\left\langle u^{\prime}v^{\prime}\right\rangle _{vw}\right)\right|$ approaches $0$, the joint p.d.f. tilts to the left, indicating negative $u^{\prime}_{O}$ in low-speed large-scale motions.
As the envelope of Reynolds stress gradually increases, the joint p.d.f. shifts, tilting to the right, which is particularly evident in the upper half of the distribution.
This pattern suggests that locations with strong residual Reynolds stress fluctuations are typically situated below large-scale high-speed regions, while areas with weaker residual Reynolds stress generally correspond to large-scale low-speed regions.
This observation is consistent with the findings illustrated in figure \ref{fig:u-xz-1000}.
As the range of blowing and suction velocities is extended, although the intensity of $-\left\langle u^{\prime}v^{\prime}\right\rangle _{vw}$ diminishes, the influence of outer large-scale structures on the distribution of Reynolds stress remains nearly unchanged. 
This further substantiates the relationship between the amplitude modulation of outer large-scale structures and the residual Reynolds stress at the virtual wall.

In summary, compared to the traditional opposition control method, the DRL-based control strategy demonstrates superior drag reduction capabilities by effectively elevating the virtual wall to a higher position.
As the range of blowing and suction velocities is expanded, the virtual wall ascends further and the residual Reynolds stress on the virtual wall decreases, both of which enhance the drag reduction rate of the DRL models.
However, as the Reynolds number increases, large-scale structures emerge in the outer region. 
Their amplitude modulation effect significantly increases the residual Reynolds stress on the virtual wall, and disrupts the virtual wall’s blockage in large-scale high-speed regions, thereby reducing the drag reduction rate of the DRL models.

\subsection{Dynamic analysis of drag reduction using budget equations}\label{sec:budget}

In the previous subsection, the drag reduction mechanism was examined from a kinematic perspective using the virtual wall theory. 
This subsection will further discuss the dynamics mechanism behind drag reduction based on the analysis of budget equations.

According to the FIK identity proposed by \cite{fukagata2002contribution}, the skin frictions in the current cases are primarily attributed to the Reynolds shear stress $\left\langle -u^{\prime}v^{\prime}\right\rangle$.
Hence, it is necessary to discuss how the DRL-based control strategies reduce the drag by altering $\left\langle -u^{\prime}v^{\prime}\right\rangle$.
The transport equation of the Reynolds stress $\left\langle -u^{\prime}v^{\prime}\right\rangle$ is written as
\begin{equation}
	\begin{split}
		\frac{\partial}{\partial t}\left\langle -u^{\prime}v^{\prime}\right\rangle=&
		\underbrace{\left\langle v^{\prime}v^{\prime}\right\rangle 	\frac{dU}{dy}}_{P_{12}}
		+\underbrace{\frac{d}{dy}\left\langle u^{\prime}v^{\prime}v^{\prime}\right\rangle }_{D_{12,t}}
		+\underbrace{\left(-\mu\frac{d^{2}\left\langle u^{\prime}v^{\prime}\right\rangle }{dy^{2}}\right)}_{D_{12,\nu}}\\
		&
		+\underbrace{\frac{1}{\rho}\left\langle v^{\prime}\frac{\partial p^{\prime}}{\partial y}+u^{\prime}\frac{\partial p^{\prime}}{\partial x}\right\rangle}_{VP_{12}}
		+\underbrace{2\mu\left\langle \frac{\partial u^{\prime}}{\partial x_{j}}\frac{\partial v^{\prime}}{\partial x_{j}}\right\rangle }_{\varepsilon_{12}},
	\end{split}
	\label{eq:budget-uv}
\end{equation}	
where $P_{12}$ is the turbulent production, $D_{12,t}$ is the turbulent diffusion, $D_{12,\nu}$ is the viscous diffusion, $VP_{12}$ is the velocity pressure-gradient term and $\varepsilon_{12}$ is the dissipation. Here, $U$ is the mean streamwise velocity.

\begin{figure}
	\centering
	
	\subfigure{
		\begin{overpic}
			[scale=0.23]{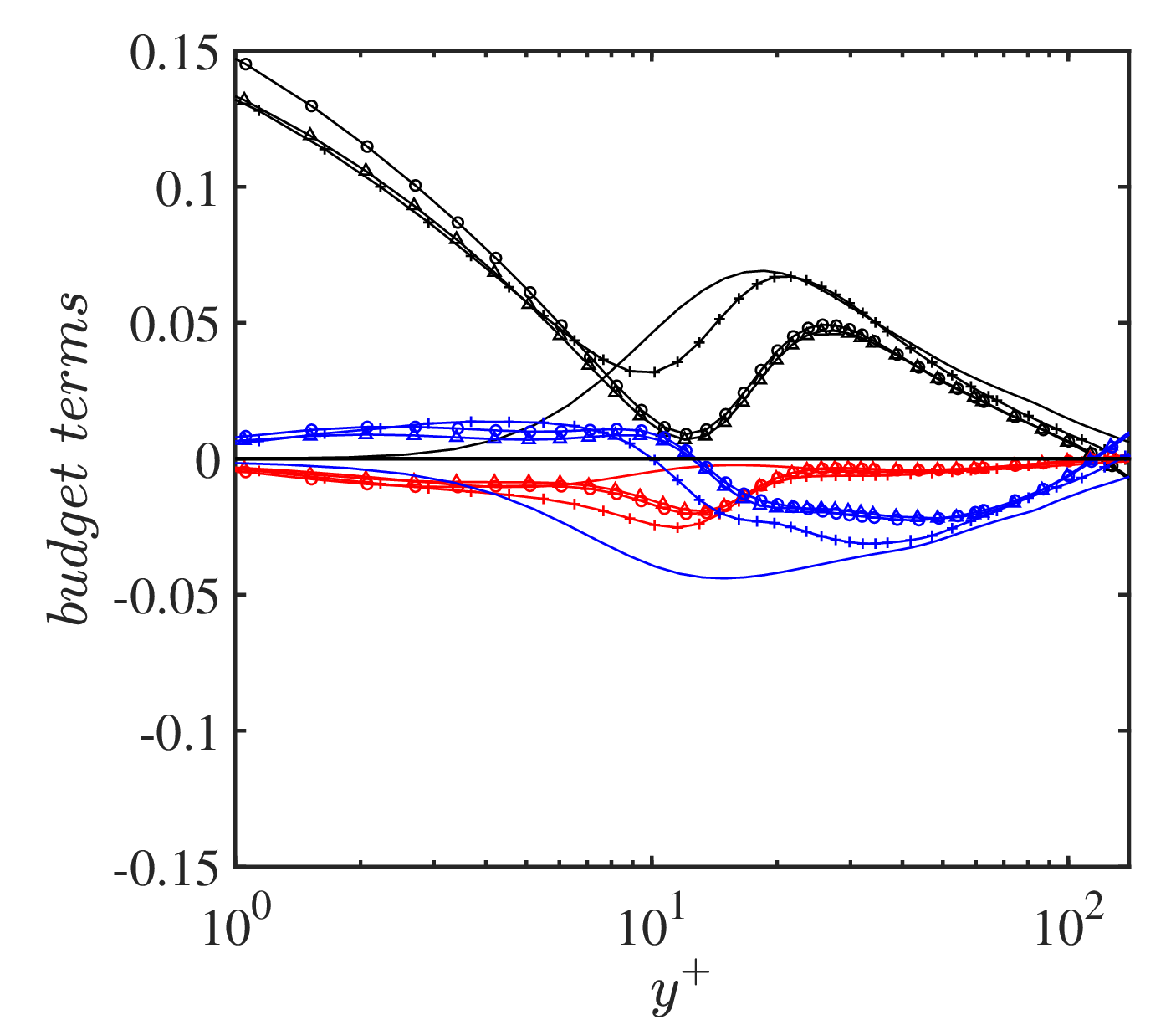}
			\put(0,81){(\textit{a})}
		\end{overpic}
		\begin{overpic}
			[scale=0.23]{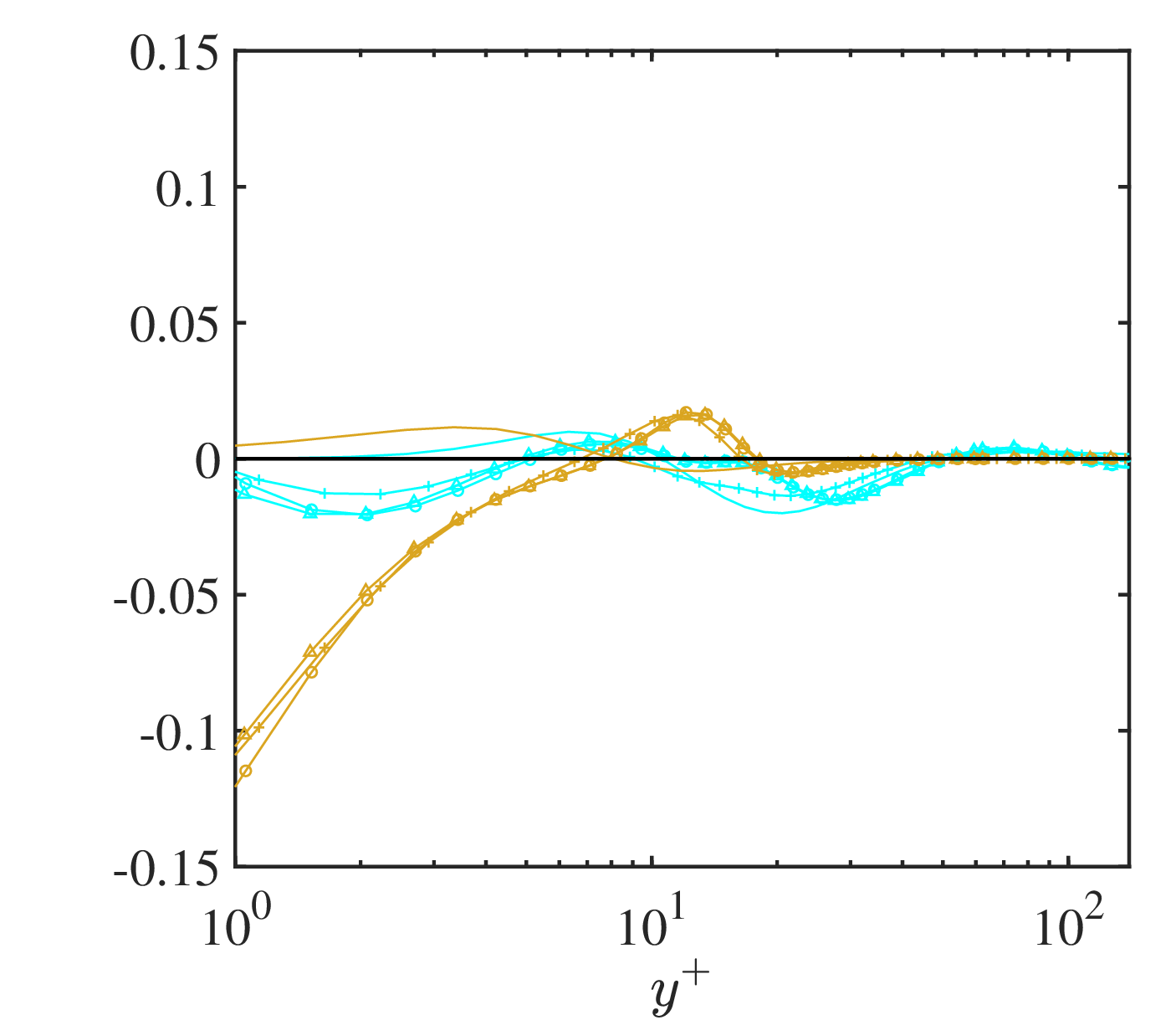}
			\put(0,81){(\textit{b})}
		\end{overpic}
	}
	\subfigure{
		\begin{overpic}
			[scale=0.23]{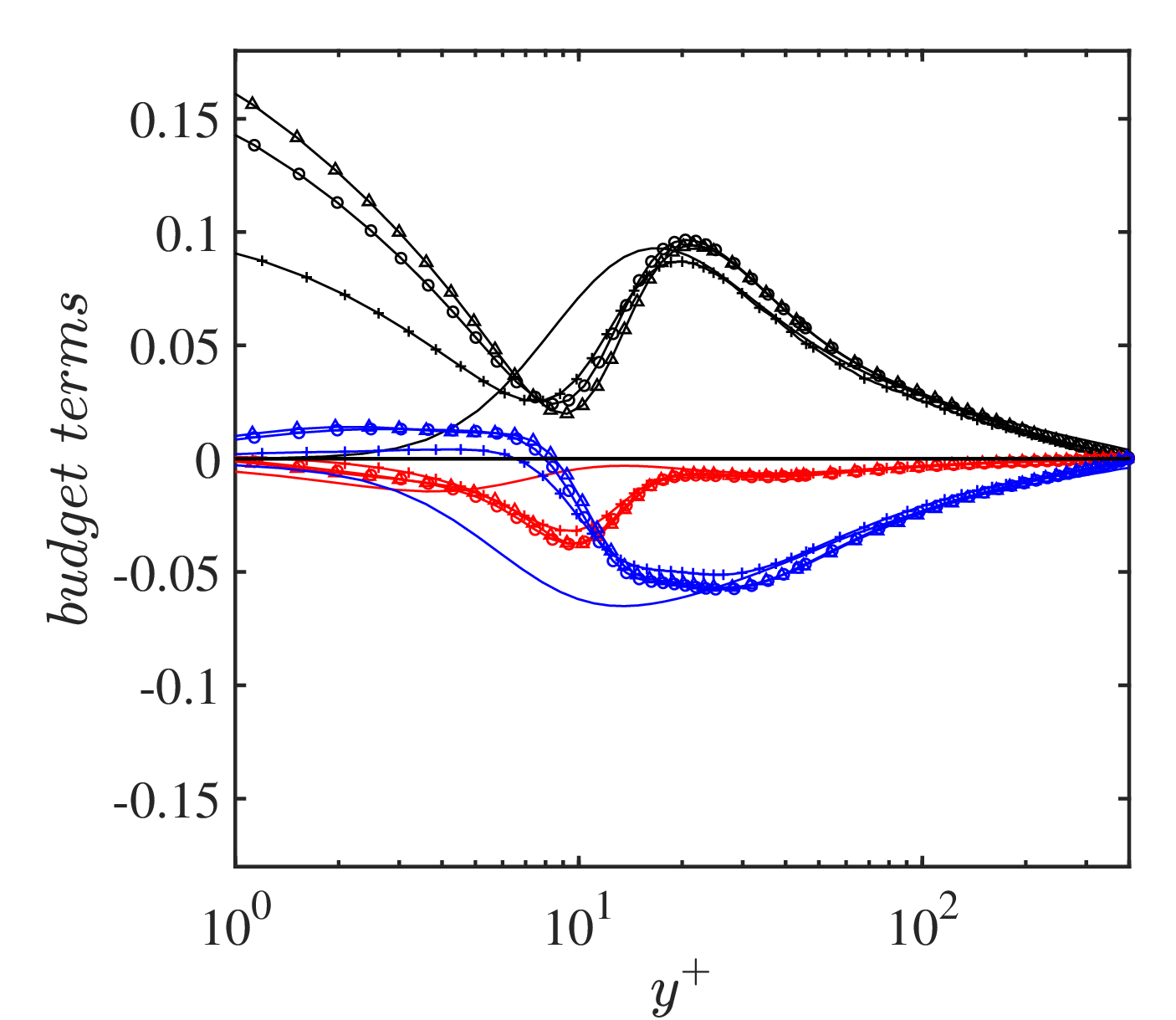}
			\put(0,81){(\textit{c})}
		\end{overpic}
		\begin{overpic}
			[scale=0.23]{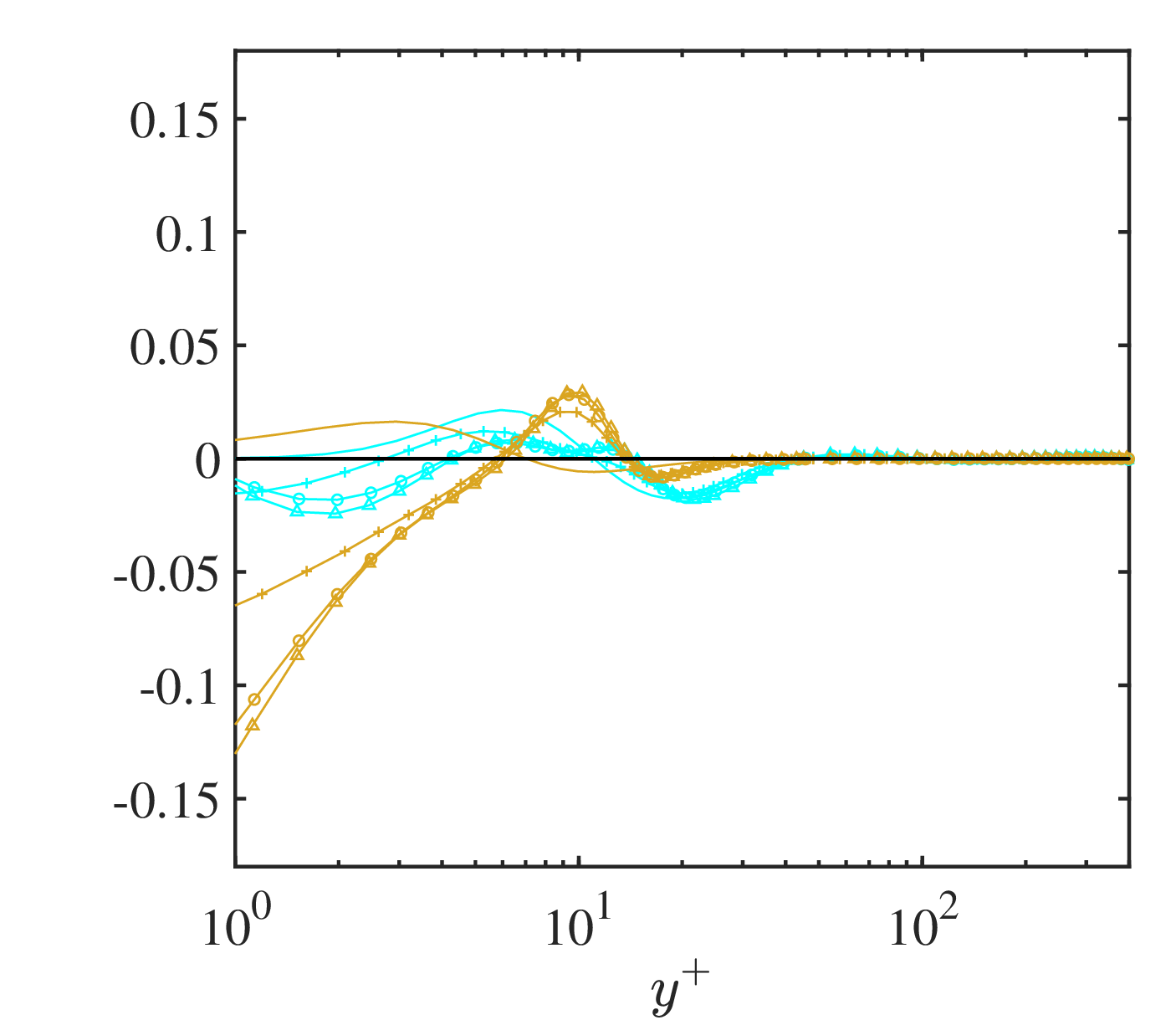}
			\put(0,81){(\textit{d})}
		\end{overpic}
	}
	\subfigure{
		\begin{overpic}
			[scale=0.23]{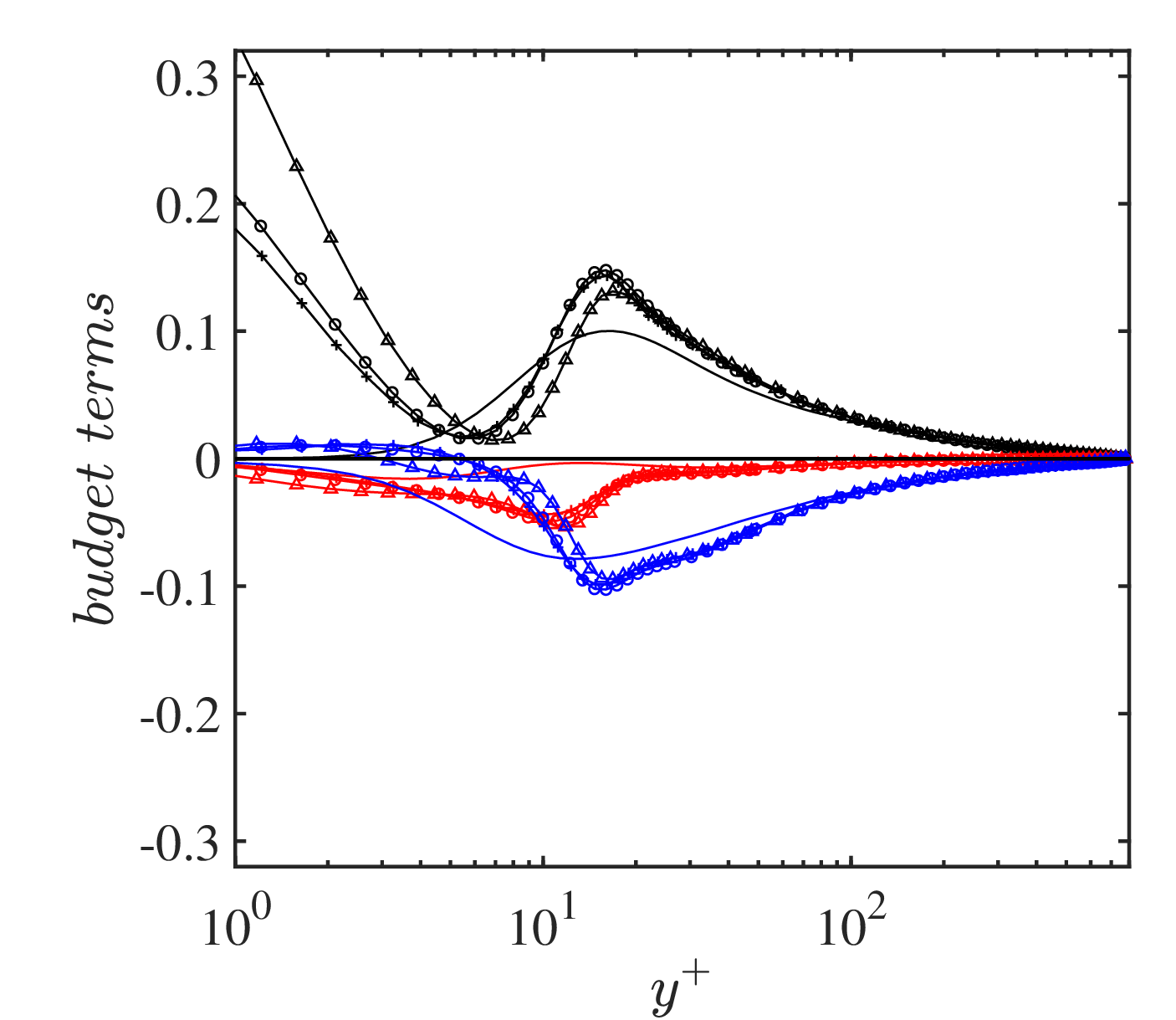}
			\put(0,81){(\textit{e})}
		\end{overpic}
		\begin{overpic}
			[scale=0.23]{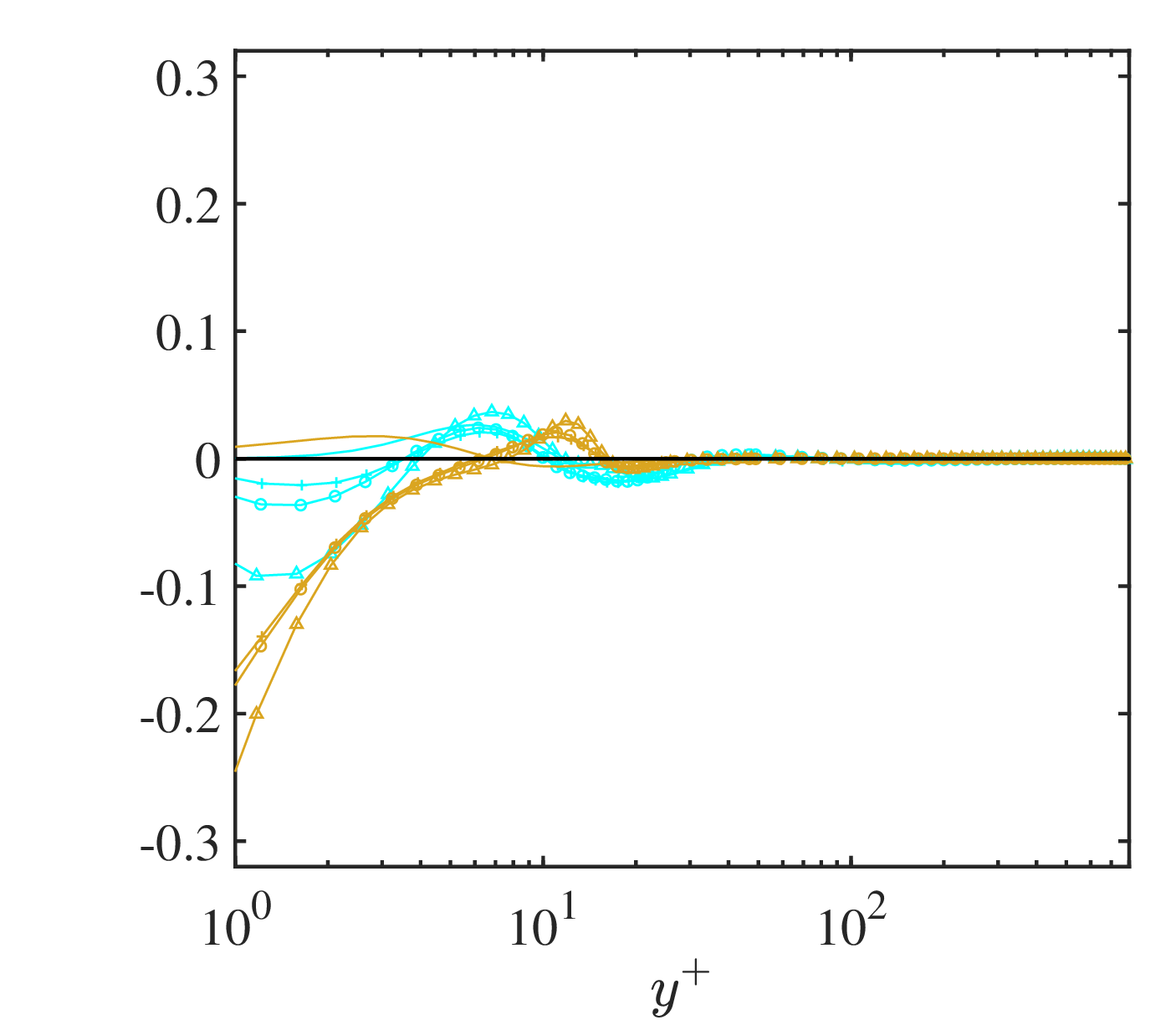}
			\put(0,81){(\textit{f})}
		\end{overpic}
	}

	\DeclareRobustCommand\mylabela{\tikz[baseline]{\draw[solid, black, thick] (0,0.5ex) -- (0.8,0.5ex);}}
	\DeclareRobustCommand\mylabelb{\tikz[baseline]{\draw[solid, red, thick] (0,0.5ex) -- (0.8,0.5ex);}}
	\DeclareRobustCommand\mylabelc{\tikz[baseline]{\draw[solid, blue, thick] (0,0.5ex) -- (0.8,0.5ex);}}
	\DeclareRobustCommand\mylabeld{\tikz[baseline]{\draw[solid, c00, thick] (0,0.5ex) -- (0.8,0.5ex);}}
	\DeclareRobustCommand\mylabele{\tikz[baseline]{\draw[solid, brown0, thick] (0,0.5ex) -- (0.8,0.5ex);}}
	\DeclareRobustCommand\mylabelf{\tikz[baseline]{\draw[solid, m00, thick] (0,0.5ex) -- (0.8,0.5ex);}}

	\caption{
		Wall-normal distributions of the budget terms of Reynolds shear stress $\left\langle -u^{\prime}v^{\prime}\right\rangle$ in eq (\ref{eq:budget-uv}).
		\mylabela, $P_{12}$; \mylabelc, $VP_{12}$; \mylabelb, $\varepsilon_{12}$; \mylabeld, $D_{12,t}$; \mylabele, $D_{12,\nu}$.
		(\textit{a})(\textit{b}) C180, (\textit{c})(\textit{d}) C550, (\textit{e})(\textit{f}) C1000.
		Lines without markers: cases with suffix '-0'; plus signs: with suffix '-1'; circles: with suffix '-2'; triangles: with suffix '-3'.
	}
	\label{fig:budget-uv}
	
\end{figure}
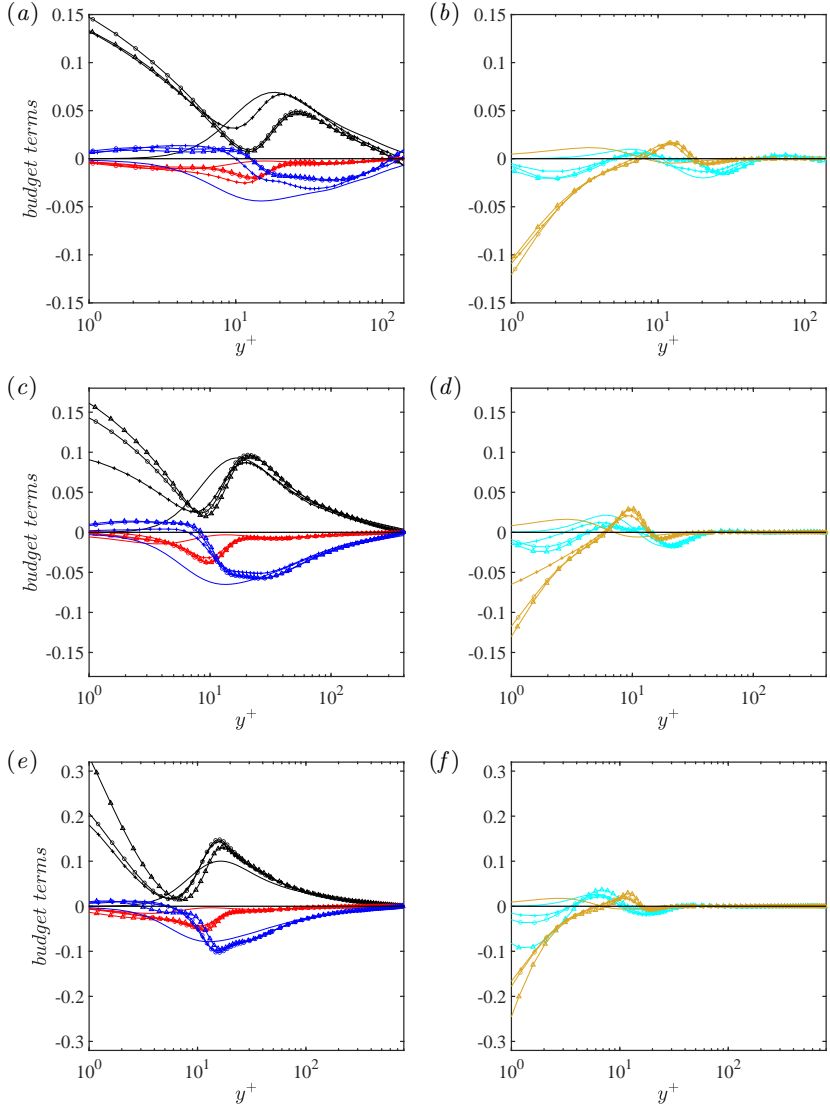

Figure \ref{fig:budget-uv} shows the wall-normal distributions of the budget terms on the right-hand side of eq (\ref{eq:budget-uv}).
In the budget terms of the Reynolds shear stress, the production $P_{12}$ and the velocity pressure-gradient term $VP_{12}$ are significantly stronger and more dominant compared to the other terms, as suggested in figure \ref{fig:budget-uv}(\textit{a})(\textit{c})(\textit{e}).
The viscous diffusion $D_{12,\nu}$, although large and negative in the viscous sublayer, rapidly decays above $y^+=5$ to become smaller than the dominant terms.
Furthermore, the dissipation $\varepsilon_{12}$ and the turbulent diffusion $D_{12,t}$ are much smaller than the other terms, and their contribution could be considered negligible.
Among the two dominant terms, the velocity pressure-gradient term $VP_{12}$, which can be further divided into the pressure diffusion and the redistribution, mainly represents the transport of Reynolds stress at different heights and the redistribution among different components caused by pressure.
And it primarily acts as a negative term to offset the production $P_{12}$, which remains positive and determines the magnitude of the Reynolds shear stress.

In the uncontrolled cases, the turbulent production $P_{12}$ increases with height, reaching a peak around $y^{+}=15\sim20$ and then continuously decreases.
After implementing wall blowing and suction, $P_{12}$ at the wall is no longer zero, leading to a significant increase in $P_{12}$ within the viscous sublayer. 
This results in a larger Reynolds stress in the viscous sublayer compared to that in the uncontrolled cases, as depicted in figure \ref{fig:uv-avg}.
However, this effect is confined to the narrow height range of the viscous sublayer and has a limited impact on the overall skin friction.
In the cases with control, $P_{12}$ decreases rapidly with height and reaches a trough around $y^{+}=10$.
In the range of $10<y^{+}<20$, $P_{12}$ is significantly smaller than in the uncontrolled case, corresponding to a lower Reynolds shear stress in figure \ref{fig:uv-avg}.
As the range of blowing and suction velocities is expanded, the height corresponding to the trough gradually increases, and $P_{12}$ at the trough further decreases, leading to a reduction in Reynolds shear stress.
Moreover, the decrease in $P_{12}$ near the trough compared with the uncontrolled case is less pronounced at higher Reynolds numbers, as indicated in figure \ref{fig:budget-uv}(\textit{e}).
This results in a reduced suppression effect on $\left\langle -u^{\prime}v^{\prime}\right\rangle$ at higher Reynolds numbers, shown in figure \ref{fig:uv-avg}(\textit{c}), further leading to a decreased drag reduction rate.
As the height increases, $P_{12}$ in the controlled case gradually rises above $y^{+}=15$, peaks, and then continuously decreases, eventually collapsing with the uncontrolled case in the outer region.

According to eq (\ref{eq:budget-uv}), the turbulent production $P_{12}$ of Reynolds shear stress consists of two parts: $\left\langle v^{\prime}v^{\prime}\right\rangle$ and $dU/dy$.
The latter could be viewed as the outcome associated with changes in Reynolds shear stress and skin friction.
Therefore, the following discussion will primarily focus on the wall-normal kinetic energy $\left\langle v^{\prime}v^{\prime}\right\rangle$ to identify the source of changes in $P_{12}$.
The wall-normal distributions of $v_{rms}$ under different control strategies have already been shown and discussed in figure \ref{fig:u-rms}.
After adopting the DRL models, the wall-normal velocity fluctuations in the buffer layer gradually decrease. 
As the range of blowing and suction velocities is extended, the $v_{rms}$ continues to decrease, but this decreasing trend slows down with increasing Reynolds numbers. 
This is similar to the evolution trend of the turbulent production $P_{12}$.
The transport equation of the wall-normal kinetic energy $\left\langle v^{\prime}v^{\prime}\right\rangle$ is written as
\begin{equation}
	\begin{split}
		\frac{\partial}{\partial t}\left\langle v^{\prime}v^{\prime}\right\rangle =&
		\underbrace{-\frac{d}{dy}\left\langle v^{\prime}v^{\prime}v^{\prime}\right\rangle }_{D_{22,t}}
		+\underbrace{\mu\frac{d^{2}\left\langle v^{\prime}v^{\prime}\right\rangle }{dy^{2}}}_{D_{22,\nu}}
		+\underbrace{\left(-\frac{2}{\rho}\frac{d}{dy}\left\langle p^{\prime}v^{\prime}\right\rangle \right)}_{D_{22,p}}\\
		&
		+\underbrace{\frac{2}{\rho}\left\langle p^{\prime}\frac{\partial v^{\prime}}{\partial y}\right\rangle }_{\Phi_{22}}
		+\underbrace{\left(-2\mu\left\langle \frac{\partial v^{\prime}}{\partial x_{j}}\frac{\partial v^{\prime}}{\partial x_{j}}\right\rangle \right)}_{\varepsilon_{22}},
	\end{split}
	\label{eq:budget-vv}
\end{equation}	
where $D_{22,t}$ is the turbulent diffusion, $D_{22,\nu}$ is the viscous diffusion, $D_{22,p}$ is the pressure diffusion, $\Phi_{22}$ is the redistribution and $\varepsilon_{22}$ is the dissipation.

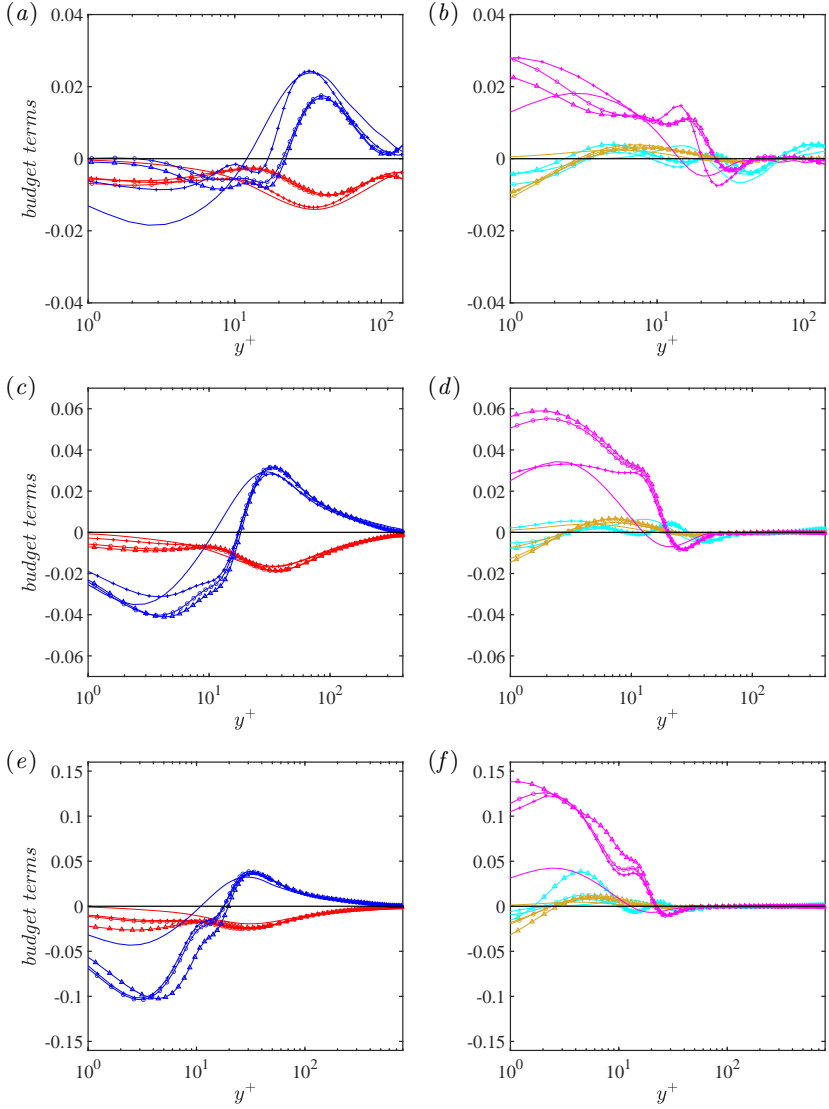
\begin{figure}
	\centering
	
	\subfigure{
		\begin{overpic}
			[scale=0.23]{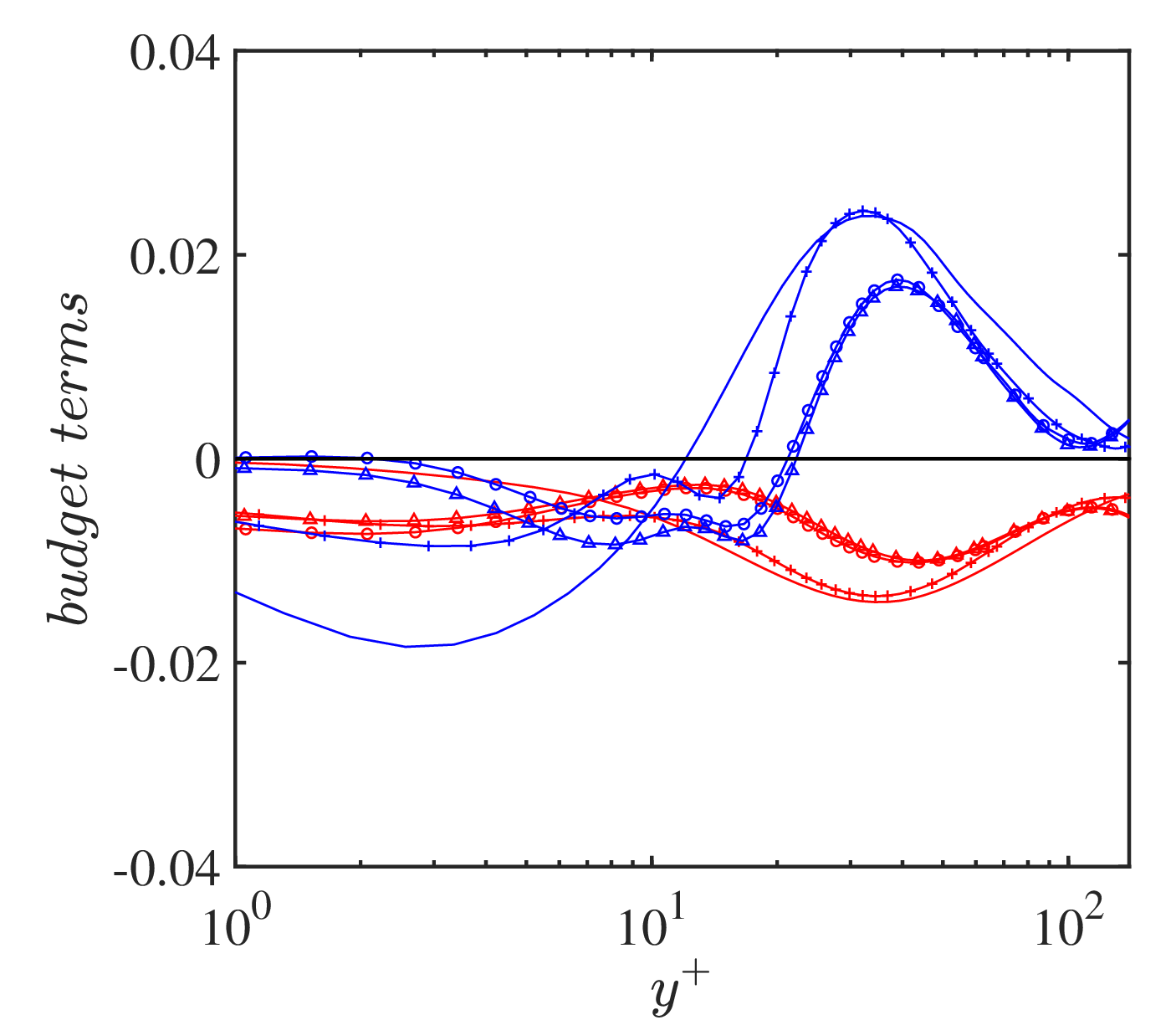}
			\put(0,81){(\textit{a})}
		\end{overpic}
		\begin{overpic}
			[scale=0.23]{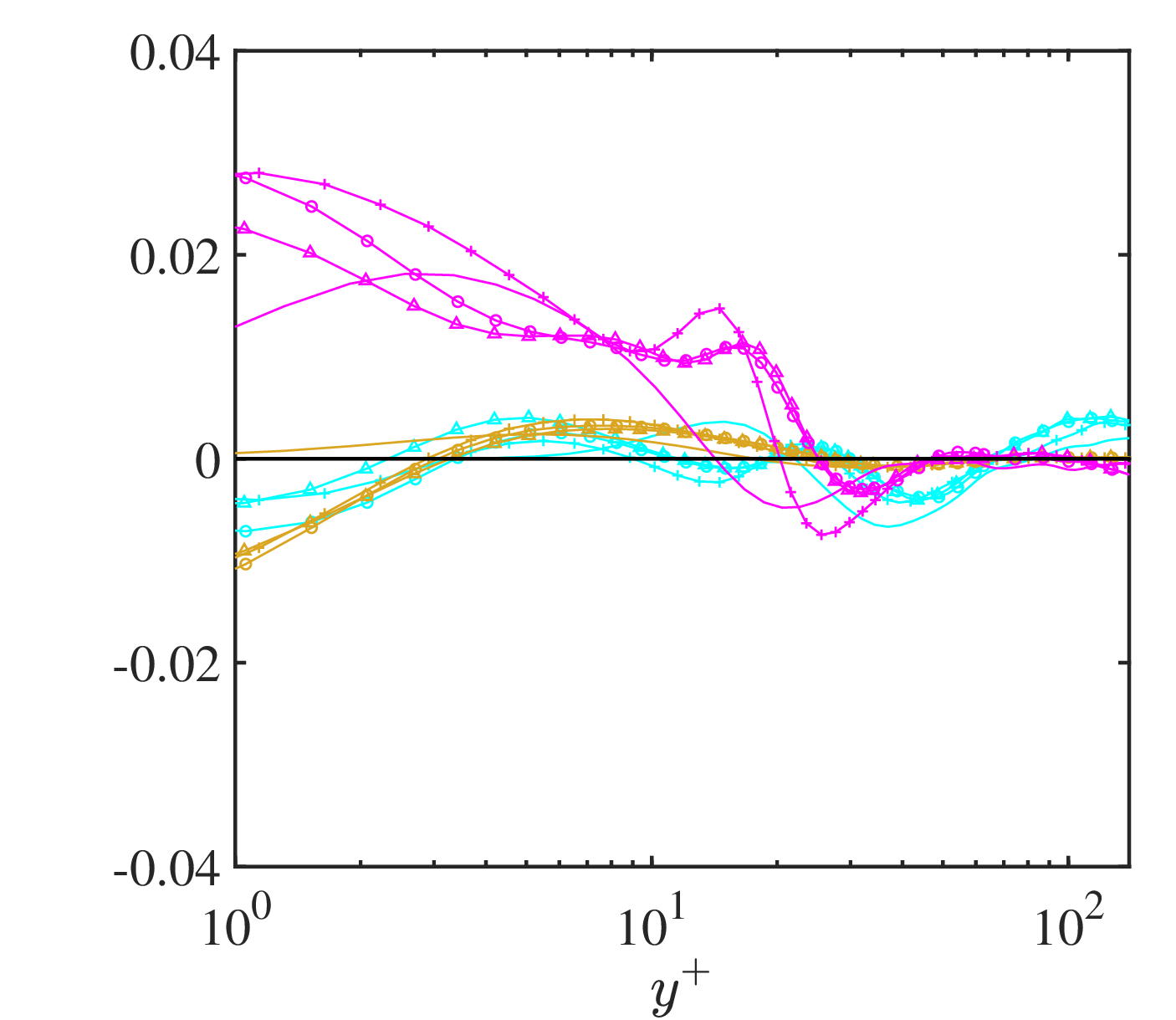}
			\put(0,81){(\textit{b})}
		\end{overpic}
	}
	\subfigure{
		\begin{overpic}
			[scale=0.23]{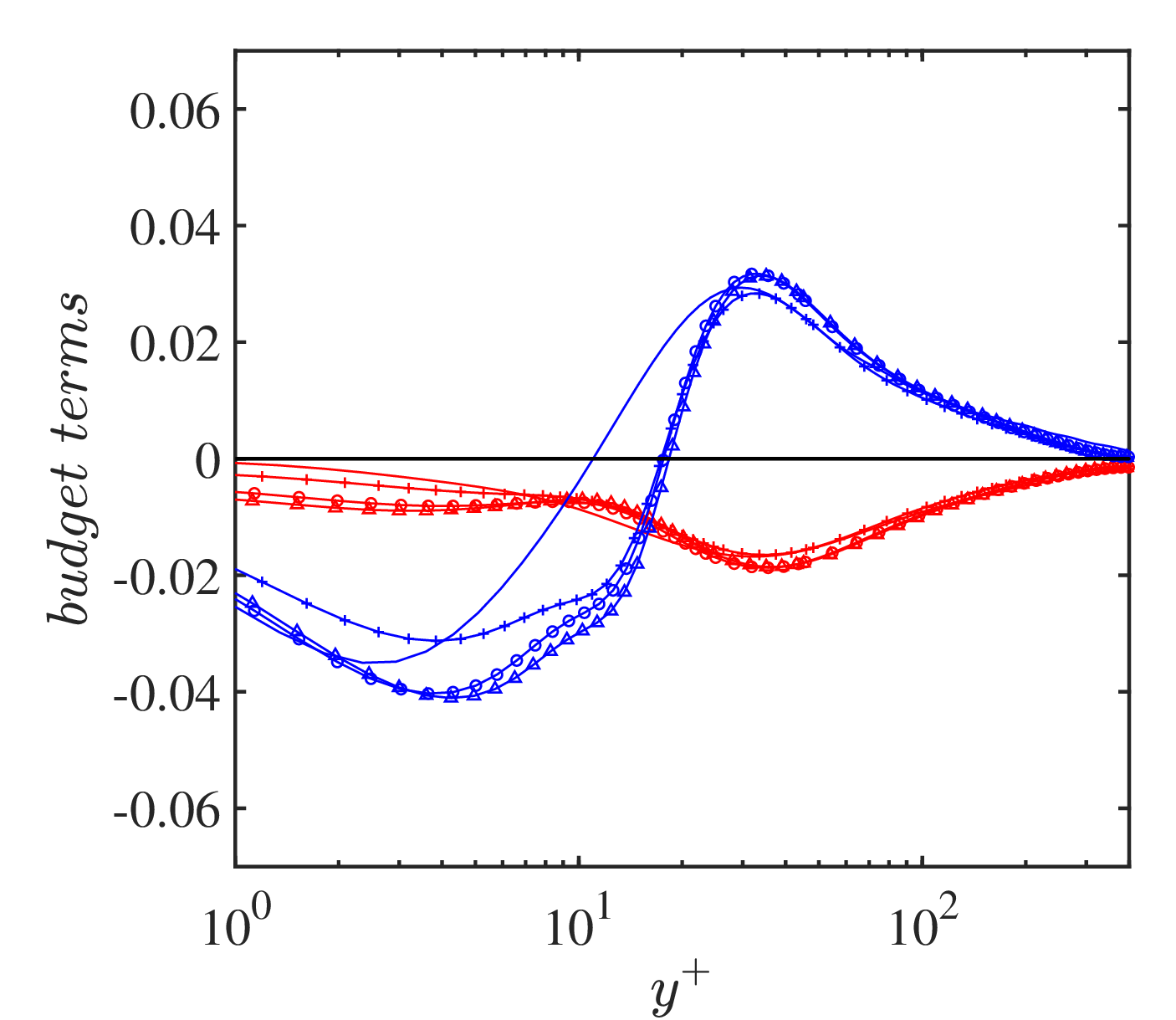}
			\put(0,81){(\textit{c})}
		\end{overpic}
		\begin{overpic}
			[scale=0.23]{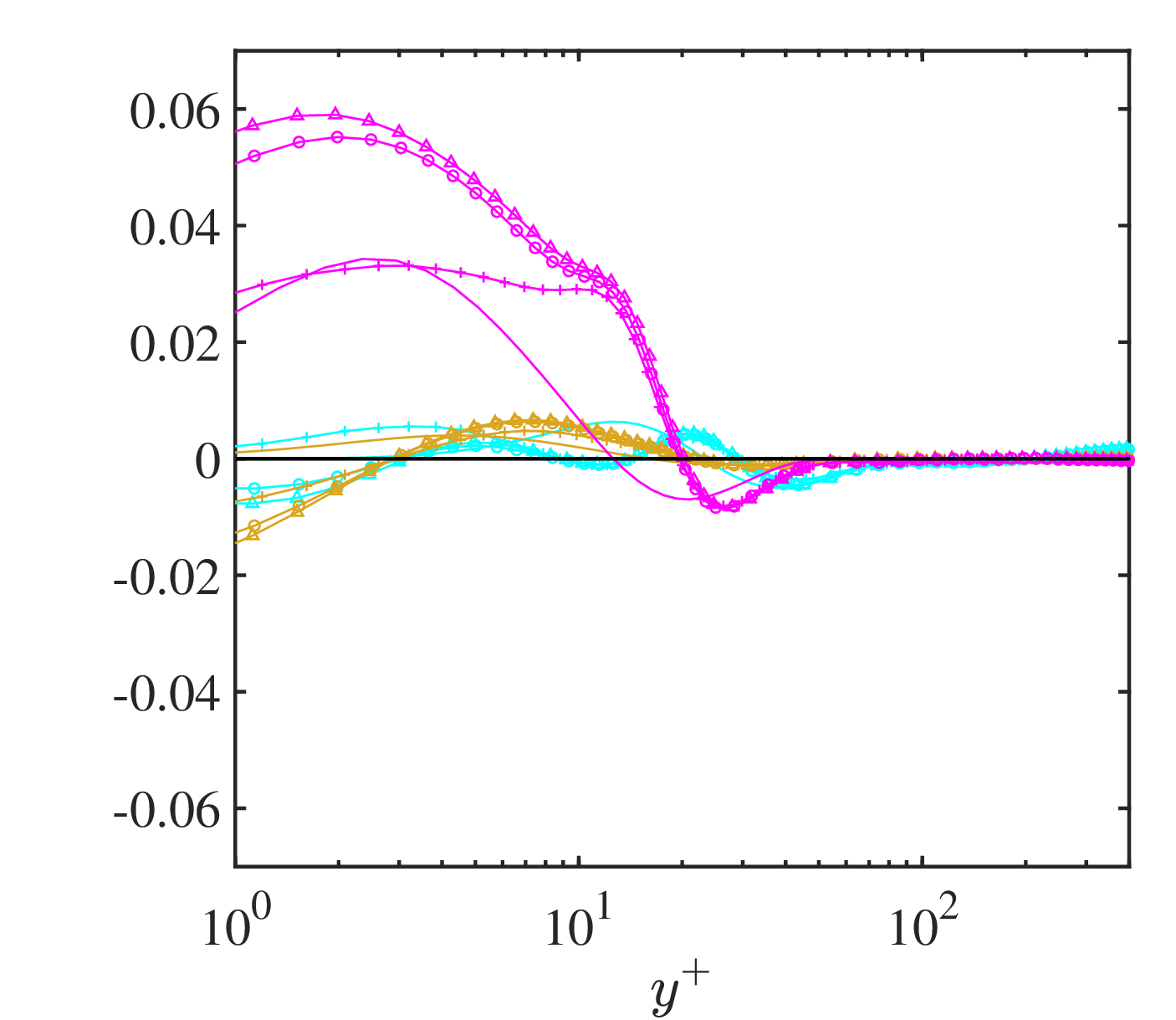}
			\put(0,81){(\textit{d})}
		\end{overpic}
	}
	\subfigure{
		\begin{overpic}
			[scale=0.23]{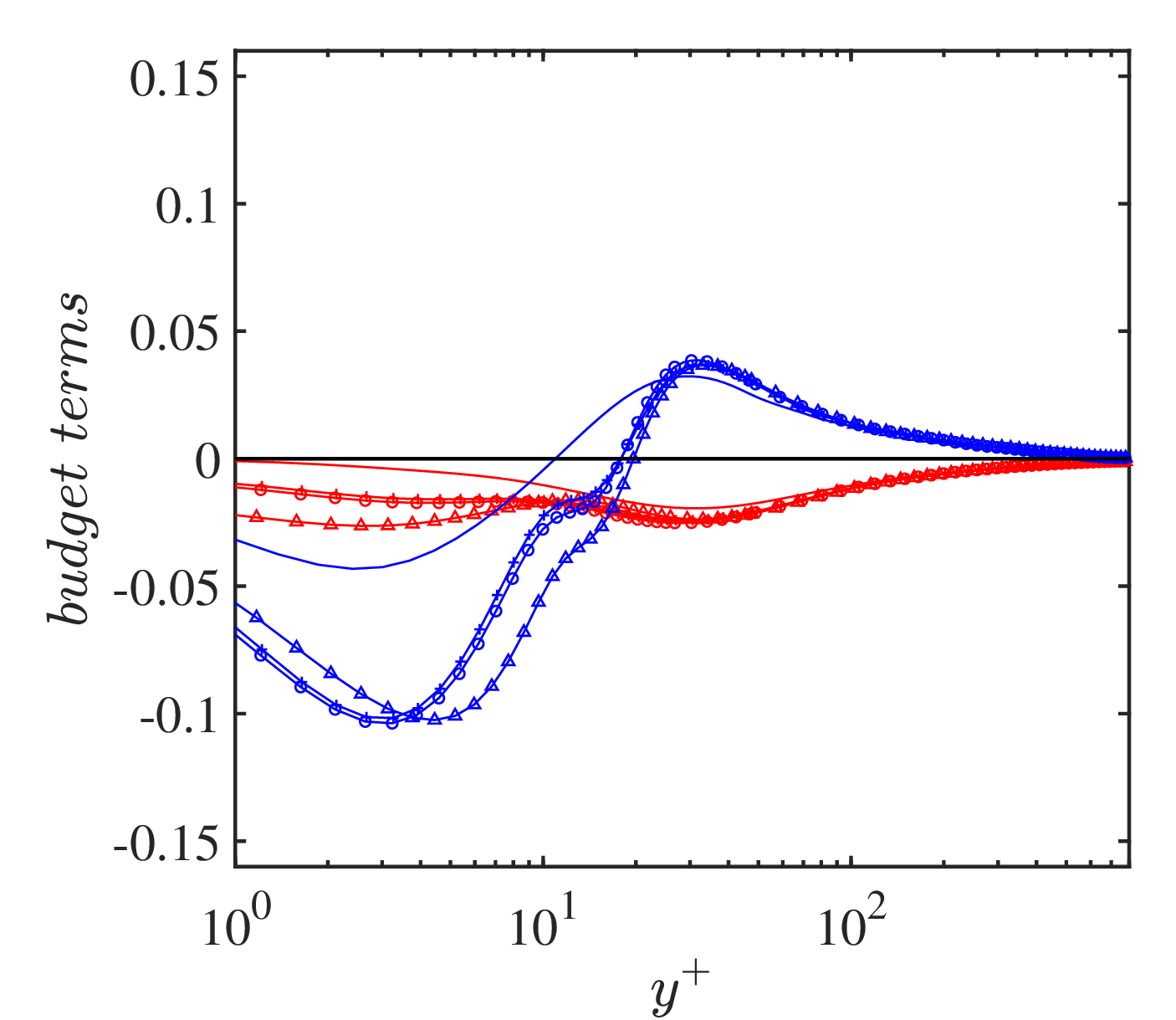}
			\put(0,81){(\textit{e})}
		\end{overpic}
		\begin{overpic}
			[scale=0.23]{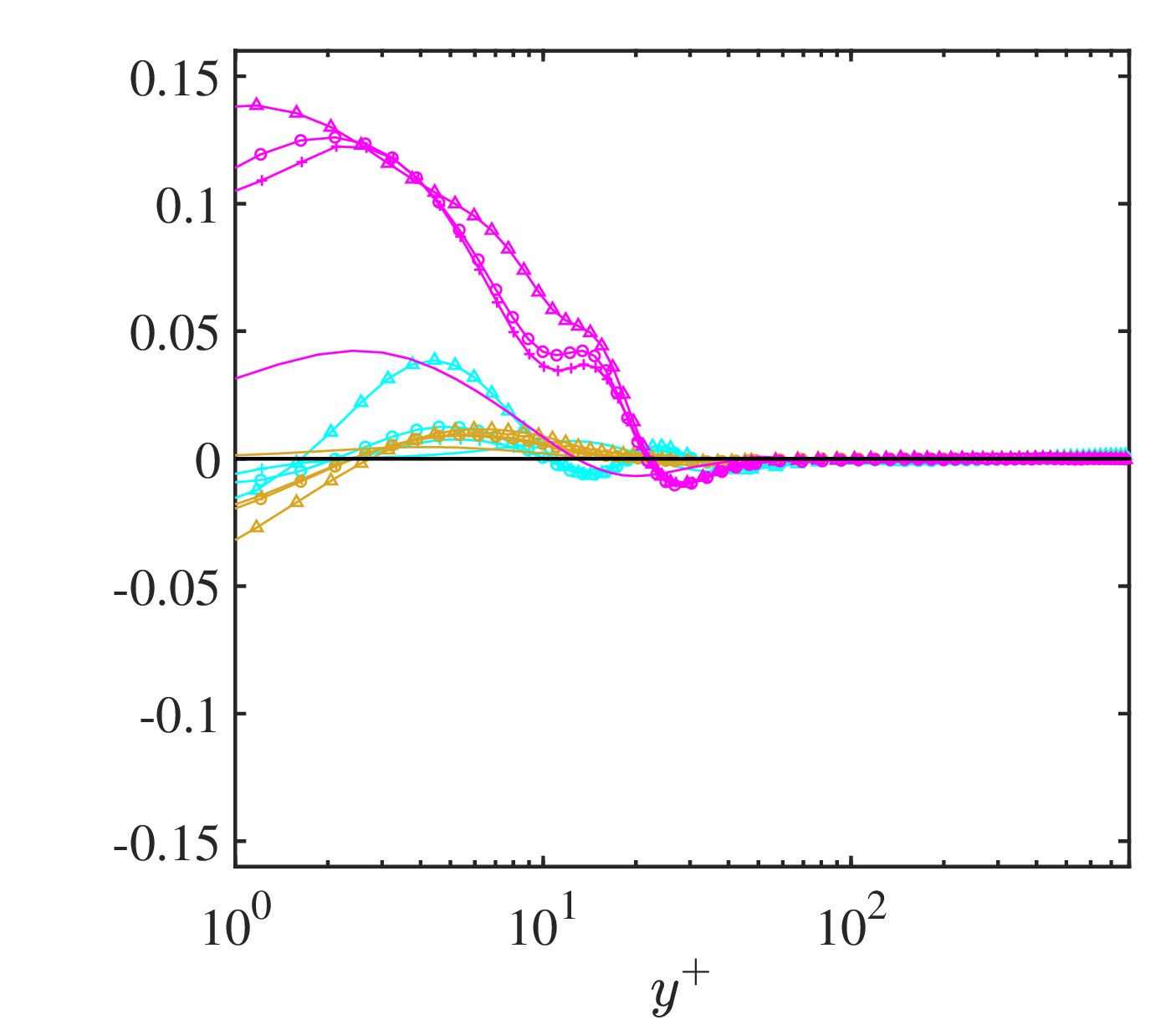}
			\put(0,81){(\textit{f})}
		\end{overpic}
	}

	\DeclareRobustCommand\mylabela{\tikz[baseline]{\draw[solid, black, thick] (0,0.5ex) -- (0.8,0.5ex);}}
	\DeclareRobustCommand\mylabelb{\tikz[baseline]{\draw[solid, red, thick] (0,0.5ex) -- (0.8,0.5ex);}}
	\DeclareRobustCommand\mylabelc{\tikz[baseline]{\draw[solid, blue, thick] (0,0.5ex) -- (0.8,0.5ex);}}
	\DeclareRobustCommand\mylabeld{\tikz[baseline]{\draw[solid, c00, thick] (0,0.5ex) -- (0.8,0.5ex);}}
	\DeclareRobustCommand\mylabele{\tikz[baseline]{\draw[solid, brown0, thick] (0,0.5ex) -- (0.8,0.5ex);}}
	\DeclareRobustCommand\mylabelf{\tikz[baseline]{\draw[solid, m00, thick] (0,0.5ex) -- (0.8,0.5ex);}}

	\caption{
		Wall-normal distributions of the budget terms of wall-normal kinetic energy $\left\langle v^{\prime}v^{\prime}\right\rangle$ in eq (\ref{eq:budget-vv}).
		\mylabelc, $\Phi_{22}$; \mylabelb, $\varepsilon_{22}$;\mylabelf, $D_{22,p}$; \mylabeld, $D_{22,t}$; \mylabele, $D_{22,\nu}$.
		(\textit{a})(\textit{b}) C180, (\textit{c})(\textit{d}) C550, (\textit{e})(\textit{f}) C1000.
		Lines without markers: cases with suffix '-0'; plus signs: with suffix '-1'; circles: with suffix '-2'; triangles: with suffix '-3'.
	}
	\label{fig:budget-vv}
	
\end{figure}

The wall-normal distributions of the budget terms on the right-hand side of eq (\ref{eq:budget-vv}) are illustrated in figure \ref{fig:budget-vv}.
Among these budget terms, the wall-normal kinetic energy $\left\langle v^{\prime}v^{\prime}\right\rangle$ is predominantly influenced by redistribution $\Phi_{22}$, pressure diffusion $D_{22,p}$, and dissipation $\varepsilon_{22}$. 
Conversely, the effects of turbulent diffusion $D_{22,t}$ and viscous diffusion $D_{22,\nu}$ are comparatively minor.
In the viscous sublayer, the redistribution $\Phi_{22}$ is primarily negative, indicating that the wall-normal velocity fluctuations are being redistributed to other directions. 
This negative contribution is mainly offset by the positive pressure diffusion $D_{22,p}$. 
As the height increases, the redistribution $\Phi_{22}$ changes from negative to positive, indicating that the wall-normal velocity fluctuations are absorbing turbulent kinetic energy from other components. 
Meanwhile, the pressure diffusion $D_{22,p}$ rapidly decreases and gradually approaches zero above $y^{+}=20$.
On the other hand, the dissipation $\varepsilon_{22}$ gradually increases, acting as a negative term to offset the positive contribution from the redistribution $\Phi_{22}$.
Based on the previous discussion, the drag reduction achieved by the DRL model mainly stems from the dynamic changes in the buffer layer, while significant changes in the viscous sublayer contribute very little to the overall drag reduction. 
Among the three dominant terms, the pressure diffusion $D_{22,p}$ primarily represents the turbulent kinetic energy transport caused by pressure at different heights, with its intensity decreasing significantly in the buffer layer compared to the viscous sublayer.
Therefore, in the following discussion, we will primarily focus on the redistribution $\Phi_{22}$, which represents the exchange mechanism between wall-normal velocity fluctuations and other velocity components.

Compared to the uncontrolled cases, the DRL-based control strategy causes a significant decrease in $\Phi_{22}$ in the buffer layer and raises the position where it changes from negative to positive to around $y^+=20$.
This leads to less kinetic energy being transferred to the wall-normal velocity fluctuations in the buffer layer, thereby suppressing the production of Reynolds stress.
The suppressing effect of the DRL models on the redistribution $\Phi_{22}$ further increases as the range of blowing and suction velocities is extended. 
At a low Reynolds number $Re_{\tau}^{0}\thickapprox180$, this decreasing trend can extend to the logarithmic layer, as illustrated in figure \ref{fig:budget-vv}(\textit{e}). 
However, as the Reynolds number increases, the reduction in the redistribution $\Phi_{22}$ above $y^+=30$ nearly vanishes, and the suppression of $\Phi_{22}$ in the buffer layer by the DRL models becomes weaker.
This change corresponds to the decrease in drag reduction rate at higher Reynolds numbers.

\begin{figure}
	\centering
	
	\subfigure{
		\begin{overpic}
			[scale=0.21]{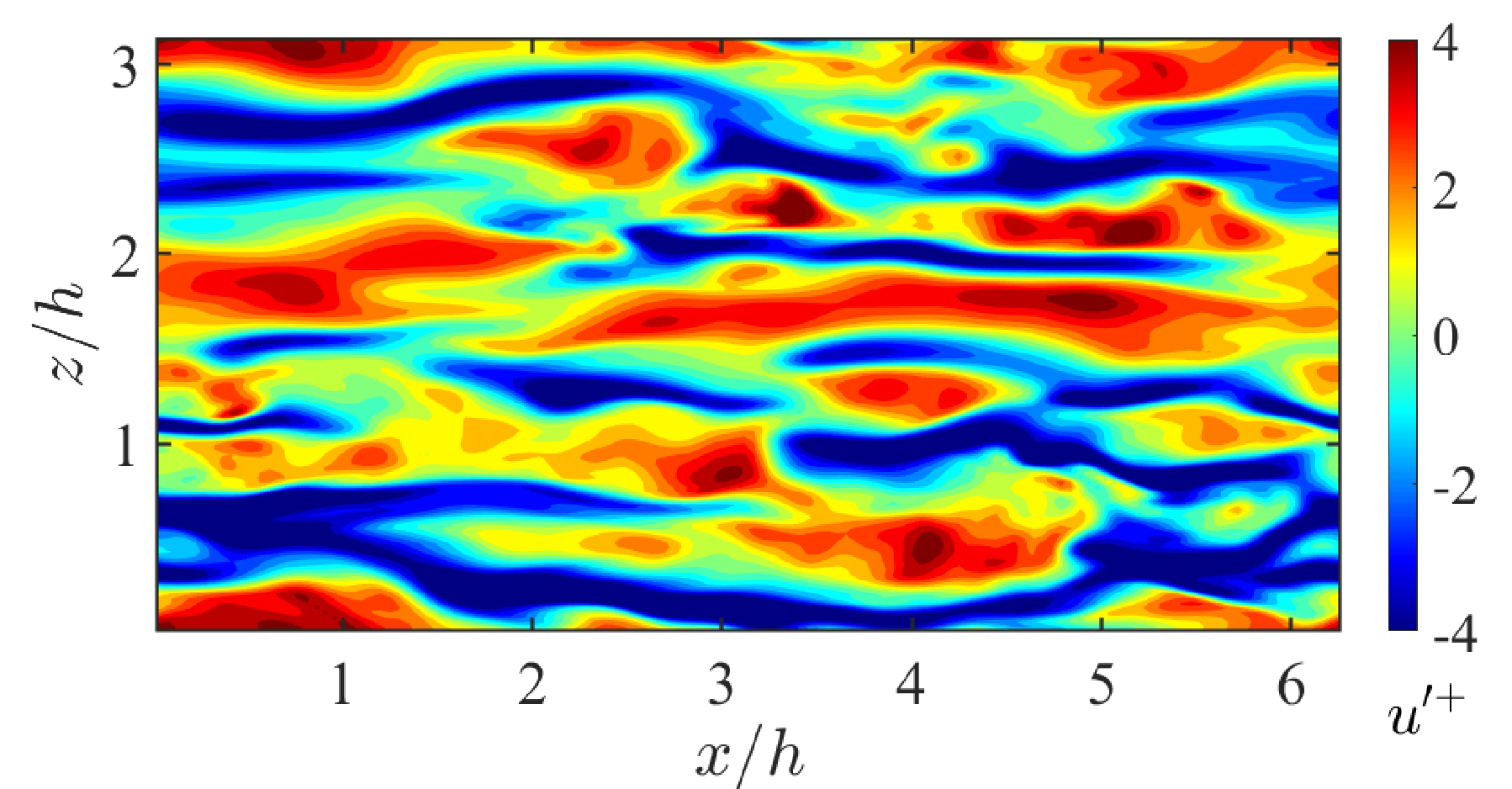}
			\put(1,52){(\textit{a})}
		\end{overpic}
		\begin{overpic}
			[scale=0.21]{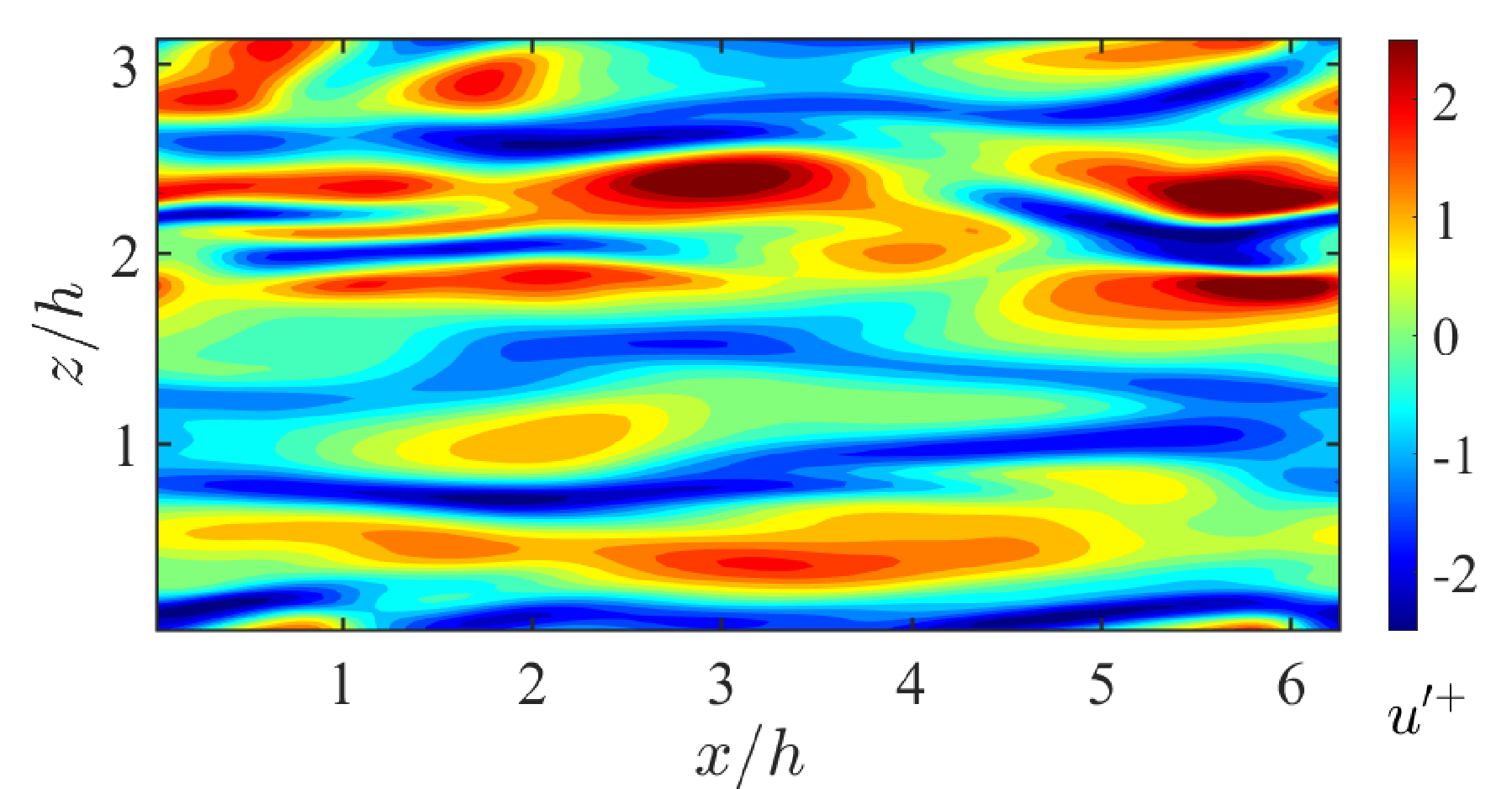}
			\put(1,52){(\textit{b})}
		\end{overpic}
	}
	
	\subfigure{
		\begin{overpic}
			[scale=0.21]{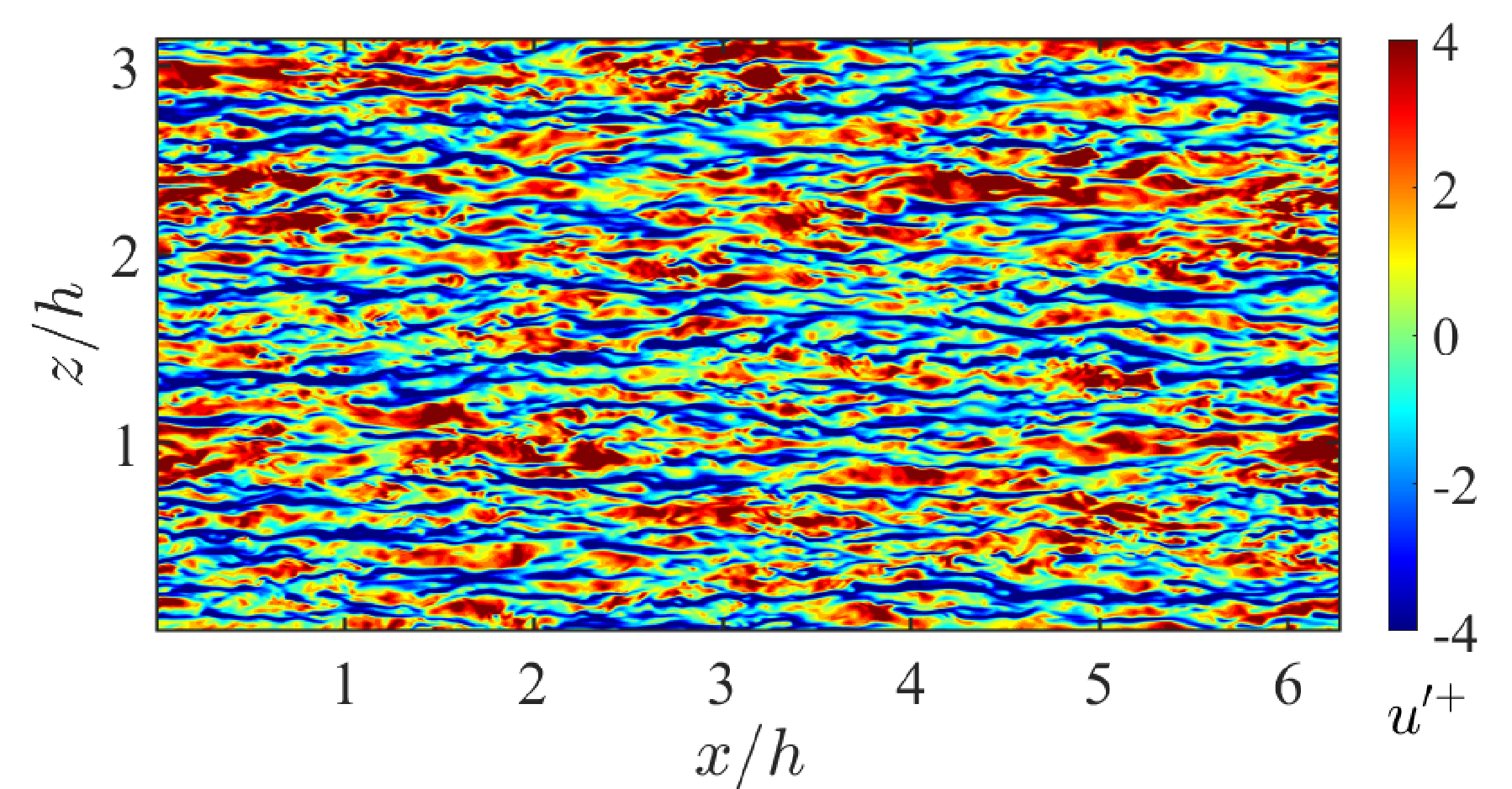}
			\put(1,52){(\textit{c})}
		\end{overpic}
		\begin{overpic}
			[scale=0.21]{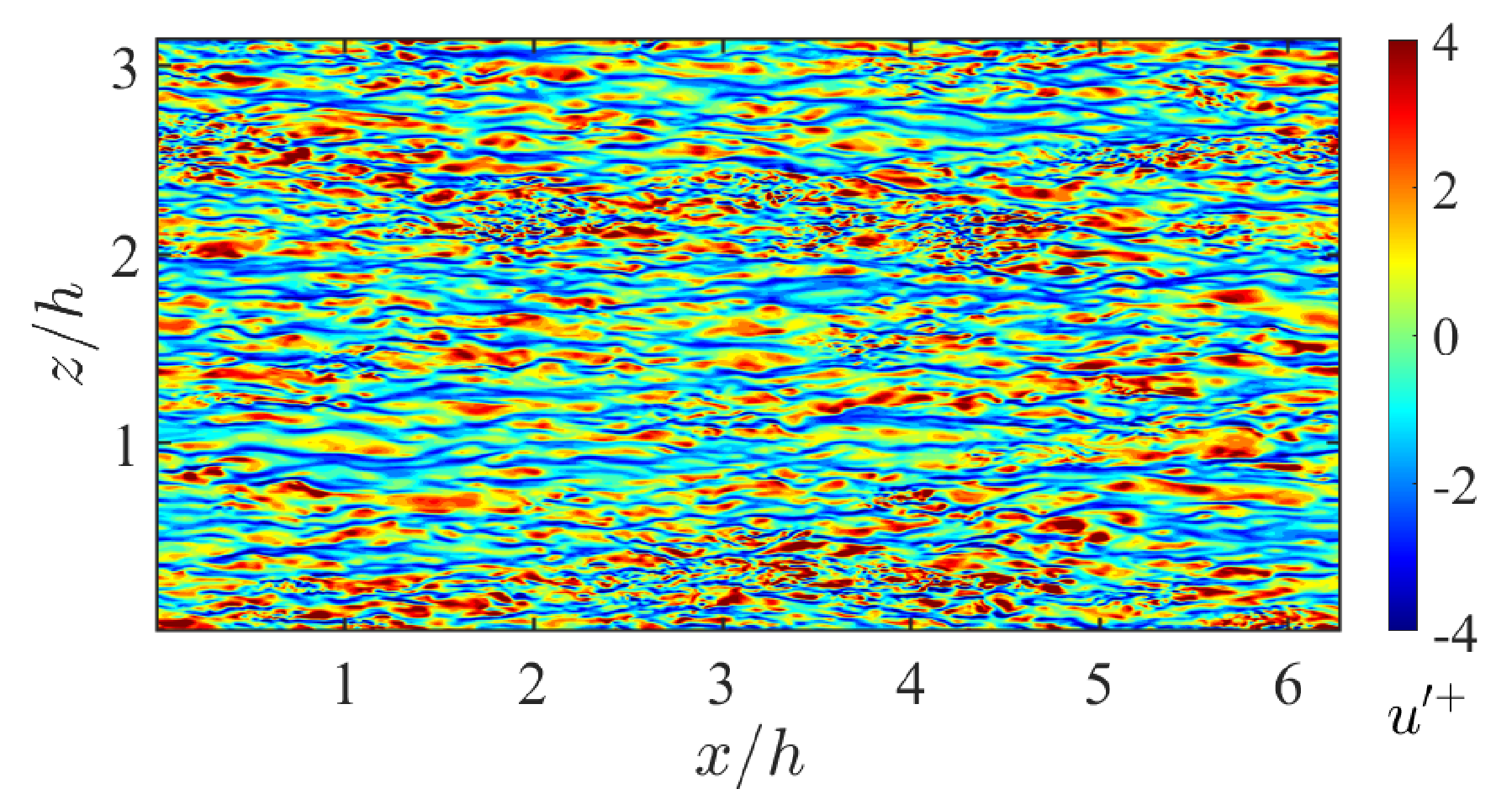}
			\put(1,52){(\textit{d})}
		\end{overpic}
	}

	\caption{
		Instantaneous distributions of $u^{\prime}$ on $(x,z)$ plane at $y^{+}=20$.
		(\textit{a}) case C180-0, (\textit{b}) C180-3, (\textit{c}) C1000-0, (\textit{d}) C1000-3.
	}
	\label{fig:u-xz-y20}
\end{figure}

It is important to note that, due to the incompressibility condition (where divergence equals zero), the sum of the redistribution terms for the three velocity components, namely $\Phi_{11}=(2/\rho)\left\langle p^{\prime}\partial u^{\prime}/\partial x\right\rangle$, $\Phi_{22}=(2/\rho)\left\langle p^{\prime}\partial v^{\prime}/\partial y\right\rangle$, and $\Phi_{33}=(2/\rho)\left\langle p^{\prime}\partial w^{\prime}/\partial z\right\rangle$, is $0$. 
Above the viscous sublayer, the turbulent kinetic energy is typically redistributed from the streamwise component to the wall-normal and spanwise components \citep{lee2019spectral}, also indicated by the positive $\Phi_{22}$ observed in figure \ref{fig:budget-vv}. 
This redistribution corresponds to the transient growth of the streamwise velocity streaks associated with $u^{\prime}$, leading to the generation of quasi-streamwise vortices associated with $v^{\prime}$ and $w^{\prime}$ in the near-wall turbulent self-sustaining cycle.
Thus, the observed weakening of $\Phi_{22}$ due to the DRL-based control strategy can be interpreted as the suppression of the near-wall self-sustaining mechanism. 
Consequently, a larger proportion of turbulent kinetic energy would remain in the streamwise component, resulting in smoother streak structures.
Figure \ref{fig:u-xz-y20} illustrates the instantaneous distributions of $u^{\prime}$ in the near-wall region. 
The controlled cases exhibit significantly smoother near-wall streaks compared to the uncontrolled results, and this trend is consistent across different Reynolds numbers. 
This observation aligns with the suppressed near-wall self-sustaining mechanism and also reflects the DRL model's influence on $\Phi_{22}$ from the perspective of flow structures.

Moreover, the streamwise velocity fluctuations merit further discussion due to their significant impact from the DRL-based control strategy, as illustrated in figure \ref{fig:u-rms}.
The transport equation of the streamwise kinetic energy $\left\langle u^{\prime}u^{\prime}\right\rangle$ is expressed as
\begin{equation}
	\begin{split}
		\frac{\partial}{\partial t}\left\langle u^{\prime}u^{\prime}\right\rangle =&
		\underbrace{-2\left\langle u^{\prime}v^{\prime}\right\rangle \frac{dU}{dy}}_{P_{11}}
		+\underbrace{\left(-\frac{d}{dy}\left\langle u^{\prime}u^{\prime}v^{\prime}\right\rangle \right)}_{D_{11,t}}
		+\underbrace{\mu\frac{d^{2}\left\langle u^{\prime}u^{\prime}\right\rangle }{dy^{2}}}_{D_{11,\nu}}\\
		&
		+\underbrace{\frac{2}{\rho}\left\langle p^{\prime}\frac{\partial u^{\prime}}{\partial x}\right\rangle }_{\Phi_{11}}
		+\underbrace{\left(-2\mu\left\langle \frac{\partial u^{\prime}}{\partial x_{j}}\frac{\partial u^{\prime}}{\partial x_{j}}\right\rangle \right)}_{\varepsilon_{11}},
	\end{split}
	\label{eq:budget-uu}
\end{equation}	
where $P_{11}$ is the turbulent production, $D_{11,t}$ is the turbulent diffusion, $D_{11,\nu}$ is the viscous diffusion, $\Phi_{11}$ is the redistribution and $\varepsilon_{11}$ is the dissipation.

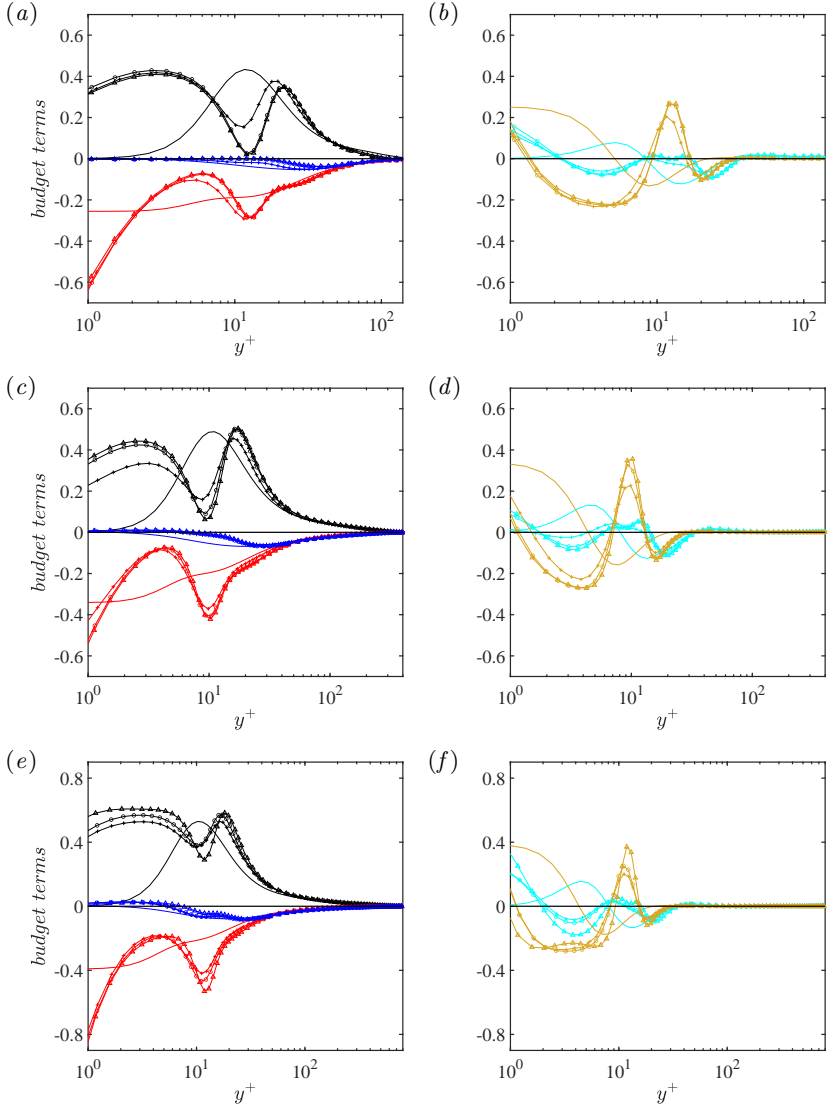
\begin{figure}
	\centering
	
	\subfigure{
		\begin{overpic}
			[scale=0.23]{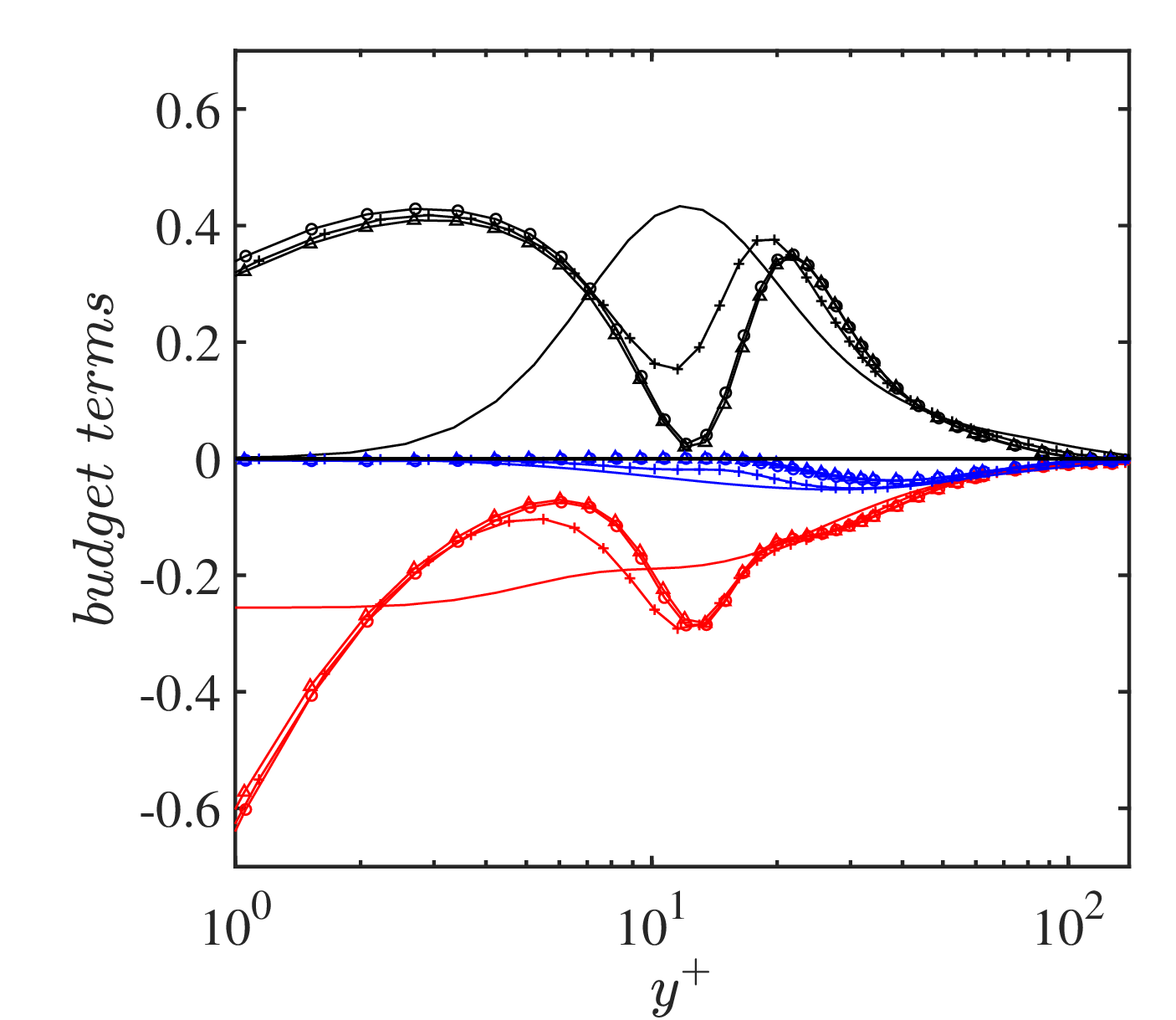}
			\put(0,81){(\textit{a})}
		\end{overpic}
		\begin{overpic}
			[scale=0.23]{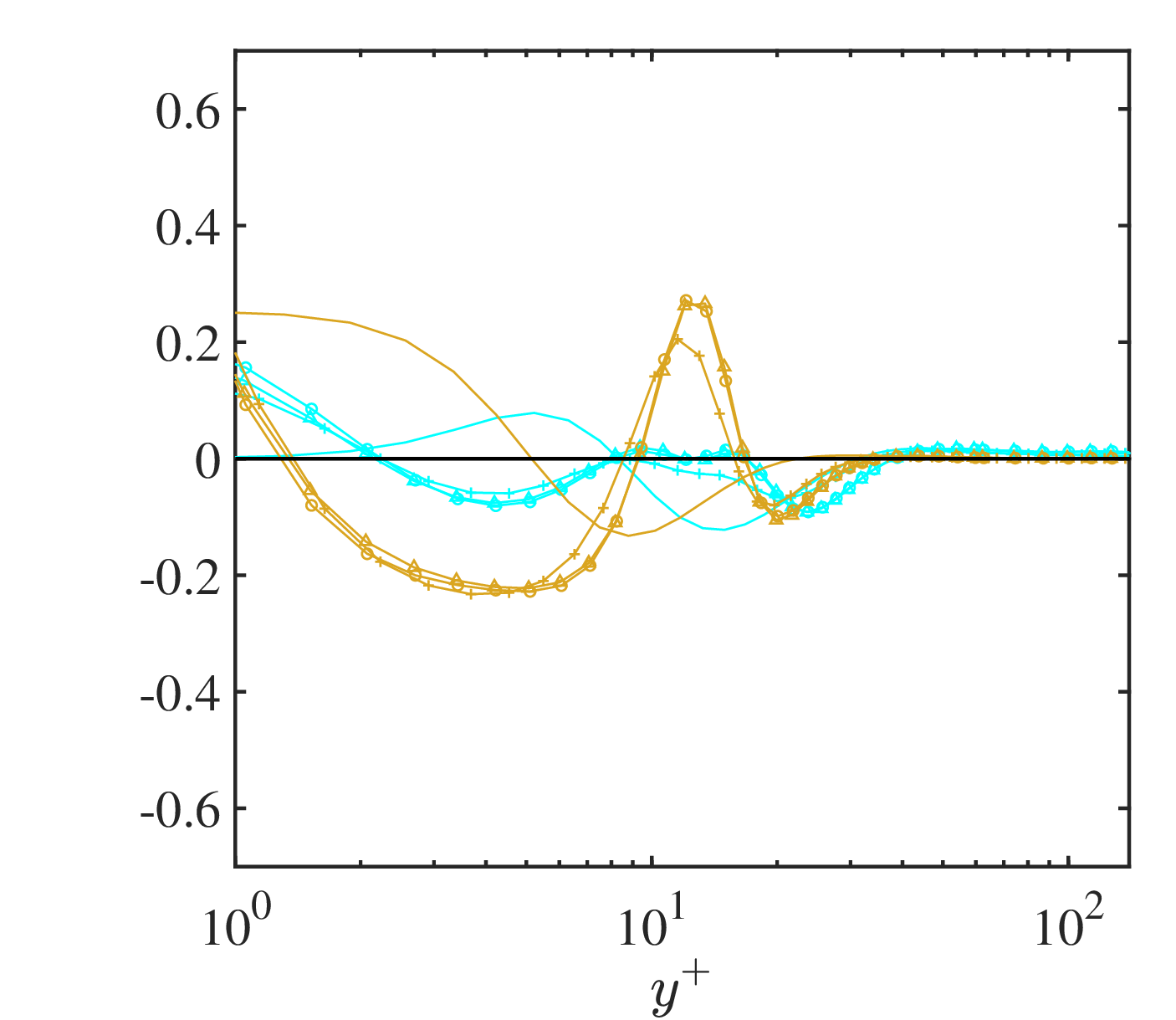}
			\put(0,81){(\textit{b})}
		\end{overpic}
	}
	\subfigure{
		\begin{overpic}
			[scale=0.23]{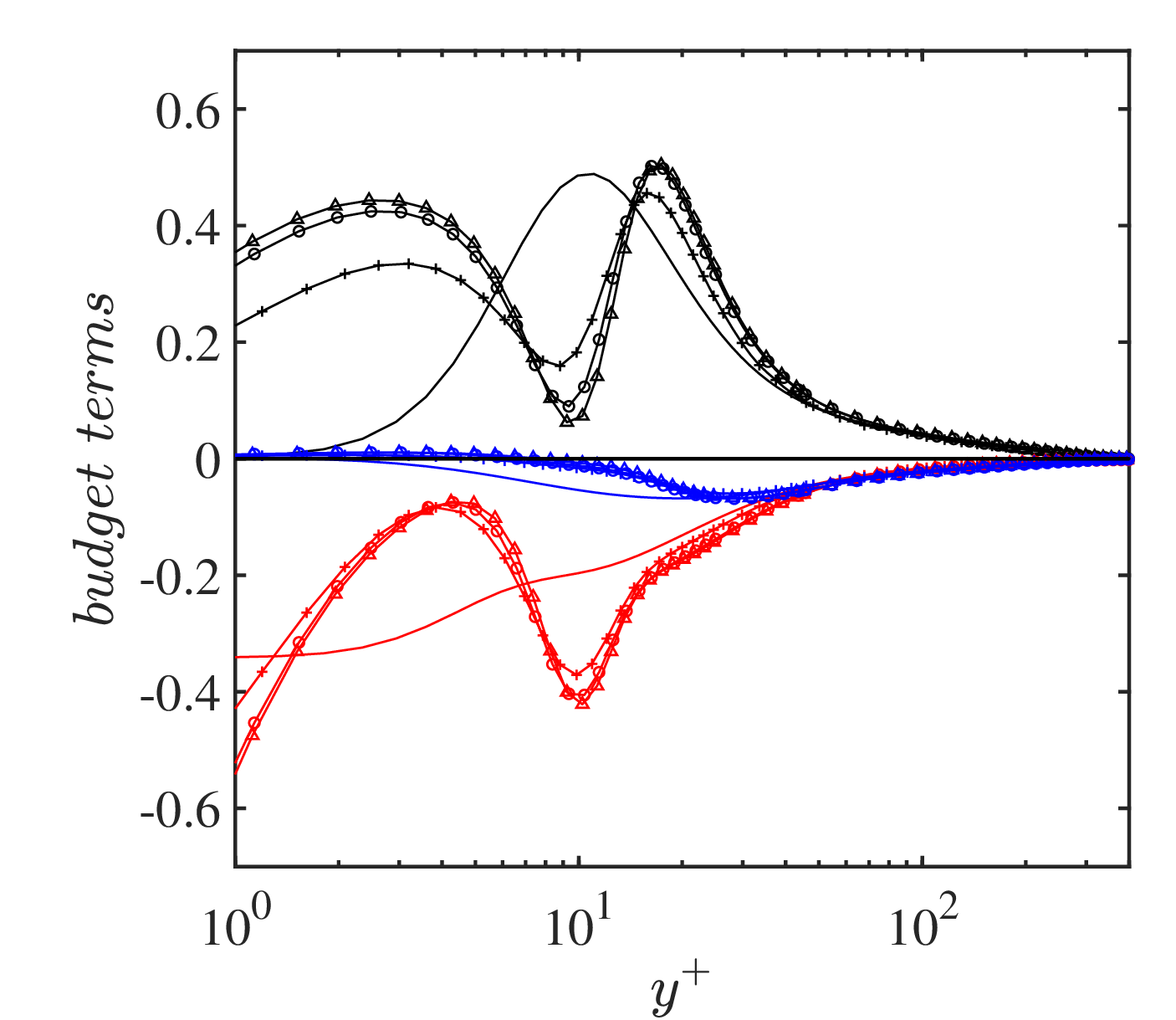}
			\put(0,81){(\textit{c})}
		\end{overpic}
		\begin{overpic}
			[scale=0.23]{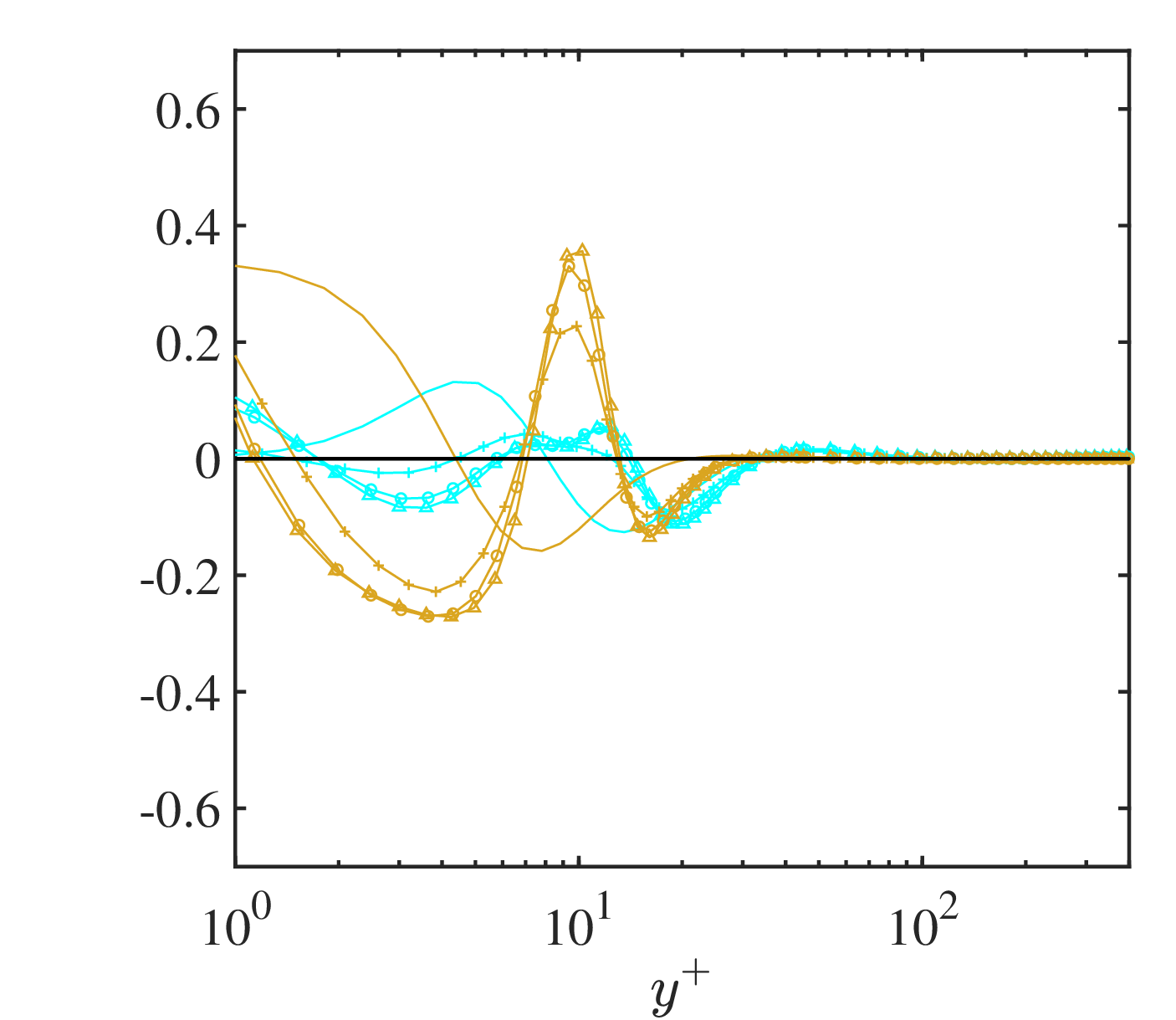}
			\put(0,81){(\textit{d})}
		\end{overpic}
	}
	\subfigure{
		\begin{overpic}
			[scale=0.23]{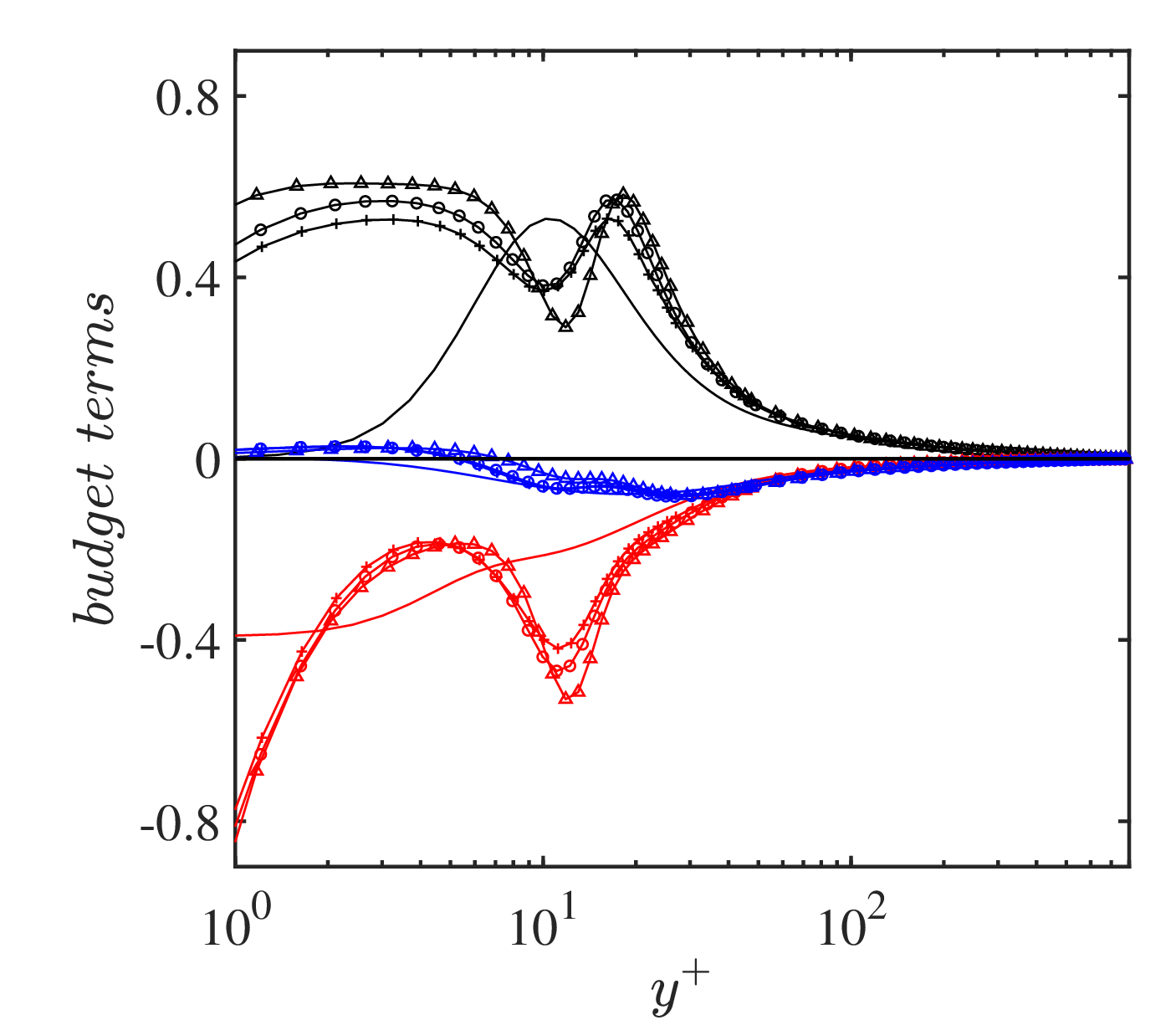}
			\put(0,81){(\textit{e})}
		\end{overpic}
		\begin{overpic}
			[scale=0.23]{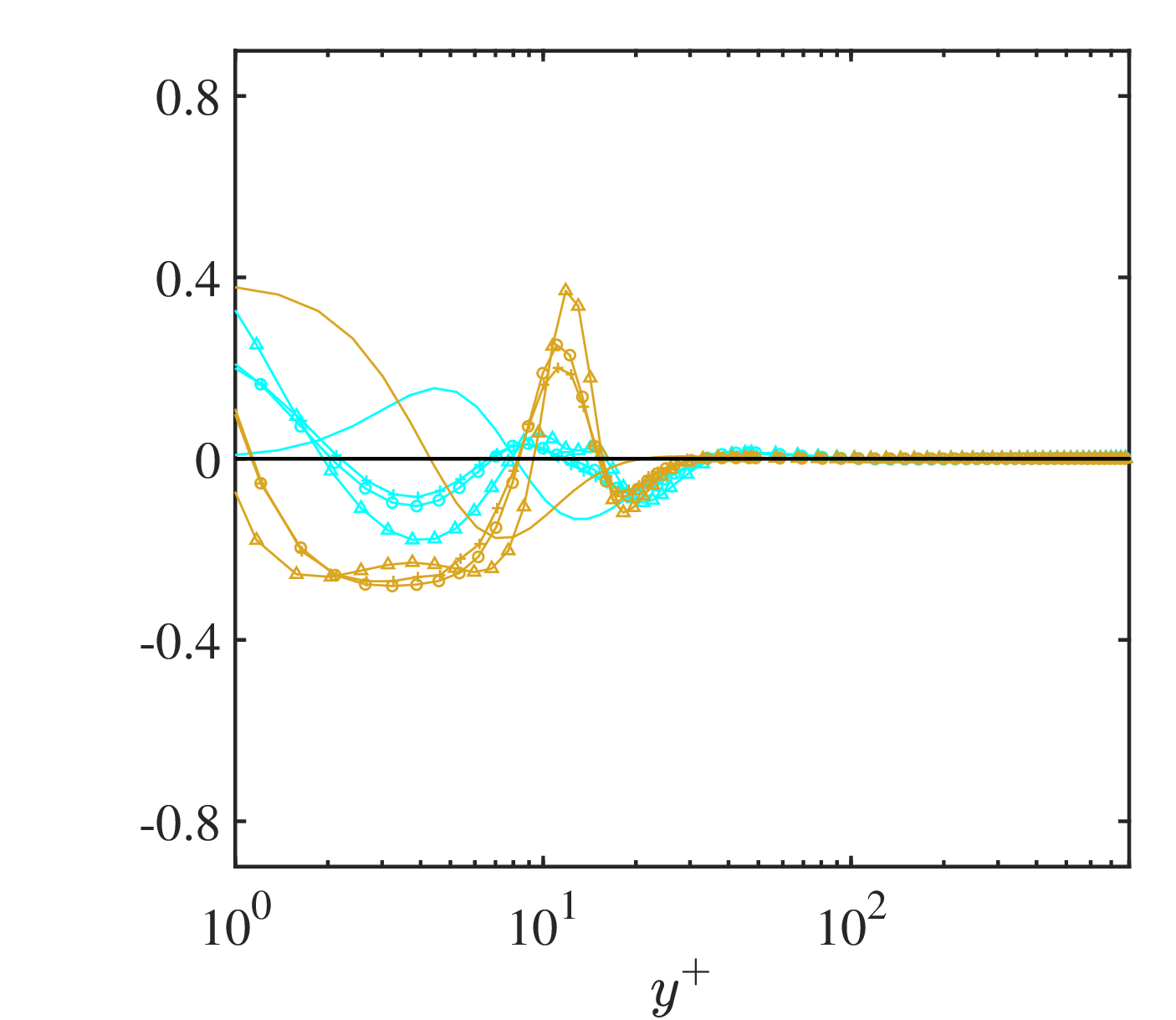}
			\put(0,81){(\textit{f})}
		\end{overpic}
	}

	\DeclareRobustCommand\mylabela{\tikz[baseline]{\draw[solid, black, thick] (0,0.5ex) -- (0.8,0.5ex);}}
	\DeclareRobustCommand\mylabelb{\tikz[baseline]{\draw[solid, red, thick] (0,0.5ex) -- (0.8,0.5ex);}}
	\DeclareRobustCommand\mylabelc{\tikz[baseline]{\draw[solid, blue, thick] (0,0.5ex) -- (0.8,0.5ex);}}
	\DeclareRobustCommand\mylabeld{\tikz[baseline]{\draw[solid, c00, thick] (0,0.5ex) -- (0.8,0.5ex);}}
	\DeclareRobustCommand\mylabele{\tikz[baseline]{\draw[solid, brown0, thick] (0,0.5ex) -- (0.8,0.5ex);}}
	\DeclareRobustCommand\mylabelf{\tikz[baseline]{\draw[solid, m00, thick] (0,0.5ex) -- (0.8,0.5ex);}}

	\caption{
		Wall-normal distributions of the budget terms of streamwise kinetic energy $\left\langle u^{\prime}u^{\prime}\right\rangle$ in eq (\ref{eq:budget-uu}).
		\mylabela, $P_{11}$; \mylabelc, $\Phi_{11}$; \mylabelb, $\varepsilon_{11}$; \mylabeld, $D_{11,t}$; \mylabele, $D_{11,\nu}$.
		(\textit{a})(\textit{b}) C180, (\textit{c})(\textit{d}) C550, (\textit{e})(\textit{f}) C1000.
		Lines without markers: cases with suffix '-0'; plus signs: with suffix '-1'; circles: with suffix '-2'; triangles: with suffix '-3'.
	}
	\label{fig:budget-uu}
	
\end{figure}

Figure \ref{fig:budget-uu} illustrates the wall-normal distributions of the budget terms on the right-hand side of eq (\ref{eq:budget-uu}).
Among these budget terms, the production $P_{11}$ plays a crucial role, consistently remaining positive across various heights.
In contrast, the terms $D_{11,t}$ and $D_{11,\nu}$ are relatively smaller and primarily represent the transport of turbulent kinetic energy in the wall-normal direction.
The dissipation term, $\varepsilon_{11}$, remains negative at different heights, counterbalancing the production $P_{11}$.
Additionally, unlike the $\Phi_{22}$ associated with wall-normal velocity fluctuations, the redistribution term $\Phi_{11}$ has a trivial impact on the streamwise turbulent kinetic energy.
In the uncontrolled cases, the production term $P_{11}$ gradually increases with height, reaching a peak around $y^{+}=10\sim15$. 
This height is similar to the $u_{rms}$ peak in the near-wall region, as depicted in figure \ref{fig:u-rms}.
After applying control, this peak disappears and transforms into a trough. 
The production term $P_{11}$ in the viscous sublayer significantly increases compared to the uncontrolled case. 
In the buffer layer, $P_{11}$ initially decreases and reaches the trough, then increases with height, with a second peak appearing around $y^{+}=20$.
It can be observed that after applying the DRL model, the trend of $P_{11}$ changes in a manner highly consistent with the changes in $u_{rms}$.
Notably, the turbulent production $P_{11}$ of the streamwise turbulent kinetic energy mainly consists of two parts: $\left\langle -u^{\prime}v^{\prime}\right\rangle$ and ${dU}/{dy}$.
The latter can be considered as a result of changes in Reynolds stress. 
Therefore, we can infer that the significant changes in $u_{rms}$ caused by the DRL-based control are primarily due to alterations in Reynolds shear stress.

\begin{figure}
	\centering
	\begin{overpic}
		[scale=0.32]{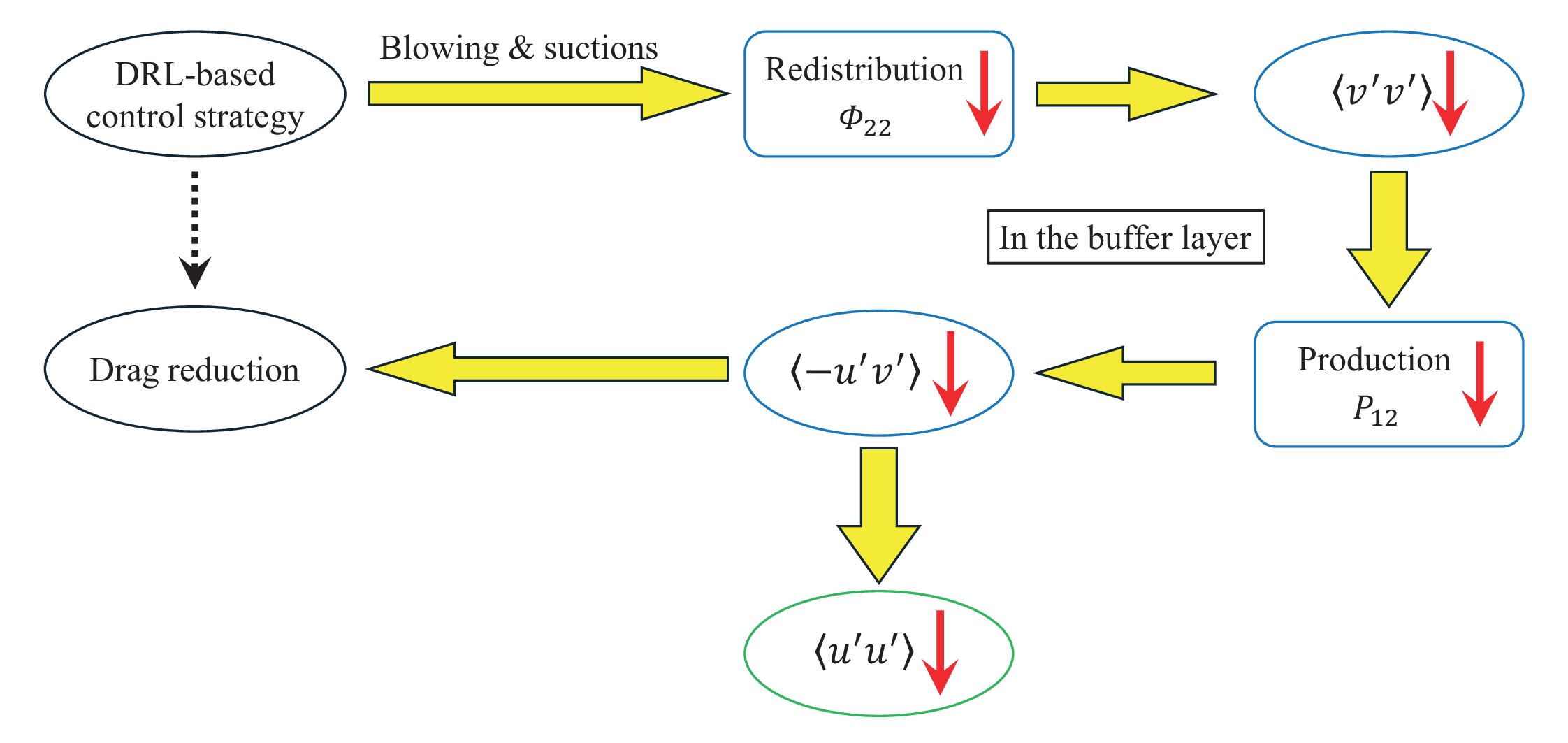}
	\end{overpic}
	\caption{
		Schematic diagram of the dynamic mechanism through which DRL-based control strategies influence skin friction.
	}
	\label{fig:sketch-budget}
\end{figure}

In summary, figure \ref{fig:sketch-budget} illustrates the dynamic mechanism through which DRL-based control strategies influence skin friction.
The application of wall blowing and suction, directed by the DRL models, effectively suppresses the near-wall self-sustaining process, thereby leading to smoother velocity streaks.
This suppression manifests as a decrease in the redistribution term of wall-normal turbulent kinetic energy within the buffer layer, consequently reducing wall-normal velocity fluctuations.
The reduction in $\left\langle v^{\prime}v^{\prime}\right\rangle $ further diminishes the production term of Reynolds stress, resulting in a decrease in $\left\langle -u^{\prime}v^{\prime}\right\rangle $.
Ultimately, this decline in Reynolds stress results in a reduction of skin friction. 
Moreover, the weakening of streamwise velocity fluctuations can also be attributed to the decrease in Reynolds stress.
When the range of blowing and suction velocities is expanded, the aforementioned effects are amplified, leading to an increase in the extent of drag reduction.
Conversely, as the Reynolds number rises and drag reduction diminishes, these trends are reversed.

\section{Summary and conclusions}\label{sec:conc}

This study employs deep reinforcement learning (DRL) to develop control strategies, aimed at reducing skin friction in DNS of turbulent channel flows at high Reynolds numbers.
Utilizing the TD3 framework, DRL predictions regulated wall blowing and suction velocities, with streamwise velocity fluctuations at $y^{+}=15$ serving as the state input for the DRL agent.

The DRL-based control strategies achieved significant drag reduction across various Reynolds numbers, with maximum reduction rates of $35.6\%$ at $Re_{\tau}\thickapprox180$, $30.4\%$ at $Re_{\tau}\thickapprox550$, and $27.7\%$ at $Re_{\tau}\thickapprox1000$.
These results demonstrate superior drag reduction compared to traditional opposition control. 
As the range of blowing and suction velocities was extended, the drag reduction rates improved.
Conversely, the effectiveness of DRL-based control decreased with higher Reynolds numbers, similar to opposition control methods. 
Further statistics indicate that the impact of DRL-based control on velocity fluctuations is limited to the near-wall region, with minimal effects on the outer region.
Unlike opposition control, the wall blowing and suction velocities are more strongly correlated with the near-wall streamwise velocity fluctuations compared to the wall-normal component, owing to the $u^{\prime}$ input state of the DRL model.

According to the virtual wall theory \citep{hammond1998observed}, the height of the virtual wall and the residual Reynolds stress on it are key indicators of drag reduction from a structural kinematic perspective. 
Compared to opposition control, the DRL model achieves higher drag reduction by elevating the virtual wall through blowing and suction.
When the range of these actions is expanded, the virtual wall height increases, and residual Reynolds stress decreases, leading to further drag reduction. 
In contrast, an increase in Reynolds number significantly raises residual Reynolds stress, disrupting the virtual wall’s effectiveness and resulting in decreased drag reduction rates.
The contribution of residual Reynolds stress mainly arises from the amplitude modulation of large-scale structures, rather than the superposition effect.
The footprint of outer large-scale structures is blocked above the virtual wall, while residual fluctuations on the virtual wall manifest as clusters of small-scale structures after DRL control.
These small-scale fluctuations tend to be distributed beneath large-scale high-speed regions, indicating that the virtual wall’s blockage is mainly disrupted in these areas.

On the other hand, analyzing the budget equations elucidates the dynamic mechanisms through which DRL-based control strategies impact skin friction.
Our observations indicate that the DRL models primarily reduce skin friction by inhibiting the redistribution term of wall-normal turbulent kinetic energy.
This effect manifests as the suppression of the near-wall self-sustaining mechanism, resulting in smoother near-wall streaks. 
The reduction in the redistribution term leads to decreased wall-normal velocity fluctuations in the buffer layer, thereby diminishing the turbulent production of Reynolds stress.
This chain of effects further weakens the Reynolds shear stress, ultimately reducing skin friction.
Notably, when the range of blowing and suction velocities is extended, these effects are amplified, leading to even greater drag reduction.
Conversely, an increase in the Reynolds number has the opposite effect, counteracting the benefits provided by the DRL-based strategies.




\backsection[Funding]{
We gratefully acknowledge the financial support from the Max Planck Society, the German Research Foundation(DFG) from grants 521319293 and 540422505, 550262949, and the Daimler and Benz foundation. 
We also thank the HPC systems of Max Planck Computing and Data Facility(MPCDF) for the allocation of the computational time.
The authors gratefully acknowledge the Gauss Centre for Supercomputing e.V. (www.gauss-centre.eu) for funding this project by providing computing time on the GCS Supercomputers SuperMUC-NG at Leibniz Supercomputing Centre (www.lrz.de) and JUWELS \citep{JUWELS} at Jülich Supercomputing Centre (JSC).
We gratefully acknowledge the grant WBS A-8001172-00-00 from the Ministry of Education, Singapore.
}

\backsection[Declaration of interests]{The authors report no conflict of interest.}

\backsection[Author ORCID]{Zisong Zhou, https://orcid.org/0000-0003-3708-1273; Mengqi Zhang, https://orcid.org/0000-0002-8354-7129; Xiaojue Zhu, https://orcid.org/0000-0002-7878-0655.}

%

\bibliographystyle{jfm}
\bibliography{main}

 -----------------------------------------------------------------------------------------------------

\appendix

\section{}\label{sec:supplementary}

This appendix discusses the drag reduction performance of the DRL-driven control strategy based on varied input states and reward, as well as the performance of the trained control strategy across different resolutions and Reynolds numbers.

To evaluate the drag reduction performance of the DRL-driven control strategy under varied input states and rewards, we conduct three test cases, summarized in table \ref{tab:supplementary_training}.
These cases are all based on C550-3, with consistent parameters except for changes in input states and rewards. 
In C550-3-drag, the reward is modified to drag reduction rate, contrasting with C550-3; in C550-3-v15, the input state is changed to $v^{\prime}$ at $y^{+}=15$; and in C550-3-u20, the input state is changed to $u^{\prime}$ at a higher position $y^{+}=20$.
We observe that each modified case converged within 10 episodes, and we select the models at 20 episodes, consistent with C550-3. 
The drag reduction results are presented in table \ref{tab:supplementary_training}. 
The drag reduction rate in C550-3-drag nearly collapses with C550-3, suggesting that similar control performance can be achieved regardless of whether TKE or total drag is used as the optimization function.
In C550-3-v15, where wall-normal velocity fluctuations are used as the input state, a slight decrease in drag reduction rate is observed. 
In C550-3-u20, with streamwise velocity fluctuations at a higher position, ${Re}_{\tau}$ increases, accompanied by a smaller drag reduction rate. 
These two cases indicate that despite variations in sensing parameters, the DRL strategy remains effective in developing flow control strategies based on the selected input variables.

\begin{table}
	\begin{center}
		\def~{\hphantom{0}}
		\begin{tabular}{lclcc}
			Cases          & Input states                       & Rewards                   & ${Re}_{\tau}$    &  $DR(\%)$           \\[4pt]
			C550-3         & $u^{\prime}\mid_{y^{+}=15}$        & TKE reduction rate        & $454.0$          &  $30.4$             \\
			C550-drag      & $u^{\prime}\mid_{y^{+}=15}$        & Drag reduction rate       & $454.4$          &  $30.3$             \\
			C550-v15       & $v^{\prime}\mid_{y^{+}=15}$        & TKE reduction rate        & $459.2$   	   &  $28.8$             \\
			C550-u20       & $u^{\prime}\mid_{y^{+}=20}$        & TKE reduction rate        & $471.7$          &  $24.9$             \\
		\end{tabular}
		\caption{Drag reduction results under different input states and rewards.}
		\label{tab:supplementary_training}
	\end{center}
\end{table}

Furthermore, we tested the performance of the trained control strategy across different resolutions and Reynolds numbers. 
In the first case, we applied the model trained in C550-3 to control a flow field around $Re_{\tau}\thickapprox550$ with the streamwise and spanwise grids refined by a factor of $2$. 
We found that, after grid refinement, the drag reduction rate collapsed with the result from case C550-3. 
This suggests that grid resolution has a trivial effect on the drag reduction performance of the DRL-derived control policy. 
In the second case, we applied the model trained in case C1000-3 to control a flow field around $Re_{\tau}\thickapprox550$. 
The resulting drag reduction rate was 28.9\%, only slightly lower than the 30.4\% achieved in C550-3. 
This indicates that the control policy trained in C1000-3 remains effective when the Reynolds number is reduced.

\end{document}